\documentclass[aps,pra,twocolumn,showpacs,superscriptaddress,floatfix,10pt]{revtex4-1}
\usepackage{framed}
\usepackage{graphicx}
\usepackage{amsmath}
\usepackage{epsfig}
\usepackage{helvet}
\usepackage{amssymb}
\usepackage{framed}
\usepackage{bbold}
\usepackage{subfigure}

\newcommand{\bra}[1]{\mbox{$\langle #1 |$}}
\newcommand{\ket}[1]{\mbox{$| #1 \rangle$}}

\newcommand{\proj}[1]{\mbox{$| #1 \rangle\langle #1 |$} }

\def\rad{R}

\allowdisplaybreaks[3]
\def\letter{paper } 
\def\appendix{Appendix }

\def\CFT{\mbox{\tiny CFT}}

\def\aUV{a_{\mbox{\tiny UV}}}
\def\inn{\mbox{\tiny in}}
\def\out{\mbox{\tiny out}}

\def\AAdS3{\mbox{\tiny AdS$_3^p$}}
\def\AdS3{\mbox{\tiny AdS$_3$}}
\def\HH2{\mbox{\tiny H$_2^p$}}
\def\ddS2{\mbox{\tiny dS$_2^p$}}
\def\LL2{\mbox{\tiny L$_2^p$}}
\def\MMplus{\mbox{\tiny $\mathcal{M}_{+}^p$}}
\def\MMmin{\mbox{\tiny $\mathcal{M}_{-}^p$}}
\def\MMzero{\mbox{\tiny $\mathcal{M}^p$}}

\def\M12{\mbox{\tiny $\mathbb{R}_{1,2}$}}
\def\H2{\mbox{\tiny H$_2$}}
\def\dS2{\mbox{\tiny dS$_2$}}
\def\L2{\mbox{\tiny L$_2$}}
\def\Mplus{\mbox{\tiny $\mathcal{M}_{+}$}}
\def\Mmin{\mbox{\tiny $\mathcal{M}_{-}$}}

\begin{document}

\title{Geometric interpretation of the multi-scale entanglement renormalization ansatz}
\author{Ashley Milsted}
\affiliation{Perimeter Institute for Theoretical Physics, Waterloo, Ontario N2L 2Y5, Canada}  
\author{Guifre Vidal}
\affiliation{Perimeter Institute for Theoretical Physics, Waterloo, Ontario N2L 2Y5, Canada}  \date{\today}

\begin{abstract}
The multi-scale entanglement renormalization ansatz (MERA) is a tensor network representation for ground states of critical quantum spin chains, with a network that extends in an additional dimension corresponding to scale. Over the years several authors have conjectured, both in the context of holography and cosmology, that MERA realizes a discrete version of some geometry. However, while one proposal argued that the tensor network should be interpreted as representing the hyperbolic plane, another proposal instead equated MERA to de Sitter spacetime. In this \letter we show, using the framework of path integral geometry [A. Milsted, G. Vidal, arXiv:1807.02501], that MERA on the real line (and finite circle) can be given a rigorous interpretation as a two-dimensional geometry, namely a light sheet (respectively, a light cone). Accordingly, MERA describes neither the hyperbolic plane nor de Sitter spacetime. However, we also propose euclidean and lorentzian generalizations of MERA that correspond to a path integral on these two geometries.
\end{abstract}


\maketitle

The multi-scale entanglement renormalization ansatz (MERA) \cite{MERA1,MERA2,MERA3,MERA4,MERA5,MERA6,MERA7} is a tensor network originally proposed as a variational ansatz for ground states of quantum spin chains that has over the years found a wider range of applications, including error correction \cite{error1,error2}, machine learning \cite{machine1, machine2, machine3, machine4}, statistical physics \cite{TNRyieldsMERA}, holography \cite{H1,H2,dS1,dS2}, and cosmology \cite{dS3,dS4}. MERA spans an additional dimension, corresponding to scale, and is particularly well-suited to describe the ground state of a critical quantum  spin chain, which we can think of as a lattice version of 1+1 conformal field theory (CFT). Moreover, the network of tensors resembles a discrete version of a hyperbolic geometry. Based on these and other observations Swingle conjectured \cite{H1,H2}, in a pioneering contribution that initiated a fruitful, lasting interdisciplinary discussion connecting questions in quantum gravity with tensor networks \cite{H1,H2, dS1,dS2, dS3,dS4, EHM1,EHM2, HAPPY, Random1, Random2, cMERA, cMERAholo1, cMERAholo2, cMERAholo3, cMERAholo4, cMERAholo5, cMERAholo6, cMERAholo7, cMERAholo8, cMERAholo9, TNR, TNRholo1, TNRholo2, TNRholo3, TNRholo4, TNRholo5, TNRholo6}, that MERA was a lattice realization of the AdS/CFT correspondence \cite{AdSCFT1,AdSCFT2,AdSCFT3}. Specifically, MERA on the real line would describe a time slice of the Poincare patch of AdS$_{3}$, which corresponds to the hyperbolic plane H$_2^p$, with line element
\begin{equation} \label{eq:H2p}
dl(\eta,r)^2
 = \left(\frac{R}{\eta}\right)^2\left(d\eta^2 + dr^2\right)
~~~\mbox{(hyperbolic plane)} 
\end{equation}
However, MERA is a quantum circuit that implements an entangling evolution (with time $\eta$ corresponding to the renormalization group or scale direction) and, as such, is naturally equipped with a causal structure and lorentzian signature \cite{MERA1,MERA2}. This prompted Beny and several other authors, both in the context of holography \cite{dS1,dS2} and cosmology \cite{dS3,dS4}, to conjecture that MERA on the real line should be interpreted instead as a Poincare patch of de Sitter spacetime dS$_2^p$, with line element 
\begin{equation} \label{eq:dS2p}
dl(\eta,r)^2
= \left(\frac{R}{\eta}\right)^2\left(-d\eta^2 + dr^2\right)
~~~~~~\mbox{(de Sitter)} 
\end{equation}

\begin{figure}
\includegraphics[width=7cm]{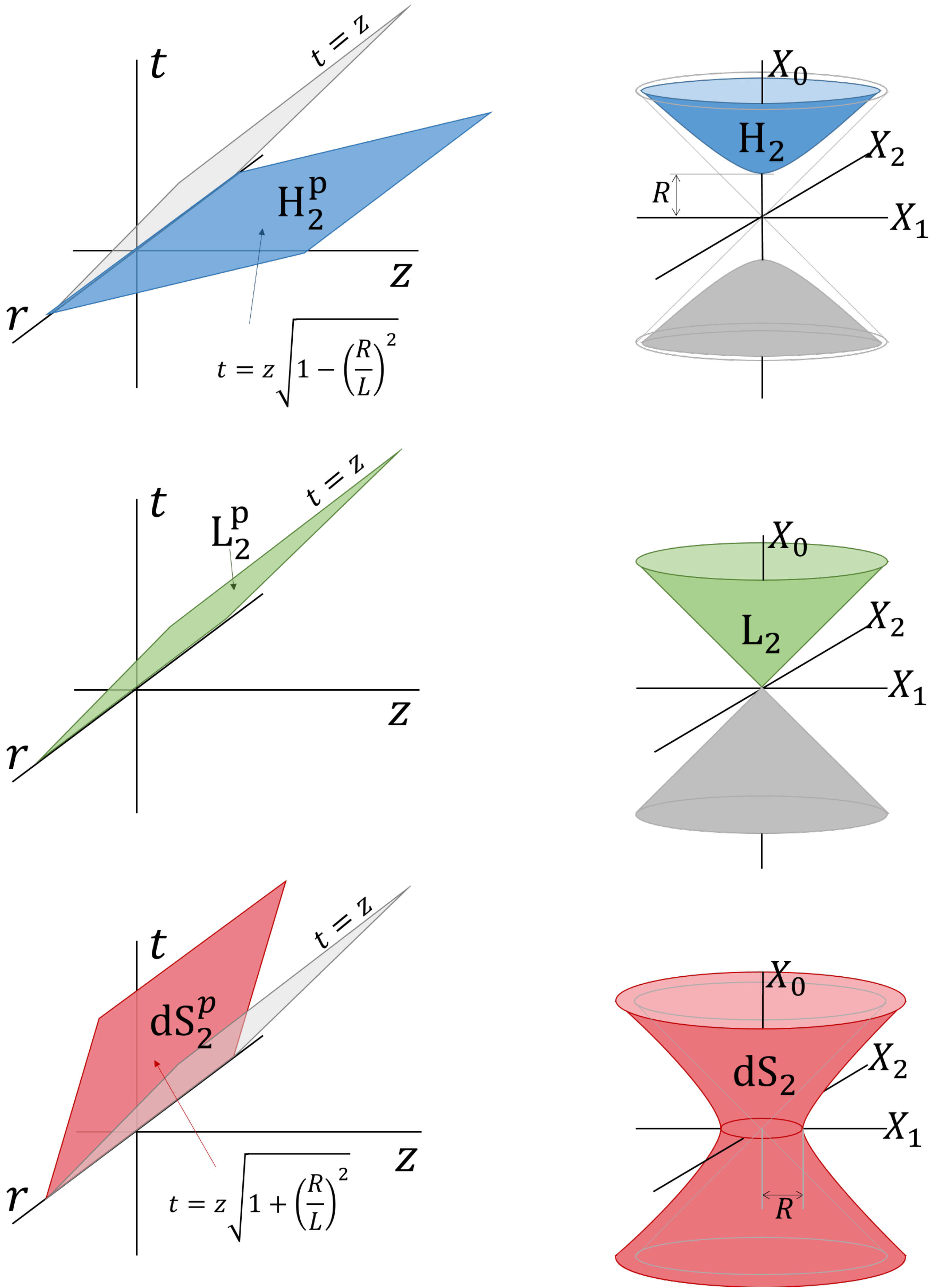}
\caption{
(Left) Three candidate geometries for MERA on the real line, as embedded in the Poincare patch of AdS$_3$ with metric $dl^2(t,z,r)=(-dt^2+dz^2+dr^2)/(z/L)^2$, namely the hyperbolic plane H$_2^p$, light sheet L$_2^p$, and Poincare de Sitter dS$_2^p$ in Eqs. \eqref{eq:H2p}-\eqref{eq:L2p} \cite{Supplemental}.
(Right) Three candidate geometries for MERA on the circle, as embedded in Minkowski $\mathbb{R}^{1,2}$, namely the hyperbolic disk H$_2$, light cone L$_2$, and de Sitter dS$_2$, see Eq. \eqref{eq:embedding1}-\eqref{eq:embedding3}.
\label{fig:geometries} 
}
\end{figure}

Determining which geometry, if any, MERA may describe is thus important in order to assess the potential of this tensor network as a theoretical and numerical framework in both quantum gravity and cosmology. In this \letter we establish that, when regarded as a discrete version of a CFT path integral, the geometry of MERA on the real line is that of a \textit{light sheet} L$_2^p$ with null scale direction $\eta$, 
\begin{equation} \label{eq:L2p}
~~dl(\eta,r)^2 
 = \frac{dr^2}{\eta^2} ~~~~~~~~\mbox{(light sheet)} 
\end{equation}
In other words, from the path integral perspective proposed in Refs. \cite{TNConfTrans, TNPathInt}, MERA describes neither the hyperbolic plane nor de Sitter spacetime, but an intermediate geometry with degenerate signature, see Fig. \ref{fig:geometries}. 
This result follows from the path integral geometry of MERA \textit{on the circle}, which we determine by studying a layer $\mathcal{W}$ of MERA optimized for the ground state of a critical quantum spin chain. Specifically, we demonstrate that $\mathcal{W}$ acts on the low energy states of a periodic spin chain simply as the identity map $\mathbb{1}$, which is also the map enacted by a path integral on an annulus of a \textit{light cone} L$_2$. In particular, $\mathcal{W}$ does not implement euclidean time evolution $e^{-\eta H}$ (nor real time evolution $e^{-i\eta H}$), as it would if it were a path integral on an annulus of the hyperbolic disk H$_2$ (respectively, of de Sitter spacetime dS$_2$). 

We then propose two generalizations of MERA that may be useful toy models in holography and cosmology: (\textit{i}) by interspersing layers $\mathcal{W}$ with layers of \textit{euclideons} (tensors that implement euclidean time evolution $e^{-\eta H}$) we obtain a new tensor network representation of the CFT ground state corresponding to a path integral in the hyperbolic plane H$_2$; (\textit{ii}) by interspersing layers $\mathcal{W}$ with layers of \textit{lorentzions} (tensors that implement real time evolution $e^{-i\eta H}$) we produce a tensor network representation of the CFT ground state corresponding to a path integral in de Sitter spacetime dS$_2$.

\textit{Strategy.---} Following Refs. \cite{TNConfTrans,TNPathInt}, our plan is to regard the MERA network for the ground state of a critical spin chain as a discrete, approximate CFT path integral on some geometry, and to then assign that geometry to MERA. Needless to say, for this strategy to make sense, the tensor network must behave as a CFT path integral in the first place. In the continuum, the path integral on a strip defines a linear map $V:\mathcal{H}^{\CFT}_{\inn} \rightarrow \mathcal{H}^{\CFT}_{\out}$ between the Hilbert spaces $\mathcal{H}^{\CFT}_{\inn}$ and  $\mathcal{H}^{\CFT}_{\out}$ at the boundaries $\Sigma_{\inn}$ and $\Sigma_{\out}$ of the strip.  This linear map $V$ enacts a conformal transformation that depends both on the geometry of the strip and on how we identified the two Hilbert spaces. 
On the other hand, a layer $\mathcal{W}$ of MERA also defines a linear map, also denoted by $\mathcal{W}$, between the Hilbert spaces of the two spin chains at its boundaries. We can then ask whether the map $\mathcal{W}$ matches the path integral map $V$ for some choice of strip geometry. It is highly non-trivial that the answer is affirmative --and not only for an optimized MERA, but also for a larger class of tensor networks described in Refs. \cite{TNConfTrans, TNPathInt}). 

The low energy states of a critical spin chain are in one-to-one correspondence with  states of a CFT \cite{Cardy, KooSaleur, SpinChain1, SpinChain2}. It is thus on these states that we want to compare the action of the linear maps $V$ and $\mathcal{W}$. 
In order to characterize the action of $\mathcal{W}$, we first need to explain how to relate the low energy states of the two spin chains in connects. Two difficulties lie ahead. First, in a numerical simulation of the real line (\textit{i.e.} infinite critical spin chain), we do not know how to reliably compute low energy states due to the absence of a mass gap in the energy spectrum. For this reason we turn to studying finite periodic spin chains instead. Second, $\mathcal{W}$ is a coarse-graining transformation that maps states of a spin chain of size $N$ to states of a spin chain with size $N/2$. It is therefore not obvious how two identify the two Hilbert spaces, which have very different dimension. Here we will use the techniques of Refs. \cite{Cardy, KooSaleur, SpinChain1, SpinChain2, SpinChain3} to relate each low energy state in the two spin chains through their corresponding low energy state in the CFT. With this identification, we will then be ready to numerically determine the action of $\mathcal{W}$ and to finally compare it to candidate CFT linear maps $V$. 

\textit{Path integral on an annulus.---} 
We start by describing the linear map $V$ for three candidate geometries for MERA on the circle. Consider Minkowski $\mathbb{R}^{1,2}$ with metric $dl^2 = -d{X_0}^2 + d{X_1}^2 + d{X_2}^2$ and the three two-dimensional manifolds given by the following constraints 
\begin{eqnarray} \label{eq:embedding1}
-{X_0}^2 + {X_1}^2 + {X_2}^2 &=& -\rad^2~~~(\mbox{H}_2)\\
-{X_0}^2 + {X_1}^2 + {X_2}^2 &=& 0~~~~~~~(\mbox{L}_2)\label{eq:embedding2}\\
-{X_0}^2 + {X_1}^2 + {X_2}^2 &=& \rad^2~~~~~(\mbox{dS}_2)\label{eq:embedding3}
\end{eqnarray}
see Fig. \ref{fig:geometries}. We introduce polar coordinates $r = \sqrt{x^2+y^2}$ and $\theta = \arctan (y/x)$ and use the above constraints to arrive at the metrics of the hyperbolic plane H$_2$, the light cone L$_2$, and the de Sitter spacetime dS$_2$, namely 
\begin{eqnarray}
dl_{\H2}(r,\theta)^2 &=& \frac{1}{(r/\rad)^2+1}dr^2 + r^2 d\theta^2, \label{eq:H2} \\
dl_{\L2}(r,\theta)^2 &=&  r^2 d\theta^2,  \label{eq:L2}\\
dl_{\dS2}(r,\theta)^2 &=& \frac{-1}{(r/\rad)^2-1}dr^2 + r^2 d\theta^2.\label{eq:dS2}  
\end{eqnarray}
Notice that the light cone L$_2$ is the limit of small radius $R$ of both H$_2$ and dS$_2$. For simplicity we specialize to the case $r \gg R$. A standard computation \cite{Supplemental} shows that the CFT path integral on an annulus with boundaries given by radial coordinate $r$ and $r/2$ produce the linear maps
\begin{eqnarray}
V_{\H2} &=& e^{-\frac{R}{r} H} ~~~ (\mbox{euclidean}) \label{eq:VH2}\\
V_{\L2} &=& e^{~0~ H} = \mathbb{1}~~~ (\mbox{null}) \label{eq:VL2}\\
V_{\dS2} &=& e^{-i\frac{R}{r} H} ~~~ (\mbox{lorentzian}) \label{eq:VdS2}
\end{eqnarray}
where $H \equiv L_0+\bar{L}_0 - c/12$ is the CFT Hamiltonian on the unit radius circle. Recall that simultaneous diagonalization of $H$ and the momentum operator $P = L_{0} - \bar{L}_{0}$,
\begin{eqnarray}
H \ket{\phi^{\CFT}_{\alpha}} = E_{\alpha} \ket{\phi^{\CFT}_{\alpha}},~~~P \ket{\phi^{\CFT}_{\alpha}} = P_{\alpha}\ket{\phi^{\CFT}_{\alpha}},~~~~~
\end{eqnarray}
yields energies $E_{\alpha}$ and momenta $P_{\alpha}$ that are given in terms of the CFT's central charge $c$, scaling dimensions $\Delta_{\alpha}$ and conformal spins $S_{\alpha}$ according to \cite{CFT1, CFT2, CFT3}
\begin{equation} \label{eq:E}
E_\alpha = \Delta_{\alpha}-\frac{c}{12}, ~~~~~~P_{\alpha} = S_{\alpha}.~~~
\end{equation}

\begin{figure}
\includegraphics[width=6cm]{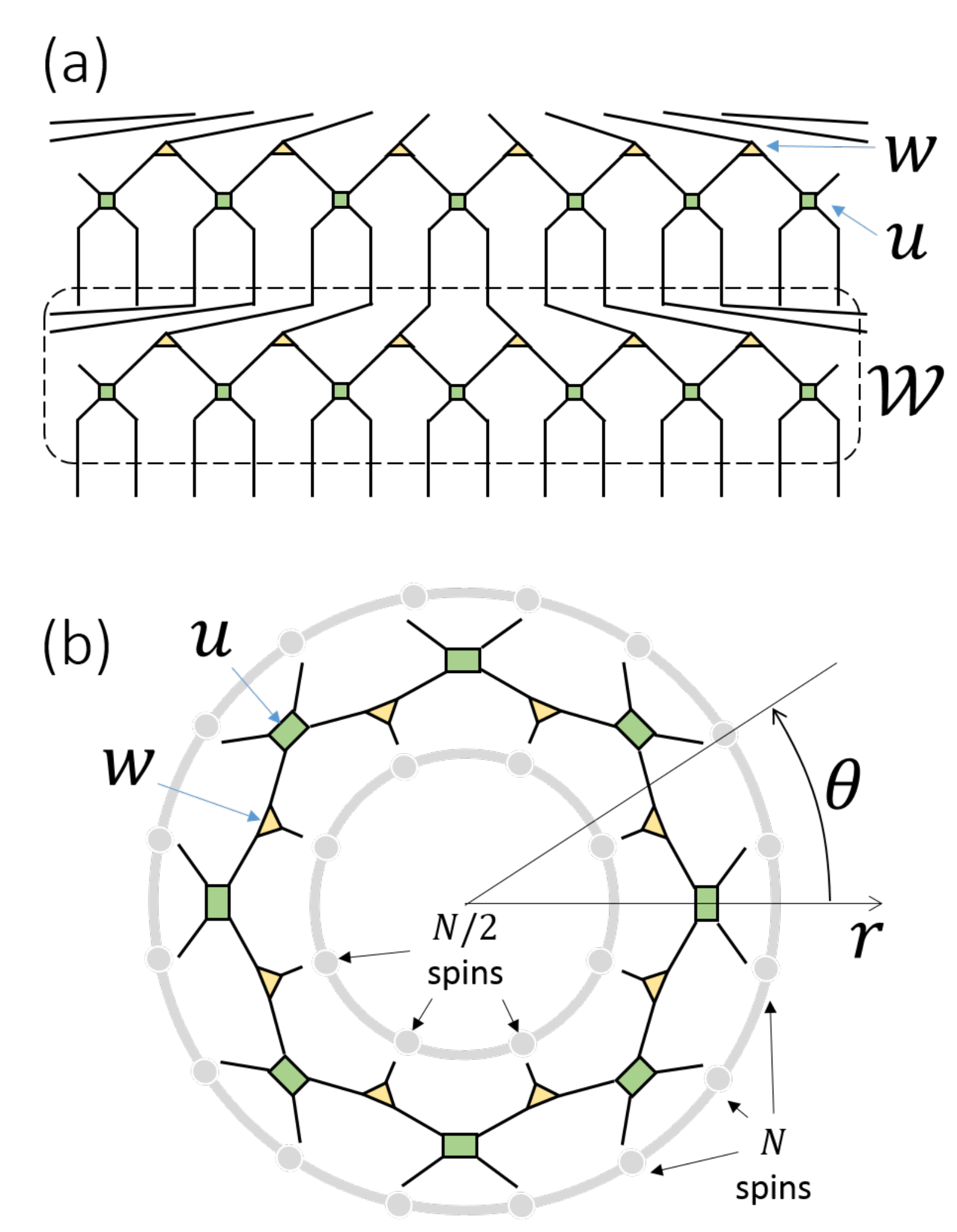}
\caption{ 
(a) The MERA is made of disentanglers $u$ and isometries $w$ organized in infinite layers $\mathcal{W}$.
(b) A finite, periodic layer $\mathcal{W}$ defines a liner map from the Hilbert space of a periodic spin chain with $N$ spins to that of a smaller periodic spin chain with $N/2$ spins.
\label{fig:layer} 
}
\end{figure}

\textit{Map $\mathcal{W}$ between low energy states of two spin chains.---} Consider now a critical spin chain Hamiltonian $H^{(\infty)} \equiv \sum_{n=\infty}^{\infty} h_n$ on the (discrete) real line, where $h_n$ is a local term that acts on a small neighbourhood of spin $n$. Following Refs. \cite{MERA5,MERA6}, we obtain an approximate, scale invariant MERA representation of the ground state of $H^{(\infty)}$. The network is made of infinite layers $\mathcal{W}$ of tensors called disentanglers $u$ and isometries $w$, see Fig. \ref{fig:layer}(a). Using the same optimized tensors $u$ and $w$, we build a finite, periodic layer $\mathcal{W}$ that defines a map between two periodic chains made of $N$ and $N/2$ spins, see Fig. \ref{fig:layer}(b).  
The Hamiltonian $H^{(N)} \equiv \sum_{n=1}^N h_n$ on the periodic chain of size $N$ and the one-site translation operator $T^{(N)}$
can be simultaneously diagonalized,
\begin{eqnarray}
H^{(N)} \ket{\phi^{N}_{\alpha}} = E_{\alpha}^{N} \ket{\phi^{N}_{\alpha}},~~~T^{(N)} \ket{\phi^{N}_{\alpha}} = e^{-i\frac{2\pi}{N}P_{\alpha}^{N}}\ket{\phi^{N}_{\alpha}}.~~~~
\end{eqnarray}
Let $\aUV$ denote the lattice spacing, so that the radius of the periodic spin chain is $r = N\aUV/2\pi$. At low energies, after suitably normalizing $h_n$ \cite{normalization}, the lattice Hamiltonian $H^{(N)}$ is a rescaled version of the CFT Hamiltonian $H$ \cite{Cardy}, namely $H^N \approx H/r$, and also $P^{N} = P$, so that
\begin{equation}
E_{\alpha}^{N} \approx \frac{1}{r} E_{\alpha}, ~~~P_{\alpha}^{N} = S_{\alpha}
\end{equation}
We can thus identify each eigenstate $\ket{\phi^{N}_{\alpha}}$ on the lattice with a corresponding CFT state $\ket{\phi^{\CFT}_{\alpha}}$, that is $\ket{\phi^{N}_{\alpha}} \sim \ket{\phi^{\CFT}_{\alpha}}$. Repeating the same procedure for the spin chain of size $N/2$, we arrive at an analogous identification $\ket{\phi^{N/2}_{\alpha}} \sim \ket{\phi^{\CFT}_{\alpha}}$, which results in an identification between the low energy states of the two spin chains, 
\begin{equation}
 \ket{\phi_{\alpha}^{N}} \sim \ket{\phi_{\alpha}^{N/2}}.
\end{equation}
Our goal is to then compute the matrix elements
\begin{equation} 
\mathcal{W}_{\alpha\beta} \equiv \bra{\phi^{N/2}_{\beta}} \mathcal{W} \ket{\phi^{N}_{\alpha}}
\end{equation}
and, by comparison with the maps in Eqs. \eqref{eq:WH2}-\eqref{eq:VdS2}, determine whether they match any of the following options
\begin{eqnarray}
(\mathcal{W})_{\alpha\beta} &\stackrel{\rm ?}{\approx} & (V_{\H2})_{\alpha\beta} = \delta_{\alpha \beta} e^{-\frac{R}{r} E_{\alpha}} ~~~~ (\mbox{euclidean}) \label{eq:WH2}\\
(\mathcal{W})_{\alpha\beta} &\stackrel{\rm ?}{\approx} & (V_{\L2})_{\alpha\beta} =\delta_{\alpha \beta}~~~~~~~~~~~~~~~~ (\mbox{null})  \label{eq:WL2}\\
(\mathcal{W})_{\alpha\beta} &\stackrel{\rm ?}{\approx} & (V_{\dS2})_{\alpha\beta} =\delta_{\alpha \beta} e^{-i\frac{R}{r} E_{\alpha}} ~~~ (\mbox{lorentzian}) \label{eq:WdS2}
\end{eqnarray}

\textit{Numerical characterization of $\mathcal{W}$.---} As a concrete example, we have considered the critical Ising model. Given the local Hamiltonian term $h'_n = -\left( \sigma_n^{x}\sigma_{n+1}^{x} + \sigma_{n}^{z}\right)$, we produce a term $h_n$ by suitably coarse-graining $h'_n$ \cite{Supplemental}. In the coarse-grained spin chain, each site is described by a vector space of dimension $\chi=8$ and $h_n$ acts on three consecutive sites. Upon diagonalization of $H^{(N)}$ for $N=8$ sites, we obtain a number of low energy states $\ket{\phi^{N}_{\alpha}}$, which we organize \cite{Cardy, KooSaleur, SpinChain1, SpinChain2} into the \textit{identity}, \textit{spin}, and \textit{energy density} conformal towers of the Ising CFT,
\begin{eqnarray}
&&\ket{\mathbb{1}^{N}},~\ket{T^{N}},~\ket{\bar{T}^N}, ~\ket{\partial T^{N}}, \cdots ~~~\mbox{(identity $\mathbb{1}$ tower)} ~~~~~~~\\
&&\ket{\sigma^N},~ \ket{\partial \sigma^N}, ~ \ket{\bar{\partial} \sigma^N}, \cdots ~~~~~~~~~~~~~~~\mbox{(spin $\sigma$ tower)} ~~~~~~~\\
&&\ket{\epsilon^N},~ \ket{\partial \epsilon^N}, ~ \ket{\bar{\partial} \epsilon^N}, \cdots ~~~~ \mbox{(energy density $\epsilon$ tower)}~~~~~~~ 
\end{eqnarray}
For instance, $\ket{\mathbb{1}^{N}}$ corresponds to the ground state of $H^{(N)}$, $\ket{\sigma^N}$ to its first excited state, and $\ket{T^N}$ is the (holomorphic) stress tensor state \cite{CFT1,CFT2,CFT3}. Proceeding similarly with a chain of $N/2=4$ sites, we obtain analogous states $\ket{\mathbb{1}^{N/2}}, \ket{T^{N/2}}, \cdots$. A first numerical check confirms that $\mathcal{W}$ is indeed diagonal in these states,
\begin{equation}
\mathcal{W}_{\alpha\beta} \approx  \delta_{\alpha\beta} f_{\alpha}
\end{equation}
in agreement with the three options \eqref{eq:WH2}-\eqref{eq:WdS2}. However, since $\mathcal{W}$ is (by construction) an isometric map, whose eigenvalues can only either vanish or be a complex phase, we have $|f_{\alpha}| = 0$ or $1$, which rules out the euclidean evolution \eqref{eq:WH2} corresponding to the hyperbolic disk H$_2$. 

To discriminate between null and lorentzian evolutions, Eqs. \eqref{eq:WL2} and\eqref{eq:WdS2}, we first need to fix the arbitrary complex phase $e^{i\varphi_{\alpha}^{N}}$ in each of state $\ket{\phi_{\alpha}^N}$ that appears when diagonalizing the Hamiltonian $H^{(N)}$ \cite{complex}, and similarly for $H^{(N/2)}$. Fortunately, we can use recently developed techniques \cite{SpinChain1, SpinChain2} based on the Koo-Saleur formula \cite{KooSaleur}, which provides a lattice version $H^{(N)} \sim L_n + \bar{L}_{-n}$ of the Virasoro generators $L_n$, $\bar{L}_{-n}$ of the CFT, to establish relations between low energy states in a given conformal tower (e.g. $H_2^{(N)}\ket{\mathbb{1}^N} \approx \sqrt{c/2} \ket{T^N}$) and in this way eliminate relative complex phases (e.g. $e^{i(\varphi_{\mathbb{1}}^{N} - \varphi_{T}^{N})}$) within each conformal tower. Moreover, relative complex phases between different towers can be eliminated using lattice version of CFT operators \cite{SpinChain3}. After carefully eliminating all these spurious complex phases \cite{Supplemental}, the resulting matrix coefficients $\mathcal{W}_{\alpha \beta}$ are seen to coincide with the identity $\delta_{\alpha \beta}$, displaying no dependence on the energies $E_{\alpha}$. For instance, on the 17 lowest energy eigenstates of the Ising model, $\mathcal{W}$ acts as the identity map up to corrections below $10^{-2}$ \cite{Supplemental}. 

\textit{Geometric interpretation of the MERA.---} We have thus numerically established that a periodic layer $\mathcal{W}$ of optimized MERA implements null time evolution on the low energy states of a periodic spin chain and therefore acts as if it was a CFT path integral on an annulus of the light cone L$_2$ in Eq. \eqref{eq:L2}. Returning to the real line, 
the path integral on a strip of the geometries \eqref{eq:H2p}-\eqref{eq:L2p} with boundaries $\eta$ and $2\eta$, where we identify the points in two boundaries points at constant value of the new space coordinate $x \equiv r R/(\aUV \eta)$,
produces the linear maps \cite{Supplemental}
\begin{eqnarray}
V_{\HH2} &=& 2^{-\frac{R}{\aUV}H+iD} ~~~ (\mbox{euclidean}) \label{eq:VHH2}\\
V_{\LL2} &=& 2^{iD} ~~~~~~~~~~~~~~~~~~~(\mbox{null}) \label{eq:VLL2}\\
V_{\ddS2} &=& 2^{-i\frac{R}{\aUV}H+iD} ~~~ (\mbox{lorentzian}) \label{eq:VddS2}
\end{eqnarray}
where now $H$ and $D$ are the CFT Hamiltonian and dilation operators on the real line \cite{Supplemental}. It is well known that an infinite layer $\mathcal{W}$ of MERA implements a rescaling transformation by a scale factor $2$ \cite{MERA1,MERA2,MERA3,MERA4,MERA5,MERA6}. Moreover, we have just seen on the circle that $\mathcal{W}$ does not implement time evolution. Therefore the infinite layer $\mathcal{W}$ of optimized MERA acts as $\mathcal{W} \approx 2^{iD}$, that is, as a CFT path integral on a strip of the light sheet L$_2^p$ of Eq. \eqref{eq:L2p}. The presence of the dilation operator $D$ in $\mathcal{W}$ can be traced back to the identification of Hilbert spaces in the real line, which proceeds by identifying one-to-one the sites on the two infinite spin chains, in close analogy to the constant-$x$ identification in the continuum that produces the maps \eqref{eq:VHH2}-\eqref{eq:VddS2} \cite{Supplemental}. Notice that on the circle, a one-to-one identification between sites was not possible due to the different sizes $N$ and $N/2$ of the spin chains, and we had to use instead the low energy spectrum identification, in close analogy to the constant-$\theta$ identification in the continuum that produces instead the maps \eqref{eq:VH2}-\eqref{eq:VdS2}.

\begin{figure}
\includegraphics[width=6cm]{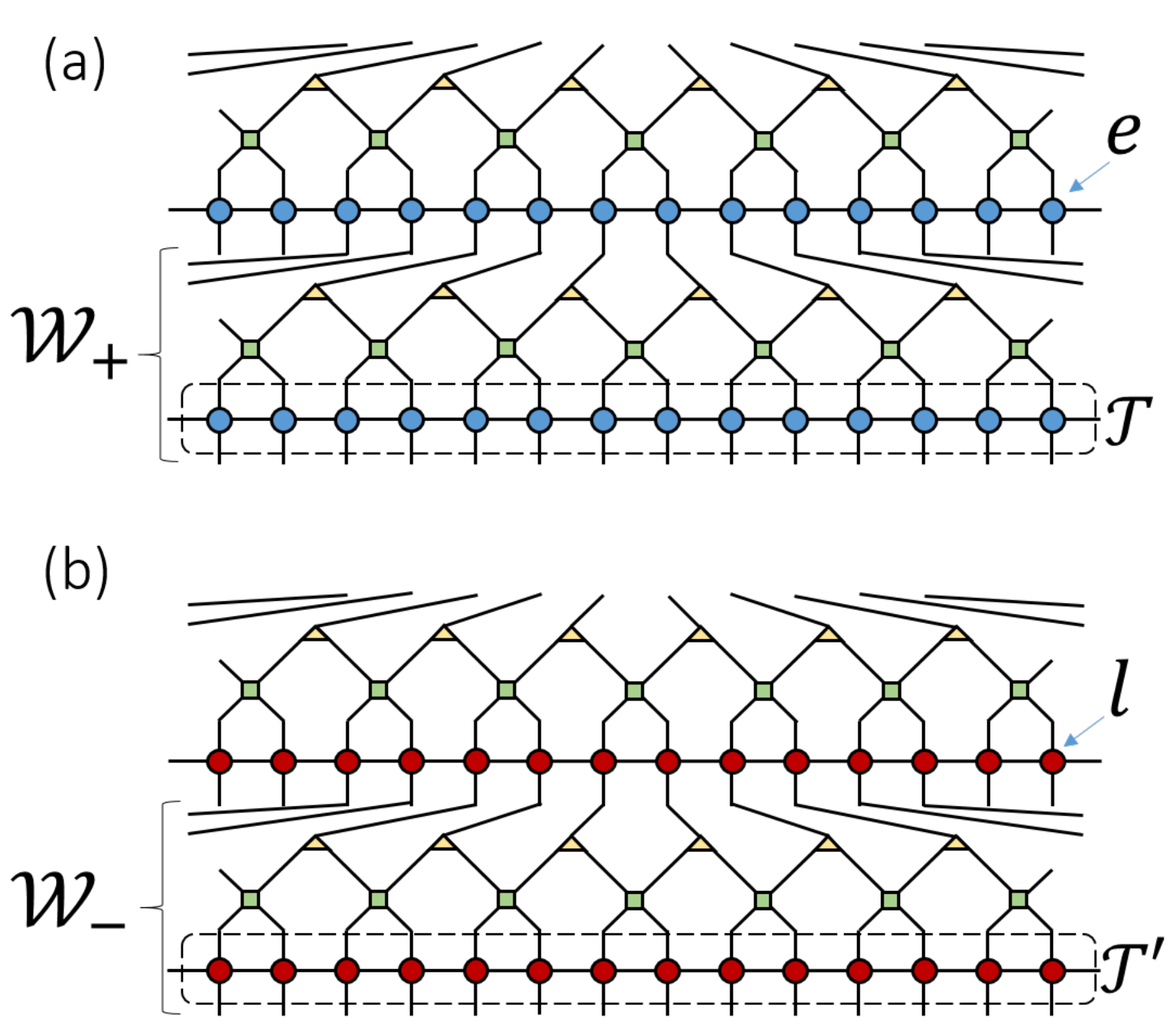}
\caption{
(a) The euclidean MERA is a tensor network made of layers $\mathcal{W}$ of optimized MERA interspersed with transfer matrices $\mathcal{T}$ made of euclideons $e$ and implementing $e^{-H}$.
(b) The lorentzian MERA is made of layers $\mathcal{W}$ of optimized MERA insterspersed with trnasfer matrices $\mathcal{T}'$ made of lorentzions $l$ and implementing $e^{-iH}$.
\label{fig:extensions} 
}
\end{figure}

\textit{Euclidean and lorentzian MERA.---} Having numerically established the linear map implemented by a layer $\mathcal{W}$ of optimized MERA, we can now modify the tensor network so that it implements other maps and in this way reproduce CFT path integrals on other geometries. For instance, an infinite layer $\mathcal{W}_{+} \equiv \mathcal{W}\mathcal{T}$ obtained by pre-multiplying $\mathcal{W}$ by a transfer matrix $\mathcal{T}$ made of a row of euclideons $e$, where $\mathcal{T}$ implements an euclidean time evolution $e^{-H}$ \cite{TNConfTrans,TNPathInt}, results in the linear map $V_{\HH2}$ of Eq. \eqref{eq:VHH2}, thus corresponding to a path integral on a strip of the hyperbolic plane H$_2^p$ (and similarly on the circle with map $V_{\H2}$ of Eq. \eqref{eq:VH2}, corresponding to a path integral on an annulus of the hyperbolic disk H$_2$). The resulting construction in Fig. \ref{fig:extensions}(a), that we call \textit{euclidean MERA}, closely realizes the scenario, envisaged by Swingle \cite{H1,H2}, of a tensor network that represents a time slice of AdS$_3$. In turn, a \textit{lorentzian MERA} is obtained using instead layers $\mathcal{W}_{-} \equiv \mathcal{W}\mathcal{T}'$, see Fig. \ref{fig:extensions}(b), where $\mathcal{T}'$ is made of a row of lorentzions $l$ and implements a lorentzian time evolution $e^{-iH}$ \cite{Supplemental}. $\mathcal{W}_{-}$ is then seen to implement the linear map $V_{\ddS2}$ of Eq. \eqref{eq:VddS2}, thus corresponding to a path integral on a strip of the Poincare de Sitter spacetime dS$_2^p$ (and similarly on the circle), thus closely realizing the scenario, envisaged by Cedric and other authors \cite{dS1,dS2,dS3,dS4}, of a tensor network that represents de Sitter spacetime.

\textit{Discussion.---} In this work we have assigned a path integral geometry to the (null, euclidean, and lorentzian) MERA starting from three candidate geometries and enforcing only rule 1 of Ref. \cite{TNPathInt}, which demands ``\textit{consistency between the path integral map $V$ and the tensor network map $\mathcal{W}$}". However, a local rescaling $g_{\mu\nu}(\eta,r) \rightarrow g'_{\mu\nu}(\eta,r)=\Omega(\eta,r)^2 g_{\mu\nu}(\eta,r)$ in a CFT leaves the map $V$ essentially unchanged, so that a geometric assignment based on rule 1 of Ref. \cite{TNPathInt} alone only determines the conformal class of the metric $g_{\mu\nu}$. However, we can now fix the scale factor $\Omega(\eta,r)$ and recover the geometries in Eqs. \ref{eq:H2p}-\ref{eq:L2p} and \ref{eq:H2}-\ref{eq:dS2} (that is, without preassuming them as candidates) by enforcing rule 2 of Ref. \cite{TNPathInt}, which states that ``\textit{the lattice spacing is the same constant $\aUV$ throughout the tensor network}" \cite{Supplemental}. 

In summary, we have numerically established that a MERA optimized to represent the ground state of a critical quantum spin chain on the circle / real line behaves as a CFT path integral on a light cone / light sheet geometry. Moreover, we have proposed two new tensor networks that correspond to hyperbolic space and de Sitter spacetime, thus realizing previously conjectured constructions. Particularly intriguing is the optimized MERA refusal to perform real time evolution (something that is in principle possible within the MERA variational class), suggesting that the optimal preparation of a CFT ground state through a path integral \cite{TNRholo1, TNRholo2, TNRholo3, TNRholo4, TNRholo5, TNRholo6, complexity1, complexity2, complexity3, complexity4} does not actually require time evolution \cite{Sully}. Perhaps even more tantalizing is the fact that the tensor networks discussed above open a new venue to numerically simulate quantum field theories in curved spacetime \cite{Adam}.

\textit{Acknowledgments}. 
The authors thank Bartlomiej Czech, Pawel
Caputa, Olalla Castro-Alvaredo, William Donnelly, 
Benjamin Doyon, Davide Gaiotto, Qi Hu, Lampros
Lamprou, Juan Maldacena, David Mateos, Samuel
McCandlish, James Sully, Vasudev Shyam,
Tadashi Takayanagi, and Xiao-liang Qi for fruitful discussions
and feedback. Very special thanks go to Ling-Yan (Janet) Hung and Rob Myers for their patient, tenacious attempt to teach us CFT and curved spacetime background material. 
The authors acknowledge support by the Simons Foundation (Many Electron Collaboration),
by NSERC (discovery grant), and by Compute Canada.
Research at Research at Perimeter Institute is supported
by the Government of Canada through the Department of
Innovation, Science and Economic Development Canada
and by the Province of Ontario through the Ministry of
Research, Innovation and Science.


\section{Appendix: MERA on the circle}
\label{sect:circle}

\begin{figure}
\includegraphics[width=8.5cm]{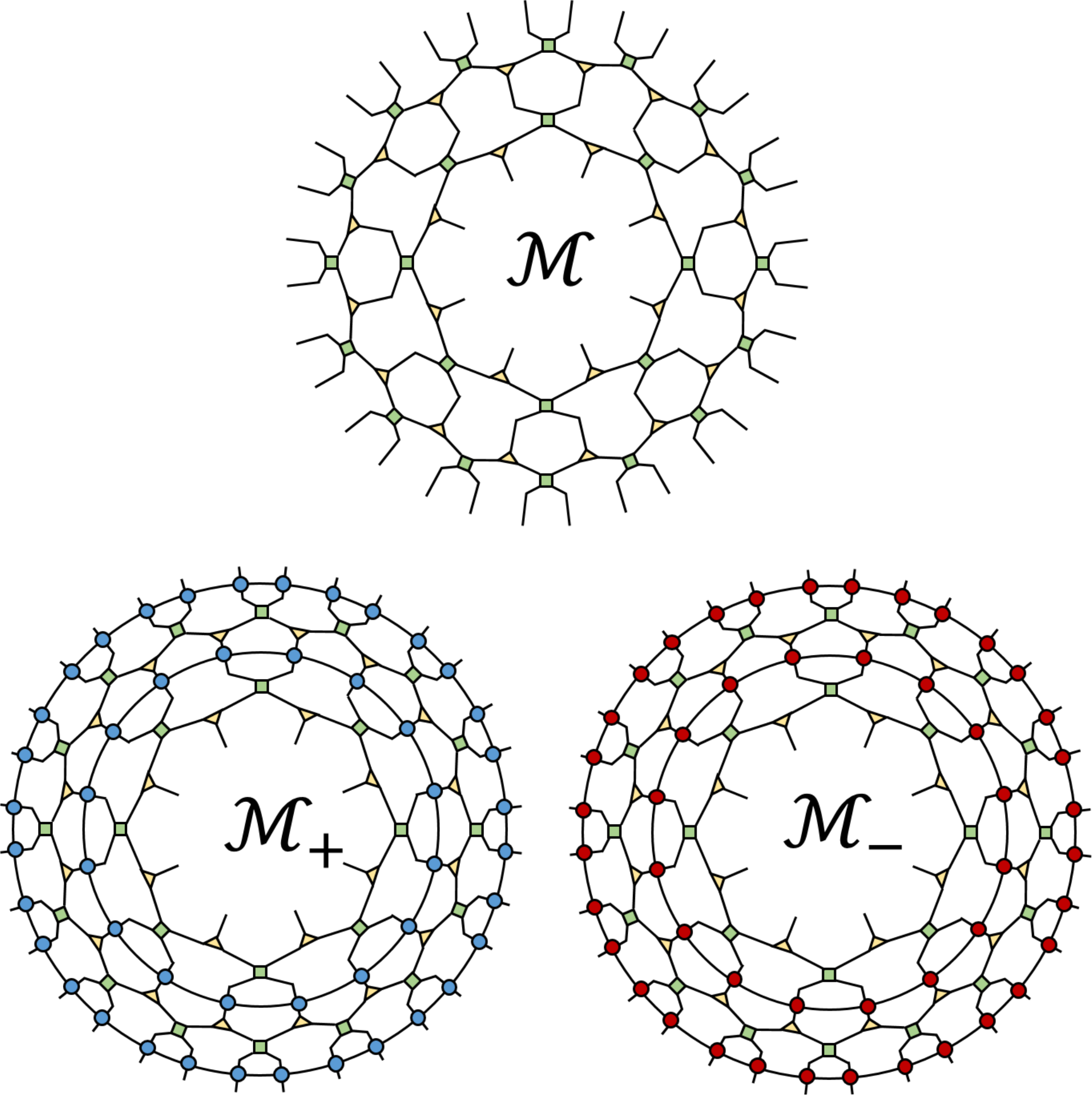}
\caption{
\label{fig:mera3circle}
Graphical representation of several MERA tensor networks on the circle (only two concentric layers of each different MERA are displayed). 
Top: null MERA $\mathcal{M}$, with layers $\mathcal{W}$ made of disentanglers and isometries.
Bottom left: euclidean MERA $\mathcal{M}_{+}$, with layers $\mathcal{W}_{+} = \mathcal{W}\mathcal{T}^{q}$ (for $q=1$), where $\mathcal{T}$ is an euclidean transfer matrix that implements euclidean time evolution.
Bottom right: lorentzian MERA $\mathcal{M}_{-}$, with layers $\mathcal{W}_{-} = \mathcal{W}\mathcal{T}_{-}^{q}$ (for $q=1$), where $\mathcal{T}_{-}$ is a lorentzian transfer matrix that implements real time evolution.}
\end{figure}

In this Appendix we review how to assign a path integral geometry to the \textit{periodic} MERA, which is a tensor network used to represent the ground state of a critical quantum spin chain \textit{on the circle}. We consider both the usual MERA (in what follows named \textit{null} MERA) as well as the two extensions proposed in the main text, namely the \textit{euclidean} MERA and the \textit{lorentzian} MERA, see Fig. \ref{fig:mera3circle}. We have argued that, from a path integral perspective, the periodic version of these tensor networks represent the following geometries:
\begin{equation}
\begin{array}{ccc}
\begin{array}{c} \mbox{null} \\ \mbox{MERA} ~\mathcal{M} \end{array} 
&\leftrightarrow& 
\begin{array}{c} \mbox{light} \\ \mbox{cone L$_2$} \end{array} \\
&&\\
\begin{array}{c} \mbox{euclidean} \\ \mbox{MERA}~\mathcal{M}_{+} \end{array}  
&\leftrightarrow& 
\begin{array}{c} \mbox{hyperbolic} \\ \mbox{space H$_2$} \end{array} \\
&&\\
\begin{array}{c} \mbox{lorentzian} \\ \mbox{MERA} ~\mathcal{M}_{-} \end{array}  
&\leftrightarrow& 
\begin{array}{c} \mbox{de Sitter} \\ \mbox{spacetime dS$_2$} \end{array}
\end{array}
\end{equation}
In euclidean and lorentzian MERA, this geometry assignment holds in the limit when the radial coordinate $r$ of H$_2$ or dS$_2$ is much larger than some radius $\rad$ (see below).

We start by reviewing how the three geometries H$_2$, dS$_2$, and L$_2$ can be embedded in a common three-dimensional ambient space, namely Minkowski $\mathbb{R}^{1,2}$ (see Appendix \ref{sect:AdS} for a review of similar embeddings when the ambient space is AdS$_3$). Then we characterize the different linear maps obtained through a CFT path integral on an annulus of the H$_2$, dS$_2$, and L$_2$ geometries. These linear maps correspond to time evolution of three types: euclidean, lorentzian, and no time evolution, respectively. Finally, using the two rules of Ref. \cite{TNPathInt} to assign a \textit{path integral} geometry to a tensor network, we establish that a periodic layer of euclidean, lorentzian, and null MERA corresponds to an annulus of the H$_2$, dS$_2$, and L$_2$ geometries, respectively. 

The geometry and path integral derivations reviewed in this Appendix are all well-known but scattered through the literature. The characterization of the linear maps, and the corresponding identification with specific tensor networks, are a particular case (translation invariant maps and networks) of the formalism recently presented in Ref. \cite{TNPathInt} (which also applies more broadly to inhomogeneous maps and networks).

Remark on notation: In this paper we use L$_2^p$, H$_2^p$, and dS$_2^p$ (that is, with the superscript \textit{p}) to denote the \textit{Poincare patch} or \textit{local coordinate} version of the L$_2$, H$_2$, and dS$_2$ geometries, namely a light sheet, the hyperbolic plane, and the Poincare patch of de Sitter spacetime. These are the geometries corresponding to null, euclidean, and lorentzian MERA when the tensor network describes the ground state of a critical spin chain \textit{on the line}, and will be addressed in Appendix \eqref{sect:line}. 

\subsection{Embeddings in Minkowski $\mathbb{R}^{1,2}$}

\begin{figure}
\includegraphics[width=\linewidth]{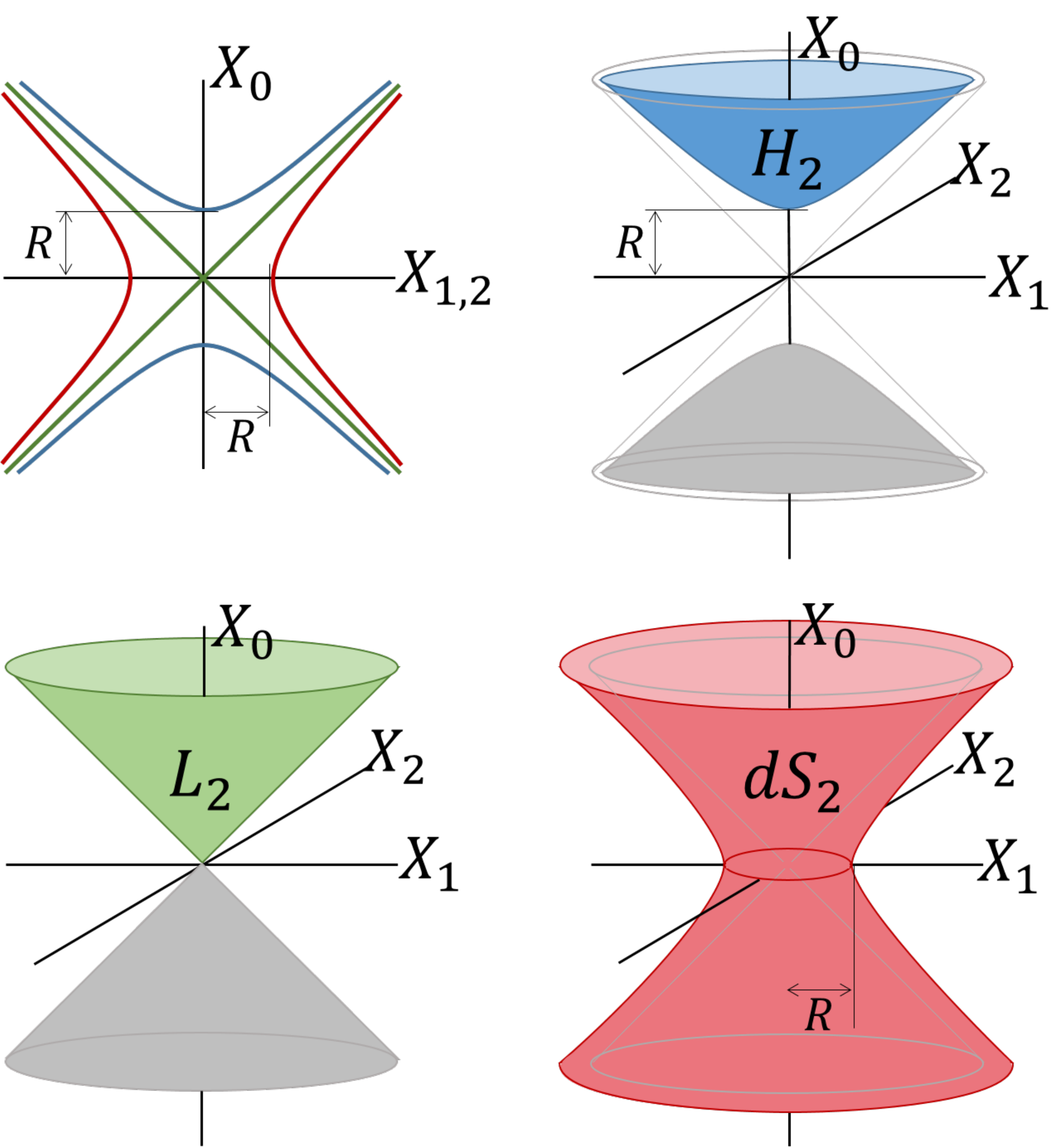}
\caption{
\label{fig:minkowski} 
The three geometries H$_2$, dS$_2$, and L$_2$ of interest are shown embedded in three-dimensional Minkowski spacetime $\mathbb{R}^{1,2}$ with time coordinate $X_0$ and space coordinates $X_1$ and $X_2$, according to the restrictions \eqref{eqap:embedding1}-\eqref{eqap:embedding3}. Notice that both H$_2$ and dS$_2$ depend on a radius $\rad$, and that L$_2$ can be understood as the limit $\rad \rightarrow 0$ of either of these two geometries, see Fig. \ref{fig:limit}.
}
\end{figure}

A useful characterization of the geometries H$_2$, dS$_2$, and L$_2$ is as two-dimensional embeddings in three-dimensional Minkowski spacetime $\mathbb{R}^{1,2}$. Let $X_0$ denote the time coordinate and $X_1$ and $X_2$ the two space coordinates, and recall that the metric of $\mathbb{R}^{1,2}$ reads, when expressed  as a squared line element $dl^2$,
\begin{equation}
dl^2 = -d{X_0}^2 + d{X_1}^2 + d{X_2}^2.
\end{equation}
Then H$_2$, dS$_2$, and L$_2$ are given by the constraints
\begin{eqnarray} \label{eqap:embedding1}
-{X_0}^2 + {X_1}^2 + {X_2}^2 &=& -\rad^2~~~(\mbox{H}_2),\\
-{X_0}^2 + {X_1}^2 + {X_2}^2 &=& \rad^2~~~~~(\mbox{dS}_2),\label{eqap:embedding2}\\
-{X_0}^2 + {X_1}^2 + {X_2}^2 &=& 0~~~~~~~(\mbox{L}_2),\label{eqap:embedding3}
\end{eqnarray}
where $\rad$ is the radius of curvature, see Fig. \ref{fig:minkowski}.

\begin{figure}
\includegraphics[width=5cm]{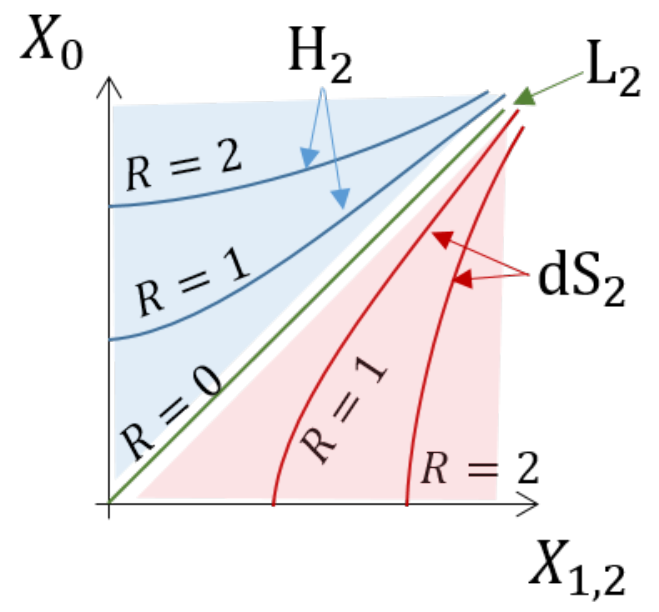}
\caption{
The light cone geometry L$_2$ can be understood as the zero radius limit $a\rightarrow 0$ of both the Poicare disk H$_2$ and global de Sitter space dS$_2$, as can be seen when considering these two-dimensional geometries as embedded in Minkowski $\mathbb{R}^{1,2}$, where $X_0$ is the time coordinateand $X_1$ and $X_2$ are space coordinates.
\label{fig:limit} 
}
\end{figure}

We can use polar coordinates 
\begin{equation}
r = \sqrt{{X_1}^2+{X_2}^2},~~~\theta = \arctan ({X_2}/{X_1}),
\end{equation}
where $r\geq 0$ and $\theta \in [0,2\pi)$, or
\begin{equation}
{X_1} = r\cos(\theta),~~~~~{X_2} = r \sin(\theta),
\end{equation}
to rewrite the metric on $\mathbb{R}^{1,2}$ and the constraints as
\begin{equation}
dl^2 = -d{X_0}^2 + dr^2 + r^2d\theta^2,
\end{equation}
and
\begin{eqnarray} \label{eqap:embedding1b}
-{X_0}^2 + r^2 &=& -\rad^2~~~(\mbox{H}_2),\\
-{X_0}^2 + r^2 &=& \rad^2~~~~~(\mbox{dS}_2),\label{eqap:embedding3b}\\
-{X_0}^2 + r^2 &=& 0~~~~~~~(\mbox{L}_2),\label{eqap:embedding3c}
\end{eqnarray}
and then use these constraints to arrive to the following induced metrics
\begin{eqnarray}
dl^2_{\H2} &=& \frac{dr^2}{(r/\rad)^2+1} + r^2 d\theta^2, ~~~~~ \label{eqap:dl1}\\
dl^2_{\dS2} &=& \frac{-dr^2}{(r/\rad)^2-1} + r^2 d\theta^2, ~~~~~\label{eqap:dl2}\\
dl^2_{\L2} &=&  r^2 d\theta^2, ~~~~~~~~~~~~~~~~~~~~~~~\label{eqap:dl3}
\end{eqnarray}
where $r^2 \geq 0$, $r^2 \geq \rad^2$, and $r^2 \geq 0$, respectively. Notice that the light cone L$_2$ is the limit of small radius $\rad$, $\lim \rad\rightarrow 0$, of both the hyperbolic plane H$_2$ and de sitter spacetime dS$_2$, see Fig. \ref{fig:limit}.

The coordinate system $(r,\theta)$ makes manifest the invariance of these three manifolds under $\theta$ rotations of the $X_1X_2$ plane. However, as the above embeddings suggest, these three two-dimensional manifolds are also invariant under the larger $SO(1,2)$ group of isometries of Minkowski $\mathbb{R}^{1,2}$, which includes two more generators (Lorentz boosts in the $X_0X_1$ and $X_0X_2$ planes). 


\begin{figure}
\includegraphics[width=6cm]{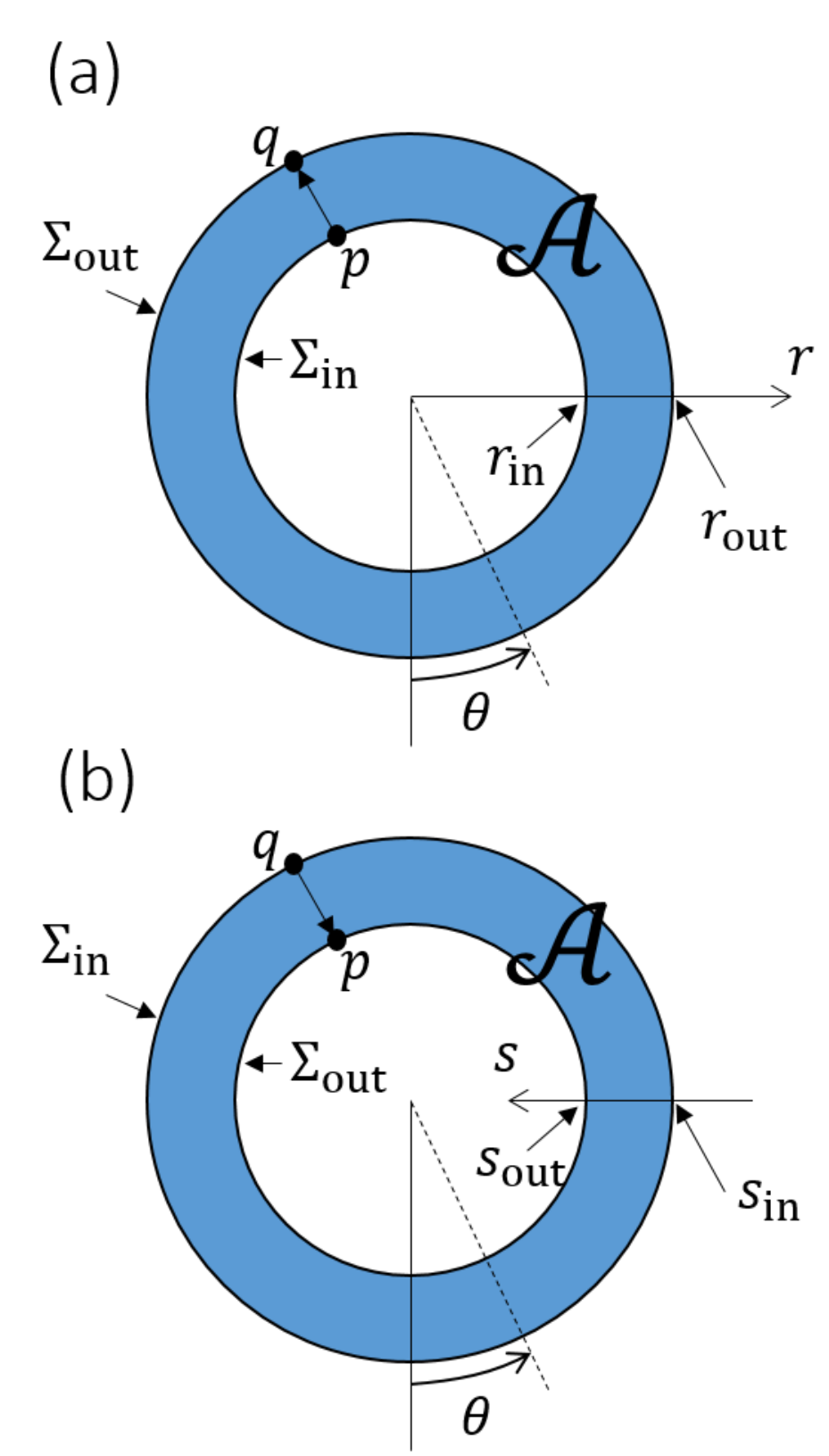}
\caption{
(a) Annulus $\mathcal{A}$ with boundaries $\Sigma_{\inn}$ and $\Sigma_{\out}$ corresponding to circles of radius $r=r_{\inn}$ and $r=r_{\out}$ where $r_{\inn} < r_{\out}$. The constant-$\theta$ identification identifies point $p \in \Sigma_{\inn}$ with point $q \in \Sigma_{\out}$. 
(b) Same annulus $\mathcal{A}$ as before but with different radial coordinate $s$, which grows when the radius of the circle decreases. Notice that the names of the boundaries of the annulus have swapped. $\Sigma_{\inn}$ and $\Sigma_{\out}$ now correspond to values $s= s_{\inn}$ and $s = s_{\out}$ with $s_{\inn} < s_{\out}$. 
\label{fig:coord_circle}
}
\end{figure}

\subsection{Linear map by path integral on an annulus} 

The above geometries H$_2$, dS$_2$, and L$_2$ can all be sliced into slices $\Sigma_r$ defined by a constant value of the radial coordinate $r$. Each slice $\Sigma_r$ corresponds to a circle of length $2\pi r$. Two concentric circles with radii $r_{\inn}$ and $r_{\out}$, with $r_{\inn}> r_{\out}$, define an annulus $\mathcal{A}$ characterized by $r \in [r_{\out}, r_{\inn}]$, $\theta \in [0,2\pi)$, see Fig. \ref{fig:coord_circle}(a).

Consider now a two-dimensional CFT with field $\phi$ and action functional $S[\phi]$. In any of the three geometries, given the circle $\Sigma_{\inn}$ for $r=r_{\inn}$, we can define the Hilbert space $\mathcal{H}(\Sigma_{\inn})$ of the CFT on that circle in terms of basis states $\ket{\varphi(\theta)}$. Here $\varphi(\theta)$, for $\theta \in [0,2\pi)$, is a field configuration that results from restricting the field $\phi(r,\theta)$ to the circle $\Sigma_{\inn}$. We can similarly define the Hilbert space $\mathcal{H}(\Sigma_{\out})$ of the CFT on a second circle $\Sigma_{\out}$ given by $r = r_{\out}$, again with basis $\ket{ \varphi( \theta)}$. We can then identify the two Hilbert spaces $\mathcal{H}_{\inn}$ and $\mathcal{H}_{\out}$ (from now on simply $\mathcal{H}$) by identifying basis vectors according to
\begin{equation}
\ket{\varphi(\theta)}_{\Sigma_{\inn}} \sim \ket{\varphi'(\theta)}_{\Sigma_{\out}}, ~\mbox{iff} ~\varphi(\theta) = \varphi'(\theta) ~~\mbox{for all} ~\theta.
\end{equation}

For concreteness, from now on we continue the discussion for the hyperbolic geometry H$_2$, and we postpone the analysis of dS$_2$ and L$_2$ to parts \eqref{subsect:circle_lorentz} and \eqref{subsect:circle_null} of this Appendix. 
We introduce the linear map $V:\mathcal{H} \rightarrow \mathcal{H}$ given by the path integral on an annulus $\mathcal{A}$ of H$_2$, that is with matrix elements
\begin{equation} \label{eqap:VS} 
\bra{\varphi'(\theta)} V \ket{\varphi(\theta)} = \int [D\phi] e^{-S[\phi(r,\theta)]}.
\end{equation}
In this expression the integral is over field configurations $\phi(r,\theta)$ restricted to the annulus $\mathcal{A}$ and with boundary conditions $\varphi(\theta)$ and $\varphi'(\theta)$, 
\begin{equation}
\phi(r_{\inn}, \theta) = \varphi(\theta),~~~~~\phi(r_{\out}, \theta) = \varphi'(\theta),
\end{equation}
whereas $[D\phi]$ and $S[\phi(r,\theta)]$ are the integration measure and the \textit{euclidean} action of the CFT. 

\subsection{Thin annulus} 

In the limit of a thin annulus, when $r_{\inn} - r_{\out} = \epsilon$ for small $\epsilon>0$, we can expand the linear map $V \approx \mathbb{1} - \epsilon Q$ in terms of a generator $Q$. An expression for $Q$ is found by specializing to the current case the general solution derived in Ref. \cite{TNPathInt} and reviewed in Appendix \eqref{sect:linear}. For a diagonal metric of the form
\begin{eqnarray} 
dl^2 = \Omega^{2}(r)\left(a(r)^2 dr^2 + d\theta^2 \right)~~~~~~\label{eqap:ge2} 
\end{eqnarray}
this generator is 
\begin{eqnarray}
Q &=&  a(r) \int_0^{2\pi} d\theta~ h(\theta) = a(r) ~H_0,\\
H_0 &\equiv&  \int_0^{2\pi} d\theta~ h(\theta) = L_0 + \bar{L}_0 - \frac{c}{12},
\end{eqnarray}
where $H_0$ is the CFT Hamiltonian $H_0$ on a circle of unit radius. Its spectrum of energies $E_{\alpha}$ is given in terms of the scaling dimensions $\Delta_{\alpha}$ of local scaling fields by $E_{\alpha} = \Delta_{\alpha} - c/12$. That is, the path integral linear map $V$ is an euclidean time evolution as generated by the CFT Hamiltonian $H_0$.


Specifically, for the hyperbolic space H$_2$ in coordinates $(r,\theta)$ we have
\begin{equation}
a(r) = \frac{\rad}{r\sqrt{r^2 + \rad^2}}.
\end{equation}
For later reference, we quote the resulting metric and generator $Q$:
\begin{eqnarray}  \label{eqap:dlH2a} 
dl^2_{\H2} &=& \frac{\rad^2dr^2}{r^2+\rad^2} + r^2d\theta^2 \\
&=& r^2\left(\frac{\rad^2(dr/r)^2}{ \left(r^2+\rad^2\right)} + d\theta^2 \right),\\
Q_{\H2} &=& \frac{\rad}{ r\sqrt{r^2+\rad^2}}~H_0.  \label{eqap:QH2} 
\end{eqnarray}

\subsection{The $r \gg \rad$ regime}

Of particular importance for our discussion is the H$_2$ geometry in the regime of large radial coordinate $r$, that is when $r \gg \rad$, since it is in this regime that the connection to euclidean MERA is most transparent. In addition, in this regime computations become simpler. To leading order in $r/\rad$, the metric and generator simplify to
\begin{eqnarray} 
dl^2_{\H2} &\approx& \frac{\rad^2dr^2}{r^2} + r^2d\theta^2 \\
&=& r^2\left(\frac{\rad^2dr^2}{r^4} + d\theta^2 \right),\\
Q_{\H2} &\approx& \frac{\rad}{ r^2}~H_0.\label{eqap:QH2regime}
\end{eqnarray}

\subsection{Thick annulus}

The path integral on a thick annulus $r\in [r_{\inn},r_{\out}]$ for $r_{\inn} \gg \rad$ corresponds to the finite linear map
\begin{eqnarray} 
V_{\H2} &=& \mathcal{P} \exp \left(-\int_{r_{\inn}}^{r_{\out}} dr ~Q_{\H2} \right)\\
&\approx& \exp \left(- H_0 \int_{r_{\inn}}^{r_{\out}} \frac{ \rad~dr}{r^2}\right)\\
&=& \exp \left(-\frac{\rad(r_{\out}-r_{\inn})}{r_{\inn}r_{\out}} H_0\right), \label{eqap:VH2}
\end{eqnarray}
where we used
\begin{eqnarray}
\int_{r_{\inn}}^{r_{\out}} \frac{\rad~dr}{r^2} = \left.\frac{-\rad}{r}\right|_{r_{\inn}}^{r_{\out}} = -\rad\left(\frac{1}{r_{\out}} - \frac{1}{r_{\inn}}\right).
\end{eqnarray}

\subsection{Another useful radial coordinate}

Let us consider a second radial coordinate $s$, given in terms of $r$ by
\begin{eqnarray}
r = r_0 e^{-s},~~~~\mbox{or} ~~~ s = \log(r_0/r),
\end{eqnarray}
where $r_0$ is a reference radius and $s$ is dimensionless.
The metric of H$_2$, the generator $Q_{\H2}$, and the finite gate $V_{\H2}$ read, in the simplifying regime $r \gg \rad$,
\begin{eqnarray}  
dl^2_{\H2} &\approx& \rad^2 ds^2  + r_0^2 e^{-2s} d\theta^2 \label{eqap:H2dls}\\
&=& \ r_0^2 e^{-2s} \left(\left(\frac{\rad}{r_0}\right)^2e^{2s}dz^2 + d\theta^2 \right),\\
Q_{\H2} &\approx& \frac{\rad}{ r_0}e^s~H_0, \\
V_{\H2} &=& \mathcal{P} \exp \left(-\int^{s_{\out}}_{s_{\inn}} ds ~Q_{\H2} \right)\\
&\approx& \exp \left(-  \int^{s_{\out}}_{s_{\inn}} ds ~  \frac{\rad}{r_0} e^s H_0  \right)\\
&=& \exp \left(-\left(e^{s_{\out}} - e^{s_{\inn}}\right)\frac{\rad}{r_0} ~H_0\right). \label{eqap:H2Vs}
\end{eqnarray}

Notice that while the radial coordinate $r$ measures the proper length of the circle $\Sigma_r$, in that the proper length of $\Sigma_r$ is $2\pi r$, see Eq. \eqref{eqap:dlH2a}, in the regime $r \gg \rad$ the merit of the radial coordinate $s$ is that it is proportional to proper euclidean time. Indeed, Eq. \eqref{eqap:H2dls} reveals that the proper euclidean time is measured by $s' \equiv \rad s$. In particular, $s$ will be seen to be a natural radial coordinate to count layers of the euclidean MERA. 


\begin{figure}
\includegraphics[width=8.5cm]{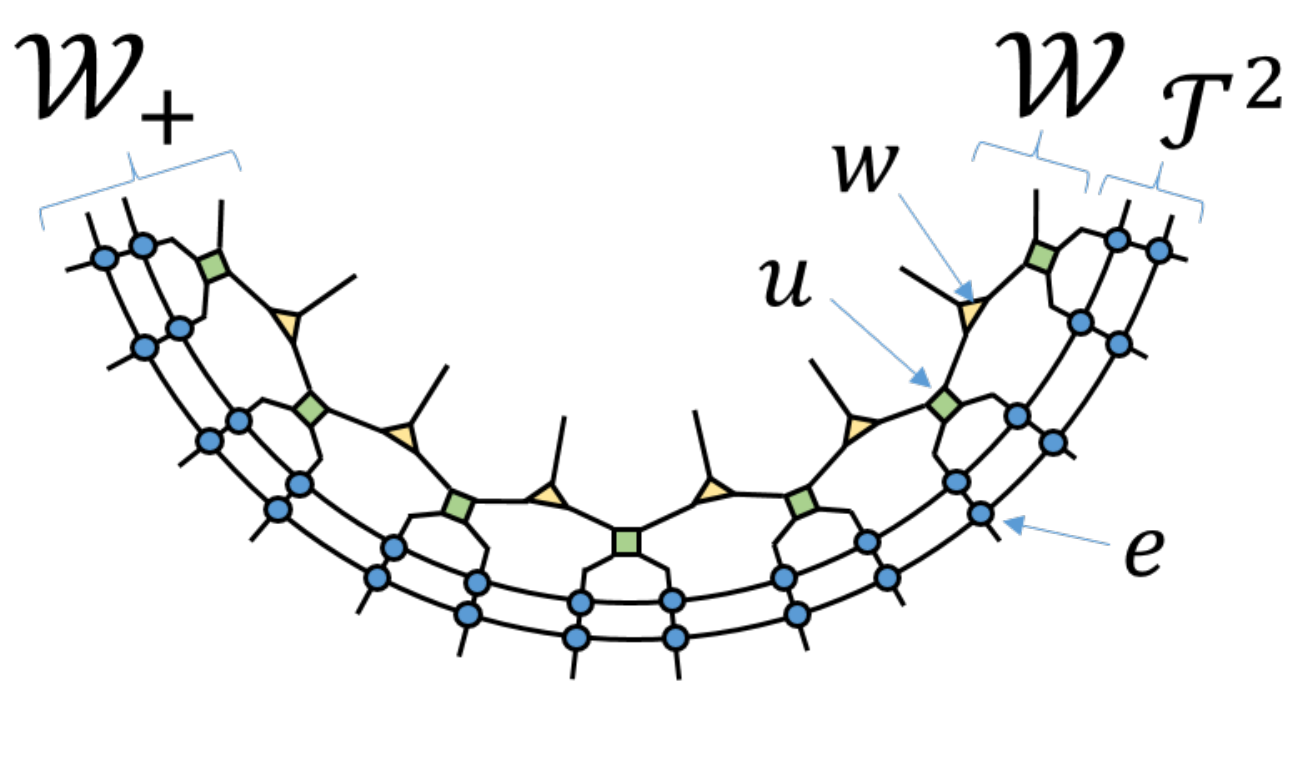}
\caption{
Layer $\mathcal{W}_{+}$ of euclidean MERA on the circle. Only part of this periodic layer is shown. $\mathcal{W}_{+}$ is made of the product of $q$ periodic euclidean transfer matrices $\mathcal{T}$ ($q=2$ in the figure) and a layer $\mathcal{W}$ of null or regular MERA, see Eq. \eqref{eqap:Wplusmap_circle}. Each euclidean transfer matrix $\mathcal{T}$ consist of the product of a periodic chain of tensors $e$ called euclideons. Each layer $\mathcal{W}$ of null MERA is made of tensors $u$ and $w$ called disentanglers and isometries.
\label{fig:layer_circle}
}
\end{figure}

\subsection{Linear map of $\mathcal{W}_{+}$ on the circle}

Next we identify what linear map is implemented by a periodic layer $\mathcal{W}_{+}$ of euclidean MERA made of $q$ copies of the periodic euclidean transfer matrix $\mathcal{T}$ and a periodic layer $\mathcal{W}$ of MERA,
\begin{equation} \label{eqap:Wplusmap_circle}
\mathcal{W}_{+} \equiv \mathcal{W} \mathcal{T}^q,
\end{equation}
see Fig. \ref{fig:layer_circle}. We think of the layer $\mathcal{W}_{+}$ as a linear map from the lower spin chain made of $N$ sites to the upper spin chain made of $N/2$ sites.
 
The euclidean transfer matrix $\mathcal{T}$ is a tensor network, consisting of a periodic row of euclideons, that acts on the Hilbert space of a periodic quantum spin chain made of $N$ spins. By construction, it implements an euclidean time evolution of the form
\begin{equation} \label{eqap:T}
\mathcal{T} \approx \exp\left( -H(N)\right) =  \exp \left(-\frac{2\pi}{N} H_0\right),
\end{equation}
where the spin chain Hamiltonian 
\begin{equation}
H(N) \equiv \sum_{l=1}^N h_l,
\end{equation}
acts approximately as $H(N) \approx (2\pi/N)(L_0 + \bar{L}_0 - c/12)$ on low energy states, up to corrections that are subleading in $1/N$.  On the other hand, as shown in this paper and further detailed in Appendix \eqref{sect:experiment}, a periodic layer $\mathcal{W}$ of null MERA defines a linear map from (the Hilbert space of a periodic spin chain made of) $N$ spins into (the Hilbert space of a periodic spin chain made of) $N/2$ spins that acts on low energy states as the identity operator,
\begin{equation} \label{eqap:W}
\mathcal{W} \approx \mathbb{1}.
\end{equation}
We conclude that a layer $\mathcal{W}_{+}$ of euclidean MERA maps states of $N$ spins into states of $N/2$ spins while implementing the linear map
\begin{eqnarray}\label{eqap:Wplus1}
\mathcal{W}_{+}&=& \mathcal{W} \mathcal{T}^q = \mathbb{1} \times \exp\left( -qH(N) \right)\\
&\approx& \exp \left(-q\frac{2\pi}{N} H_0\right)
\end{eqnarray} 
on low energy states.

\begin{figure}
\includegraphics[width=8.5cm]{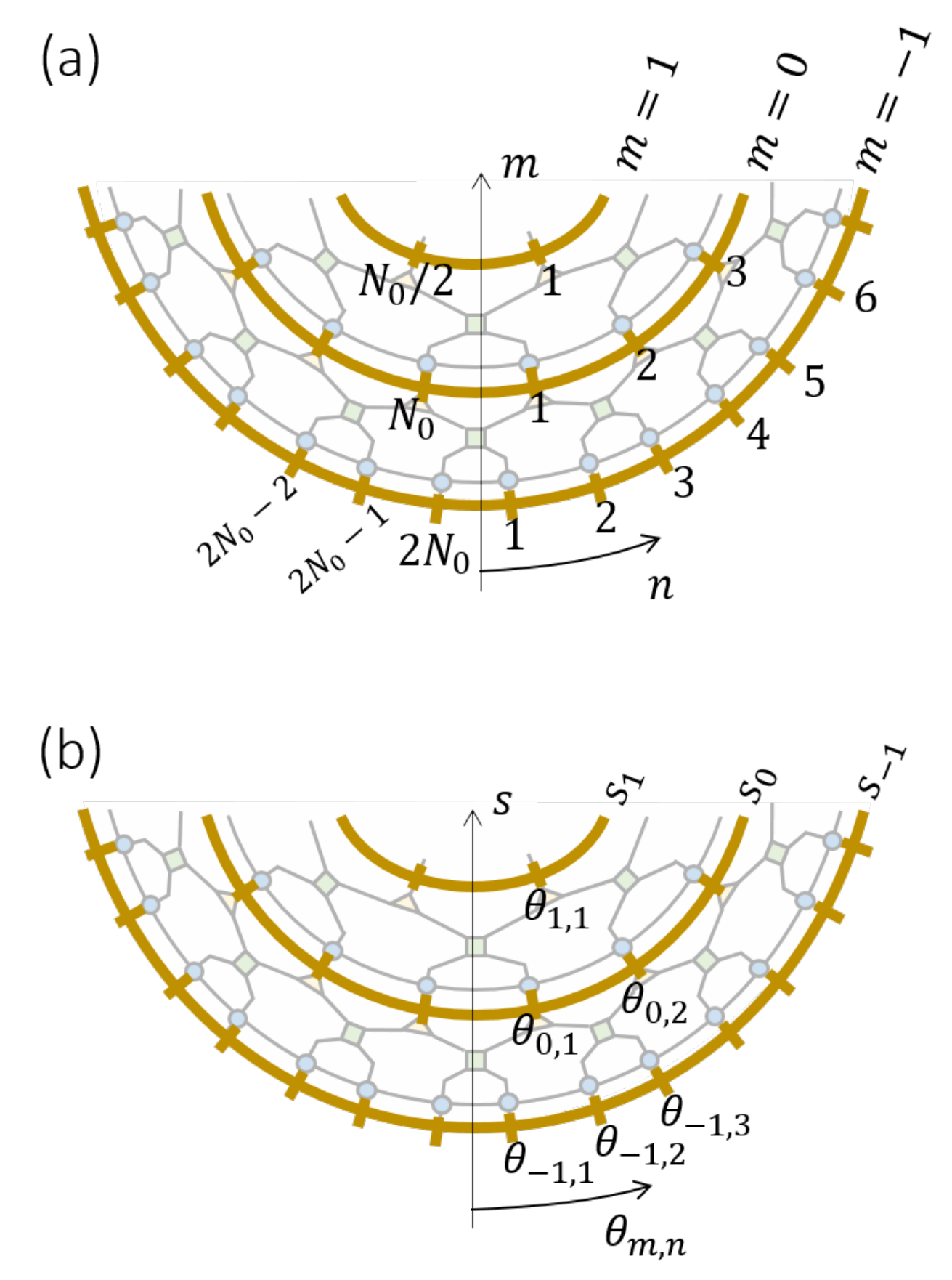}
\caption{
(a) We label each spin chain between two layers of euclidean MERA by an integer $m$, and each site within that spin chain by the pair $(m,n)$.
(b) We then assign to site $(m,n)$ discrete coordinates $(s_m, \theta_{m,n})$ as given by \eqref{eqap:discretecoor}. 
\label{fig:stheta}
}
\end{figure}

\subsection{Discrete coordinates on the euclidean MERA}

Consider the euclidean MERA on the circle. We label the spin chains between layers of euclidean MERA by an integer $m$. If the spin chain $m=0$ has $N_0$ sites (which we assume to be divisible by a large power of $2$), then the spin chain $m=1$ has $N_1 = N_0/2$ sites. More generally, the spin chain $m$ has $N_m = N_0 2^{-m}$ sites. We label the sites in the spin chain $m$ by a second integer $n \in \{1, \cdots, N_m \}$. Therefore each site of any spin chain is assigned a unique pair $(m,n)$ of integers, see Fig. \ref{fig:stheta}(a). We then further assign discrete radial and angular coordinate ($s_m,\theta_{m,n}$) to site $(m,n)$ according to
\begin{equation} \label{eqap:discretecoor}
 s_m \equiv m \log 2,~~~\theta_{m,n} \equiv \frac{2\pi}{N_m}\left(n-\frac{1}{2}\right),
\end{equation} 
see Fig. \ref{fig:stheta}(b).
 
Let us consider a candidate geometry for the euclidean MERA in the regime $ N_m \gg 2\pi q$ (which will turn out to correspond to the regime $r\gg \rad$ or $s \ll \log(r_0/\rad)$ in H$_2$). It is initially given by a generic metric in radial and angular coordinates $(s,\theta)$, which can always be written as (see Appendix \eqref{sect:linear})
\begin{eqnarray}
&& dl^2 = \Omega(s,\theta)^2 \times \nonumber \\
&&~~ \left(~[\pm a(s,\theta)^2 +  b(s,\theta)^2] ~ds^2  + 2b(s,\theta) ds d\theta + d\theta^2 ~\right).~~~~~\label{eqap:genericdl}
\end{eqnarray}

\subsection{Rules 1 and 2 of the path integral geometry}

Our goal is to constrain the functions $a(s,\theta)$, $b(s,\theta)$ and $\Omega(s,\theta)$ using rules 1 and 2 of Ref. \cite{TNPathInt}, introduced as necessary conditions for assigning a \textit{path integral} geometry to a tensor network. 

Rule 1 of Ref. \cite{TNPathInt} (\textit{compatibility with path integral}) states that the linear map implemented by an annulus $s\in [s_{\inn}, s_{\out}]$ of such geometry, namely
\begin{equation}
V(s_{\inn},s_{\out}) \equiv \mathcal{P} \exp \left(-\int_{s_{\inn}}^{s_{\out}} ds~Q(s) \right),
\end{equation}
where the generator reads (see Appendix \eqref{sect:linear})
\begin{eqnarray}
Q(s) &\equiv& \int_{0}^{2\pi}d\theta ~\Big(a(s,\theta) h(\theta) - i b(s,x) p(x) \Big)
\end{eqnarray}
for euclidean signature ($+$ sign) and
\begin{eqnarray}
Q(s) &\equiv& \int_{0}^{2\pi}d\theta ~\Big(i a(s,\theta) h(\theta) - i b(s,\theta) p(\theta) \Big)
\end{eqnarray}
for lorentzian signature ($-$ sign), 
should match, for $s_{\inn} = s_m$ and $s_{\out} = s_{m+1}$, the linear map \eqref{eqap:Wplus1} implemented by the layer of euclidean MERA between spin chains $m$ and $m+1$, namely
\begin{equation}
\mathcal{W}_{+}^{(m,m+1)} \approx \exp \left(-\frac{2\pi }{N_0}q2^m~ H_0 \right).
\end{equation}
That is, $V(s_m,s_{m+1}) = \mathcal{W}_{+}^{(m,m+1)}$ or
\begin{eqnarray}
\label{eqap:rule1}
&&\mathcal{P} \exp \left(-\int_{s_m}^{s_{m+1}} ds~Q(s) \right) \nonumber \\
&&~~~~~~~~~~~~~~~~~~~= \exp \left(-\frac{2\pi}{N_0} q2^m~H_0 \right).~~~~\mbox{(rule 1)}~~~~~~
\end{eqnarray}
This results in a constraint on the signature ($\pm$ sign) and the functions $a(s,\theta)$ and $b(s,\theta)$, but not on the scale factor $\Omega(s,\theta)$.

Rule 2 of Ref. \cite{TNPathInt} (\textit{constant lattice spacing}) states that the proper distance between nearest neighbor sites $(m,n)$ and $(m,n+1)$ in spin chain $m$ is the same constant $\aUV$ for all $m$ and $n$. Then the length of the periodic spin chain $m$ is $N_m\aUV$ and therefore its radius $r_m$ is
\begin{eqnarray}
r_m &=& \frac{N_m \aUV}{2\pi} = \frac{N_0 \aUV}{2\pi}2^{-m} = r_0 e^{-m},\\
r_0 &\equiv& \frac{N_0 \aUV}{2\pi},~~~~ \label{eqap:r0N0}
\end{eqnarray}
where $r_0$ is a reference radius.

In a constant $s$ cut, the metric \eqref{eqap:genericdl} reads $dl^2 = \Omega(s,\theta)^2 d\theta^2$ and therefore the distance (within the cut) between points $(s,m)=(s_m,\theta_{m,n})$ and $(s,m)=(s_m,\theta_{m,n+1})$ is
\begin{equation}
\int_{\theta_{m,n}}^{\theta_{m,n+1}} \sqrt{dl^2(s_m,\theta)} = \int_{\theta_{m,n}}^{\theta_{m,n+1}} d\theta ~\Omega(s_m,\theta).
\end{equation}
Rule 2 then implies the constraint
\begin{equation} \label{eqap:rule2}
\int_{\theta_{m,n}}^{\theta_{m,n+1}} d\theta ~\Omega(s_m,\theta) = \aUV.~~~~~~~~~~~\mbox{(rule 2)} ~~~~~~ 
\end{equation}
This is a constraint on the scale factor $\Omega(s,\theta)$, and not on the signature or functions $a(s,\theta)$ and $b(s,\theta)$ of metric \eqref{eqap:genericdl}.

There are now two possible routes to assigning a (path integral) geometry to the euclidean MERA. The first route simply identifies the metric H$_2$ with radius $\rad=\aUV q$ in the $r \gg \rad$ regime as one that satisfies rules 1 and 2. The second route first restricts the possible geometries on the grounds of discrete symmetries of the network. After that, requiring rules 1 and 2 is seen to completely specify the metric, which is again that of H$_2$ with radius $\rad= \aUV q$.

\subsection{Path integral geometry of euclidean MERA on the circle}

The first option is to provide an explicit example of a metric that fulfils rules 1 and 2. That is, to provide functions $a(s,\theta)$, $b(s,\theta)$ that lead to a generator $Q(s)$ compatible with rule 1 as expressed by condition \eqref{eqap:rule1}, as well as to provide a scale factor $\Omega(s,\theta)$ compatible with rule 2 as expressed by condition \eqref{eqap:rule2}.

One such example is given by
\begin{eqnarray} \label{eqap:ABOmega} 
a(s,\theta) = \frac{\aUV}{r_0}qe^{s},~~~
b(s,\theta) = 0,~~~
\Omega(s,\theta) = r_{0}e^{-s},~~~~~
\end{eqnarray}
and the choice of euclidean signature ($+$ sign) in metric \eqref{eqap:genericdl}, which leads to the metric and generator
\begin{eqnarray} \label{eqap:gMplus}
dl^2_{\Mplus} &\equiv&  \aUV^2 q^2ds^2 + r_0^2e^{-2s}d\theta^2 \\
&=& r_0^2e^{-2s}  \left(\left(\frac{\aUV}{r_0}\right)^2q^2e^{2s}ds^2 +d\theta^2 \right),\\
Q_{\mathcal{M}_{+}} &\equiv& \frac{\aUV}{r_0}qe^s H_0.
\end{eqnarray}

Indeed, condition \eqref{eqap:rule1} for $Q(s) = Q_{\mathcal{M}_{+}}$ becomes
\begin{eqnarray}
 \exp \left(-\frac{\aUV}{r_0}qH_0\int_{s_m}^{s_{m+1}} \!\!e^sds  \right) = \exp \left(-\frac{2\pi}{N_0} q2^m~H_0 \right),~~~
\end{eqnarray}
which is fulfilled since 
\begin{eqnarray}
\int_{s_m}^{s_{m+1}} e^s ds~ =  \left. e^{s}\right|_{2^m}^{2^{m+1}} = 2^m,
\end{eqnarray}
and \eqref{eqap:r0N0} implies $\aUV/r_0 = 2\pi/N_0$,  whereas condition \eqref{eqap:rule2} is fulfilled because
\begin{eqnarray}
\int_{\theta_{m,n}}^{\theta_{m,n+1}} d\theta ~r_0e^{-s} &=& r_0 e^{-s} \int_{\theta_{m,n}}^{\theta_{m,n+1}} d\theta = r_0 e^{-s}\frac{2\pi}{N_m} ~~~~~\\
&=& r_0 e^{-s}\frac{2\pi}{N_0}e^{s} = r_0 \frac{2\pi}{N_0} \\
&=& \frac{\aUV N_0}{2\pi} ~ \frac{2\pi}{N_0}= \aUV.
\end{eqnarray}

We immediately recognize $dl^2_{\mathcal{M}_{+}}$ above as the metric $dl_{\H2}^2$ of the hyperbolic space H$_2$ with radius $\rad=\aUV q$, in the regime $r\gg \rad$, see Eq. \eqref{eqap:H2dls}. We also have
\begin{eqnarray}
\frac{N_m}{2\pi q} = \frac{\aUV N_m}{2\pi \aUV  q} = \frac{r_m}{\rad},
\end{eqnarray}
so that indeed, the regime $r \gg \rad$ corresponds to $N_m \gg 2\pi q$ as anticipated earlier on.

\subsection{First symmetries, then rules 1 and 2}

Alternatively, we can start again with a general metric specified by three generic functions $a(s,\theta)$, $b(s,\theta)$, and $\Omega(s,\theta)$ in Eq. \eqref{eqap:genericdl} and first impose rotation invariance and (apparent) scale invariance, then rules 1 and 2.

\begin{figure}
\includegraphics[width=7cm]{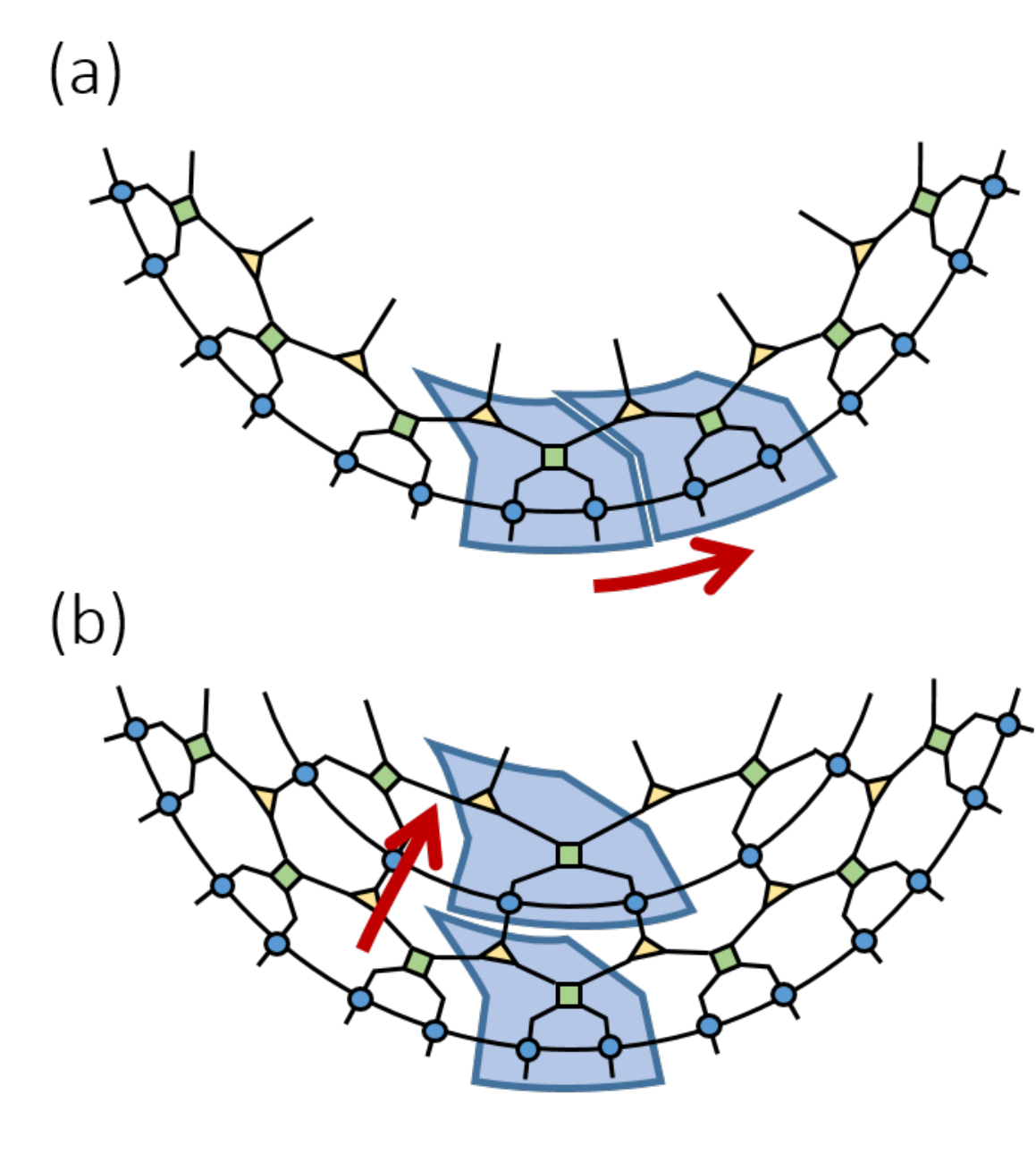}
\caption{
(a) A layer of periodic euclidean MERA, obtained by multiplying a small unit cell of tensors, is invariant under discrete rotations.
(b) Since each layer is made of copies of the same unit cell, there is also an apparent scale symmetry (or translation symmetry when moving in scale $s$).  
\label{fig:sym_circle}
}
\end{figure}

\textit{Rotation symmetry.---} A layer $\mathcal{W}_{+}^{(m,m+1)}$ of euclidean MERA (between spin chains $m$ and $m+1$) is a tensor network constructed by multiplying $N_{m+1}$ times a basic building block in a way that it is explicitly invariant under rotations by an angle $\Delta \theta = 2\pi/N_{m+1} = 2\pi 2^{m+1}/N_0$, see Fig. \ref{fig:sym_circle}(a). Notice that this \textit{network} rotation invariance of the layer \textit{implies} that the linear map implemented by the layer is rotation invariant. In addition, we may choose (as we do next) to interpret this \textit{network} rotation invariance as also implying that the distance between nearest neighbour spins is a constant ${\aUV}_{,m}$ within each spin chain (which may still depend on the integer $m$ that labels different spin chains). 
 
We then \textit{promote} the discrete rotation symmetry to continuous rotation symmetry of the metric, 
\begin{eqnarray}
a(s,\theta) &\rightarrow& a(s), \\
b(s,\theta) &\rightarrow& b(s), \\
\Omega(s,\theta) &\rightarrow& \Omega(s). 
\end{eqnarray}

This promotion is compatible with, but not implied by, rules 1 and 2. It is an additional  constraint on the metric, motivated by the discrete network rotation symmetry of $\mathcal{W}_{+}$. We emphasize that layer $W^{(m+1,m+2)}_{+}$ is only invariant under rotations by discrete angles $2\pi 2^{m+2}/N_0$, that is, by angles that are twice as large as those for layer $W^{(m,m+1)}_{+}$. The euclidean MERA as a whole is then only invariant under rotations by the angles allowed by the discrete rotation symmetry of its top layer.

\textit{Scale symmetry.---} The fact that all layers $\mathcal{W}_{+}$ of euclidean MERA are made from the same building block suggests an \textit{apparent} symmetry of the network under simultaneous discrete rescaling in both the radial and angular directions,
\begin{eqnarray}
s_m &\rightarrow& s_m + \log 2 = s_{m+1} ,~~~\\
\theta_{m,n} &\rightarrow&  2 \theta_{m,n} = \theta_{m+1,n},
\end{eqnarray}
see Fig. \ref{fig:sym_circle}(b). This is an \textit{apparent} (as opposed to actual) scale symmetry because it only applies in a local neighbourhood, as opposed to globally. Indeed, globally the radius of the spin chain $m+1$ is only half of the radius of the spin chain $m$. (On the Poincare patch, analyzed in Appendix II, the analogous rescaling will be a true discrete symmetry.) We can promote this (apparent) discrete scale symmetry to an (apparent) continuous scale symmetry by requiring that the metric be invariant under 
\begin{equation}
s \rightarrow s + \log \lambda,~~~~\theta \rightarrow \lambda \theta,
\end{equation}
which implies
\begin{eqnarray}
 \Omega(s) = \Omega e^{-s},~~~~a(s)=ae^{s},~~~~b(s)=be^s.
\end{eqnarray} 
for arbitrary constants $a$, $b$, $\Omega$.

In conclusion, imposing rotation and (apparent) scale symmetries we have arrived at the metric
\begin{equation}
dl^2 = \Omega^2 e^{-2s}\left([\pm a^2 + b^2] e^{2s} ds^2 + 2be^s ds d\theta + d\theta^2 \right),
\end{equation}
with unknown signature ($\pm$ sign) and constants $a$, $b$, and $\Omega$. This corresponds to a generator
\begin{eqnarray}
Q(s) &=& e^{s}a \int d\theta ~h(\theta) - i e^{s}B\int d\theta~p(\theta)\\
&=& e^{s}\left(a H - i b P \right)
\end{eqnarray}
for euclidean signature and
\begin{eqnarray}
Q(s) = ie^{s}\left(a H - b P \right)
\end{eqnarray}
for lorentzian signature (see Appendix \eqref{sect:linear}).

Imposing rule 1 we then obtain (i) euclidean signature ($+$ sign), (ii) $a = q(2\pi/N_0)$, and (iii) b = 0, whereas imposing rule 2 implies $\Omega = r_0$ and $a = q (\aUV /r_0)$. This uniquely leads to $dl_{\Mplus}^2$ in Eq. \eqref{eqap:gMplus}, corresponding to H$_2$ with radius $\rad=\aUV q$ for $r\gg \rad$.

\subsection{Regime $r \sim \rad$}

Recall that the above geometric assignment is only valid when the size of the spin chain $N_m$ is much larger than the width $q$ of euclideons in $\mathcal{W}_{+}$ or, equivalently, the radial coordinate $r$ is much larger than the radius $\rad$. If our goal is for the tensor network to still be a discretized version of H$_2$ also for $r$ on the order of the radius $\rad$, this can be accomplished by adjusting the width $q_m$ of layer $\mathcal{W}_{+}^{(m,m+1)}$ of the tensor network as a function of $m$, so that the layer $\mathcal{W}_{+}^{(m,m+1)}$ implements a finite linear map as generated by $Q_{\H2}$ in \eqref{eqap:QH2}, instead of the asympotic $Q_{\H2}$ in \eqref{eqap:QH2regime}. 

\subsection{Lorentzian signature: real time evolution on the circle}
\label{subsect:circle_lorentz}

Next we study the linear map $V$ obtained from a path integral on an annulus of de Sitter spacetime dS$_2$ and the tensor network geometry of the lorentzian MERA. The analysis is very similar to the one above for hyperbolic space H$_2$ and the euclidean MERA. Accordingly, we proceed by only sketching the argument and highlighting the main differences with the previous case, to which we refer for further details.

As in Eq. \eqref{eqap:VS}, the linear map $V$ is again defined in terms of a path integral on an annulus $\mathcal{A}$ with boundaries at $r = r_{\inn}$ and $r=r_{\out}$ (now for $ r_{\inn}, r_{\out} > a$) according to
\begin{equation} \label{eqap:VS_dS2} 
\bra{\varphi'(\theta)} V \ket{\varphi(\theta)} = \int [D\phi] e^{-iS[\phi(r,\theta)]},
\end{equation}
where now $S[\phi(r,\theta)]$ is the \textit{lorentzian} action. For a diagonal metric of the form
\begin{equation}
dl^2 = \Omega(r)^2 \left(-a(r)^2dr^2 + d\theta^2  \right)
\end{equation}
the generator $Q$ of the linear map $V \approx \mathbb{1} - \epsilon Q$ for a thin cylinder is now given by
\begin{equation}
Q = ia(r) \int_0^{2\pi} d\theta~h(\theta) = ia(r) H_0,
\end{equation}
see Ref. \cite{TNPathInt} and Appendix \eqref{sect:linear}. Specifically, for de Sitter spacetime dS$_2$ in coordinates $(r,\theta)$ we have
\begin{equation}
a(r) = \frac{\rad}{r\sqrt{r^2 + \rad^2}}
\end{equation}
Thus the metric and generator $Q$ read:
\begin{eqnarray}  \label{eqap:dldS2} 
dl^2_{\dS2} &=& \frac{-\rad^2dr^2}{r^2-\rad^2} + r^2d\theta^2 \\
&=& r^2\left(\frac{-\rad^2(dr/r)^2}{ \left(r^2-\rad^2\right)} + d\theta^2 \right),\\
Q_{\dS2} &=& i\frac{\rad}{ r\sqrt{r^2-\rad^2}}~H_0,  \label{eqap:QdS2} 
\end{eqnarray}
with $r\geq \rad$. When the radial coordinate $r$ is much larger than the radius $\rad$, $r \gg \rad$ (this is the regime where this geometry is more straightforwardly connected to the lorentzian MERA), the metric and generator simplify to
\begin{eqnarray} 
dl^2_{\dS2} &\approx& \frac{-\rad^2dr^2}{r^2} + r^2d\theta^2 \\
&=& r^2\left(\frac{-\rad^2dr^2}{r^4} + d\theta^2 \right),\\
Q_{\dS2} &\approx& i\frac{\rad}{ r^2}~H_0,\label{eqap:QdS22} 
\end{eqnarray}
and the path integral on a thick annulus $r\in [r_{\inn} , r_{\out}]$ for $r_{\inn} \gg \rad$ corresponds to the finite linear map
\begin{eqnarray} 
V_{\dS2} &=& \mathcal{P} \exp \left(-\int_{r_{\inn}}^{r_{\out}} dr ~Q_{\dS2} \right)\\
&\approx& \exp \left(- iH_0 \int_{r_{\inn}}^{r_{\out}} \frac{\rad~dr}{r^2} \right) \label{eqap:VdS2q}\\
&=& \exp \left( -i \frac{\rad(r_{\out}-r_{\inn})}{r_{\out} r_{\inn}} H_0 \right). \label{eqap:VdS2qq}
\end{eqnarray}
We again introduce the scale radial coordinate $s \equiv \log (r_0/r)$, where $r_0$ is some reference radius with $r_0 \geq \rad$ such that the origin $s=0$ of $s$ corresponds to $r=r_0$. 
The metric of dS$_2$, the generator $Q_{\dS2}$, and the finite gate $V_{\dS2}$ read, in the simplifying regime $r \gg \rad$,
\begin{eqnarray}  
dl^2_{\dS2} &\approx& -\rad^2 ds^2  + r_0^2 e^{-2s} d\theta^2 \label{eqap:dldS2ass} \\
&=& \ r_0^2 e^{-2s} \left(-\left(\frac{\rad}{r_0}\right)^2e^{2s}ds^2 + d\theta^2 \right),\label{eqap:dS2dls}\\
Q_{\dS2} &\approx& i\frac{\rad}{ r_0}e^s~H_0, \label{eqap:QdS2b}\\
V_{\dS2} &=& \mathcal{P} \exp \left( -\int^{s_{\out}}_{s_{\inn}} ds ~Q_{\H2} \right)\\
&\approx& \exp \left(-i  \int^{s_{\out}}_{s_{\inn}} ds ~  \frac{a}{r_0} e^s H_0  \right)\\
&=& \exp \left(-i\left(e^{s_{\out}} - e^{s_{\inn}}\right)\frac{\rad}{r_0} ~H_0\right). \label{eqap:dS2Vs}
\end{eqnarray}

We note that the above two sets of expressions for $Q_{\dS2}$ and $V_{\dS2}$, first in terms of the radial coordinate $r$ and later in terms of the radial coordinate $s$, are mutually inconsistent. Indeed, in both cases we assume that time increases as we increase the radial coordinate: the first set assumes that time grows with $r$ and the second set assumes that time grows with $s$. However, the radial coordinate $r$ decreases as $s$ increases, and therefore evolving forward in time $s$ corresponds to evolving backwards in time $r$. From now on we follow the convention that time increases monotonically with $s$, that is, as the radial coordinate $r$ decreases. The 
general formalism in Appendix \eqref{sect:linear} assumes that time progresses when the time coordinate increases. Accordingly, we should correct Eqs. \eqref{eqap:QdS2} and \eqref{eqap:QdS22} for $Q_{\dS2}$ by adding a minus sign, and Eqs. \eqref{eqap:VdS2q} and \eqref{eqap:VdS2qq} for $V_{\dS2}$ by removing the minus sign in the exponential. These expressions will not be used in the rest of this Appendix. (We chose to first write Eqs. \eqref{eqap:QdS2}, \eqref{eqap:QdS22}, \eqref{eqap:VdS2q}, and \eqref{eqap:VdS2qq} inconsistently, and then explain how to correct them, in order not to obscure their relation to the general formalism in Appendix \eqref{sect:linear}). 
 
Notice that in the euclidean signature case, addressed earlier on, the change in time direction does not affect the generator. This is because there the generator of time translations is Hermitian, not anti-Hermitian as in the lorentzian signature case.  considerations would apply in the euclidean signature case. Indeed, in changing the orientation of the radial direction, we need to take the hermitian conjugate of the linear map. With lorentzian signature we have $V_{\dS2}(-\alpha)\equiv e^{-i\alpha H}$ for $\alpha \in \mathbb{R}$ and $H^{\dagger}=H$, and therefore $V_{\dS2}(\alpha)^{\dagger} =  V_{\dS2}(-\alpha)$. Instead, with euclidean signature we have $V_{\H2}(\alpha) \equiv e^{-\alpha H}$ (again with $\alpha \in \mathbb{R}$ and $H^{\dagger}=H$) and thus $V_{\H2}(\alpha)^{\dagger} = V_{\H2}(\alpha)$.

A periodic layer $\mathcal{W}_{-}$ of lorentzian MERA is made of $q$ copies of the periodic lorentzian transfer matrix $\mathcal{T}_{-}$ and a periodic layer $\mathcal{W}$ of MERA,
\begin{equation}
\mathcal{W}_{-} \equiv \mathcal{W} \left(\mathcal{T_{-}}\right)^q.
\end{equation}
The lorentzian transfer matrix $\mathcal{T}_{-}$ is a tensor network, made of tensors called lorentzions, that acts on the Hilbert space of a periodic quantum spin chain made of $N$ spins. By construction, it implements a real time evolution of the form
\begin{equation} \label{eqap:T}
\mathcal{T}_{-} = \exp\left( -iH(N)\right) \approx  \exp \left(i\frac{2\pi}{N} H_0\right).
\end{equation}
Recall once more that the periodic layer $\mathcal{W}$ maps low energy states on $N$ spins to low energy states on $N/2$ spins as the identity operator $\mathbb{1}$. Therefore a periodic layer of lorentzian MERA implements the low energy linear map
\begin{equation} \label{eqap:Wminus1}
\mathcal{W}_{-} \approx \exp\left( -iq\frac{2\pi}{N}H_0 \right) = \exp\left(-iq\frac{\aUV}{r}H_0 \right).
\end{equation}
between states of the CFT from a circle of radius $r=N\aUV/2\pi$ to a smaller circle of radius $r/2$. 

Given the discrete coordinates $(s_m,\theta_{m,n})$ from Eq. \eqref{eqap:discretecoor} applied now to label sites in the spin chains of a lorentzian MERA, and the candidate metric of Eq. \eqref{eqap:genericdl}, rule 1 implies
\begin{equation} \label{eqap:rule1dS}
\mathcal{P} \exp \left(-i\int_{s_m}^{s_{m+1}} ds~Q(s) \right) = \exp \left(-iq\frac{2\pi}{N_0} 2^m~H_0 \right).
\end{equation}
whereas rule 2 implies Eq. \eqref{eqap:rule2} as in the euclidean MERA. Then the same choice of functions $a(s,\theta)$, $b(s,\theta)$, $\Omega(s,\theta)$ as in Eq. \eqref{eqap:ABOmega} as in the euclidean case, but with the choice of lorentzian signature ($-$ sign) in metric \eqref{eqap:genericdl}, leads to a solution of the constraints \eqref{eqap:rule1dS} (rule 1) and \eqref{eqap:rule2} (rule 2) with metric and generator
\begin{eqnarray} \label{eqap:gMmin}
dl^2_{\Mmin} &\equiv& - \aUV^2 p^2ds^2 + r_0^2e^{-2s}d\theta^2 \\
&=& r_0^2e^{-2s}  \left(-\left(\frac{\aUV}{r_0}\right)^2p^2e^{2s}ds^2 +d\theta^2 \right),\\
Q_{\Mmin} &\equiv& i \frac{\aUV}{r_0}qe^s H_0.
\end{eqnarray}
The metric corresponds to de Sitter spacetime dS$_2$ with radius $\rad = \aUV q$, see metric $dl^2_{\dS2}$ in Eq. \eqref{eqap:dldS2ass}.

\subsection{Null signature: no time evolution on the circle}
\label{subsect:circle_null}

To compute the linear map $V$ corresponding to a path integral on an annulus of the light cone L$_2$, we will take the zero radius limit, $\rad\rightarrow 0$, of the derivation for either H$_2$ or dS$_2$. We just list the resulting metric $dl^2_{\L2}$, generator $Q_{\L2}$, and linear map $V_{\L2}$,
\begin{eqnarray}
dl_{\L2}^2 &=& r^2 d\theta^2 = r_0^2 e^{-s} d\theta^2,\\
Q_{\L2} &=& 0,\\
V_{\L2} &=& \mathbb{1},
\end{eqnarray}
as well as tensor network linear map $\mathcal{W}$ and proposed tensor network geometry $dl^2_{\mathcal{M}}$
\begin{eqnarray}
\mathcal{W} &=& \mathbb{1},\\
dl_{\mathcal{M}}^2 &=& (r_0)^2 e^{-2s} d\theta^2.
\end{eqnarray}

Since a layer $\mathcal{W}$ of null MERA does not implement either euclidean nor lorentzian time evolution, from a path integral perspective the geometry of the null MERA is neither the hyperbolic space H$_2$ nor the de Sitter spacetime dS$_2$.

\subsection{Discrete sequence of tensor network geometries }

\begin{figure}
\includegraphics[width=8.5cm]{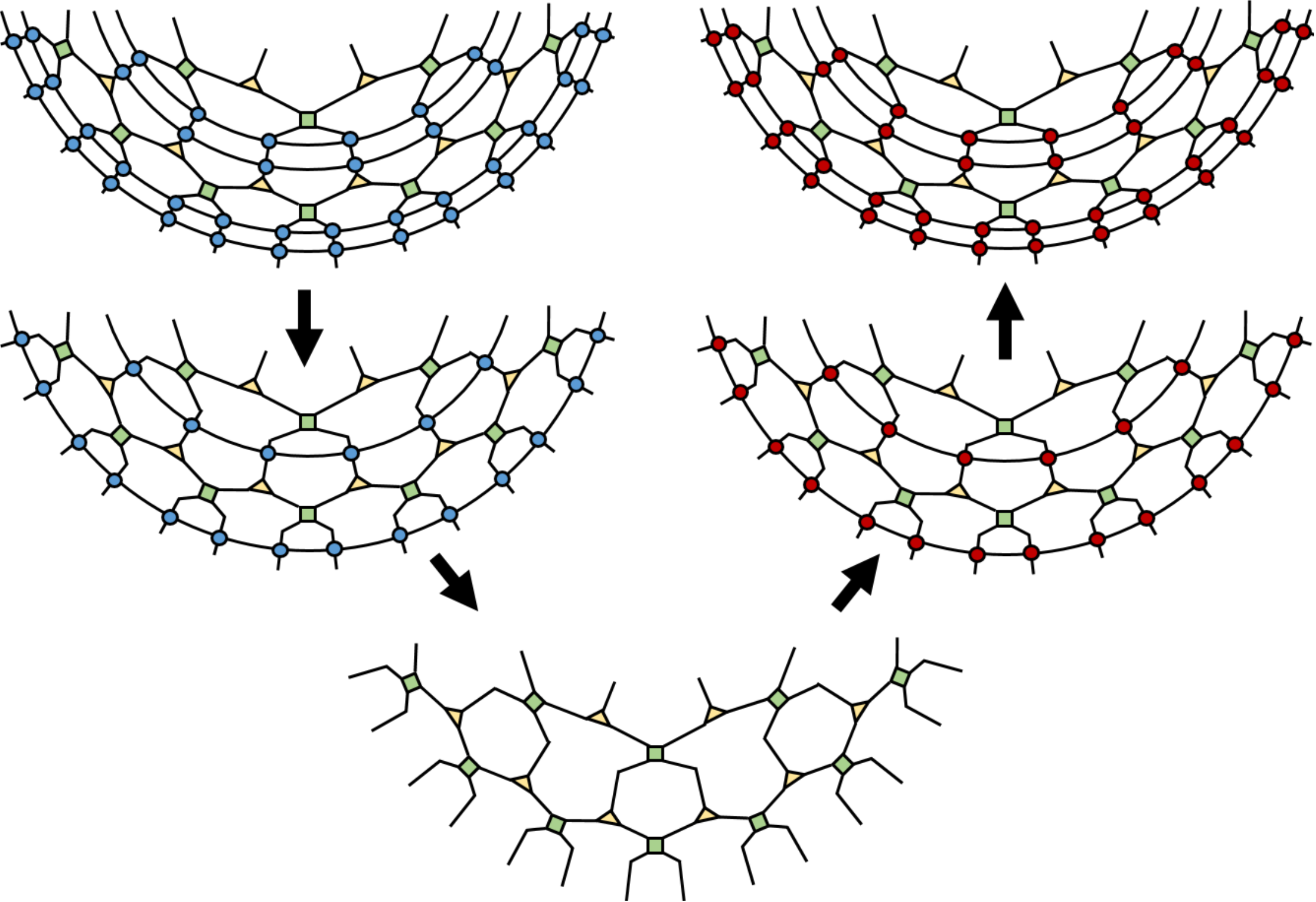}
\caption{
We can consider a sequence of hyperbolic spaces H$_2$ for decreasing values of the (continuous) radius $\rad$, which has L$_2$ in the limit $\rad \rightarrow 0$, then continue with de Sitter spacetimes dS$_2$ for increasing values of the radius $\rad$, see Fig. \ref{fig:limit}. The above discrete sequence of tensor networks mimics that for discrete values $\rad = \aUV q$ of the radius $\rad$, for $q = 0,1,2,\cdots$.
\label{fig:sequence_circle}
}
\end{figure}

To summarize, we have seen that the euclidean, and lorentzian MERA tensor networks correspond to a discrete version of a CFT path integral over the hyperbolic space H$_2$ and the de Sitter spacetime dS$_2$ in the $r\gg \rad$ regime, whereas the null MERA corresponds to a CFT path integral over the light cone L$_2$. These three geometries can be embedded in the same ambient space $\mathbb{R}^{1,2}$, where both H$_2$ and dS$_2$ become L$_2$ in the limit of a small radius $\rad \rightarrow 0$. Analogously, the null MERA $\mathcal{M}$ is a particular case of either the euclidean MERA $\mathcal{M}_{+}$ or lorentzian MERA $\mathcal{M}_{-}$, namely when there are $q=0$ transfer matrices $\mathcal{T}$ or $\mathcal{T}_{-}$ (implementing euclidean or real time evolution) between layers $\mathcal{W}$, see Fig. \ref{fig:sequence_circle}.

\section{Appendix: MERA on the real line}
\label{sect:line}

\begin{figure}
\includegraphics[width=7cm]{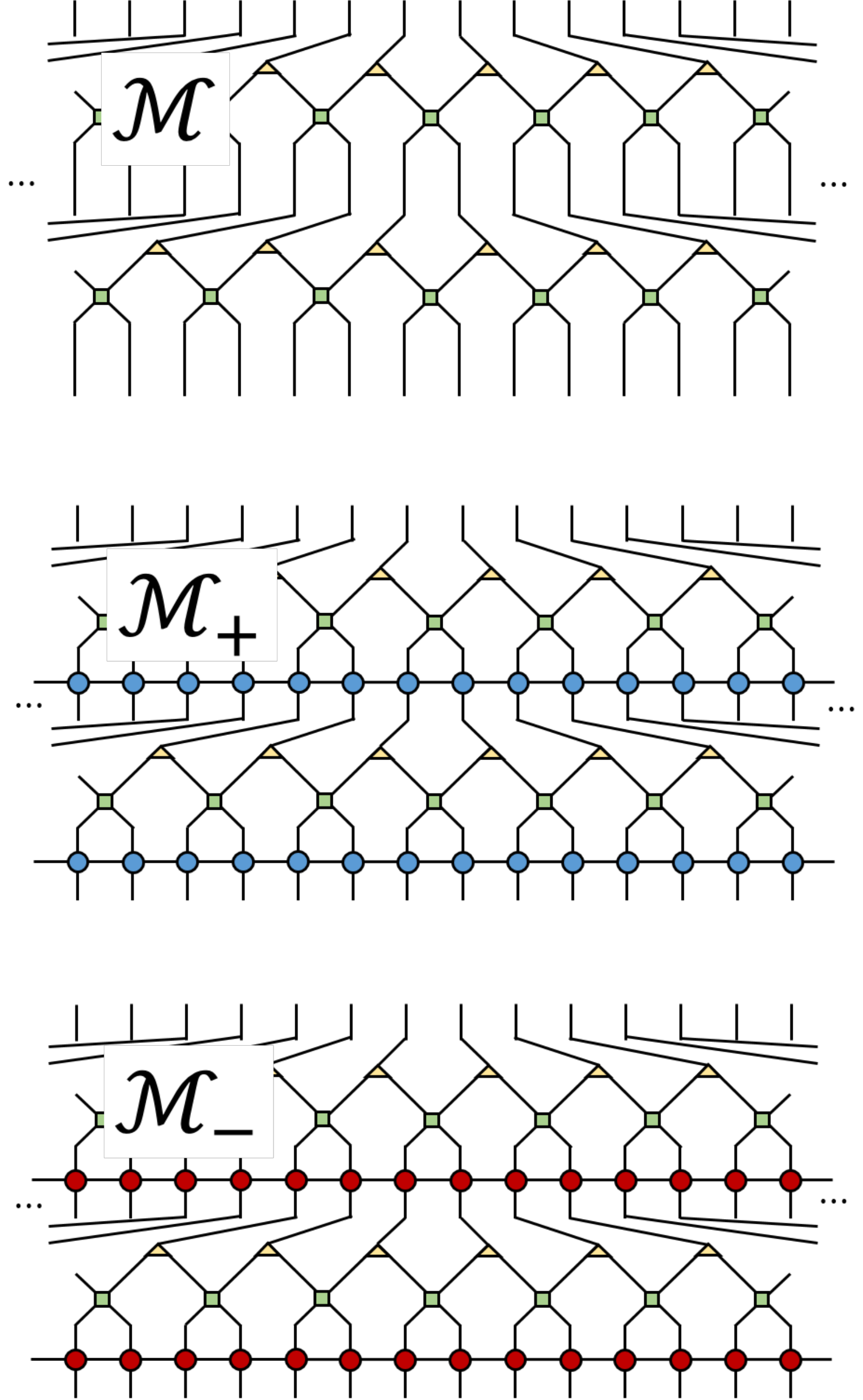}
\caption{
\label{fig:mera3line}
Graphical representation of several MERA tensor networks on the line (only a finite part of two layers of each different MERA are displayed). 
Top: null MERA $\mathcal{M}$, with layers $\mathcal{W}$ made of disentanglers and isometries.
Bottom left: euclidean MERA $\mathcal{M}_{+}$, with layers $\mathcal{W}_{+} = \mathcal{W}\mathcal{T}^{q}$ (for $q=1$), where $\mathcal{T}$ (made of euclideons, blue colour) is an euclidean transfer matrix that implements euclidean time evolution.
Bottom right: lorentzian MERA $\mathcal{M}_{-}$, with layers $\mathcal{W}_{-} = \mathcal{W}\mathcal{T}_{-}^{q}$ (for $q=1$), where $\mathcal{T}_{-}$ (made of lorentzions, red colour) is an lorentzian transfer matrix that implements real time evolution.}
\end{figure}

In this Appendix we review how to assign a path integral geometry to the \textit{null}, \textit{euclidean}, and \textit{lorentzian} MERA on the real line, see Fig. \ref{fig:mera3line}. In this paper we denote L$_2^p$, H$_2^p$, and dS$_2^p$ the light sheet, hyperbolic plane, and Poincare de Sitter spacetime geometries, respectively. Then the assignment of geometries to tensor networks is as follows: 

\begin{equation}
\begin{array}{ccc}
\begin{array}{c} \mbox{null} \\ \mbox{MERA}~\mathcal{M}  \end{array} 
&\leftrightarrow& 
\begin{array}{c} \mbox{light} \\ \mbox{sheet L$_2^p$} \end{array} \\
&&\\
\begin{array}{c} \mbox{euclidean} \\ \mbox{MERA}~\mathcal{M}_{+}  \end{array}  
&\leftrightarrow& 
\begin{array}{c} \mbox{hyperbolic} \\ \mbox{plane H$_2^p$} \end{array} \\
&&\\
\begin{array}{c} \mbox{lorentzian} \\ \mbox{MERA}~\mathcal{M}_{-} \end{array}  
&\leftrightarrow& 
\begin{array}{c} \mbox{Poincare de Sitter} \\ \mbox{spacetime dS$_2^p$} \end{array}
\end{array}
\end{equation}

First we review how H$_2^p$, dS$_2^p$, and L$_2^p$ can be embedded in a common three-dimensional ambient space, namely Poincare AdS$_3$, denoted AdS$_3^p$. Then we characterize the different linear maps obtained through a CFT path integral on a strip of the H$_2^p$, dS$_2^p$, and L$_2^p$ geometries. Finally, using the two rules of Ref. \cite{TNPathInt}, we establish that an infinite layer of euclidean, lorentzian, and null MERA corresponds to a strip of H$_2^p$, dS$_2^p$, and L$_2^p$ geometries, respectively.

\subsection{Embeddings in Poincare anti de Sitter AdS$_3^p$}
 
\begin{figure}
\includegraphics[width=\linewidth]{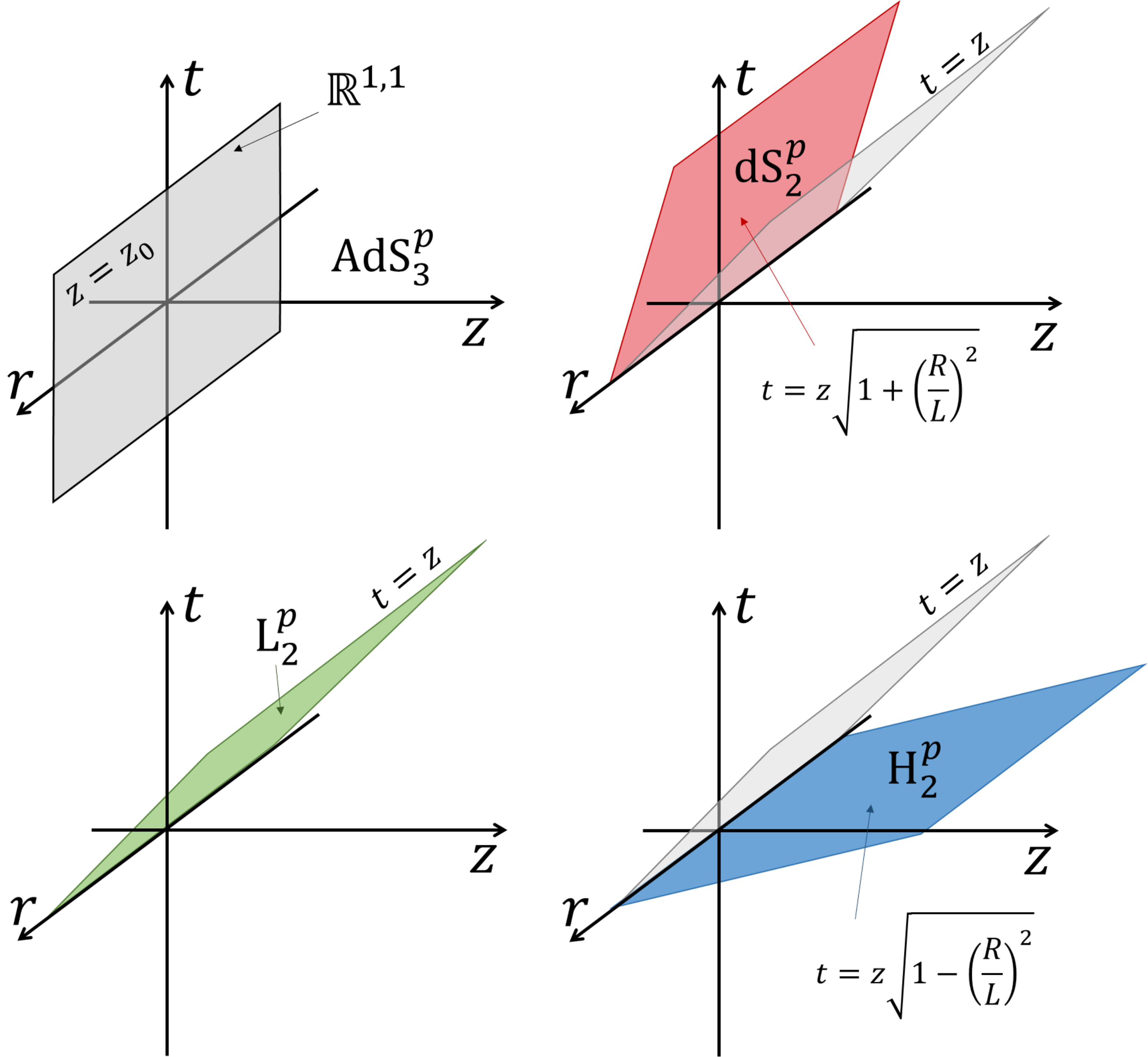}
\caption{
\label{fig:patch} Poincare patch of anti de Sitter spacetime AdS$_3$ with radius $L$ and coordinates $(t,z,r)$. For each fixed value of $z$ we obtain a copy of Minkowski $\mathbb{R}^{1,1}$. By constraining $t$ and $z$ through $t = z\sqrt{1+\sigma \rad^2/L^2}$ with $\sigma = +1,0,-1$ we obtain the two-dimensional geometries H$_2$, L$_2$, and dS$_2$, respectively.
}
\end{figure}

The Poincare patch of AdS$_3$ with radius $L$ has metric
\begin{equation} \label{eqap:AdS3p}
dl_{\AAdS3}^2 = \frac{-dt^2 + dz^2 + dr^2}{(z/L)^2}
\end{equation}
for $r,t \in \mathbb{R}$ and $z\geq 0$. It is a useful ambient space for the three geometries under consideration, see Fig. \ref{fig:patch}. Indeed, each of the H$_2^p$, dS$_2^p$, and L$_2^p$ geometries can be regarded as residing in an \textit{upper half plane} $(\eta,r)$ for $\eta>0$, where $\eta$ is some linear combination of time coordinate $t$ and scale coordinate $z$. The euclidean, lorentzian, or null signature of the induced two-dimensional metric is determined by the ratio of time $t$ versus scale $z$ in this linear combination.

Firstly, a hyperbolic plane H$_2^p$ with radius $\rad$, where $L \geq  \rad > 0$, and metric
\begin{equation}
dl_{\HH2}^2 = \frac{d\eta^2 + dr^2}{(\eta/\rad)^2} 
\end{equation}
can be obtained through the embedding
\begin{eqnarray}
t = \sqrt{\left(\frac{L}{\rad}\right)^2-1} ~\eta, ~~~z = \frac{L}{\rad} \eta 
\end{eqnarray}
for all $r$. 
Secondly, the (upper half plane) de Sitter spacetime dS$_2^p$ with radius $\rad > 0$ and metric
\begin{equation}
dl_{\ddS2}^2 = \frac{-d\eta^2 + dr^2}{(\eta/\rad)^2}  
\end{equation}
is obtained through the embedding
\begin{eqnarray}
t = \sqrt{\left(\frac{L}{\rad}\right)^2+1} ~\eta, ~~~z = \frac{L}{\rad} \eta 
\end{eqnarray}
for all $r$.  (Notice that for $a=\infty$ we obtain $t = \eta$, $z=0$ -- that is, Minkowski $\mathbb{R}^{1,1}$). Finally, the light sheet L$_2^p$ with metric
\begin{equation}
dl_{\LL2}^2 = \frac{dr^2}{(\eta/\rad)^2}
\end{equation}
is given by the embedding
\begin{eqnarray}
t = \frac{L}{a} ~\eta, ~~~z = \frac{L}{\rad} \eta 
\end{eqnarray}
for all $r$, where $\rad>0$ is here some (geometrically meaningless) constant that can be changed by rescaling $\eta$.

\begin{figure}
\includegraphics[width=5cm]{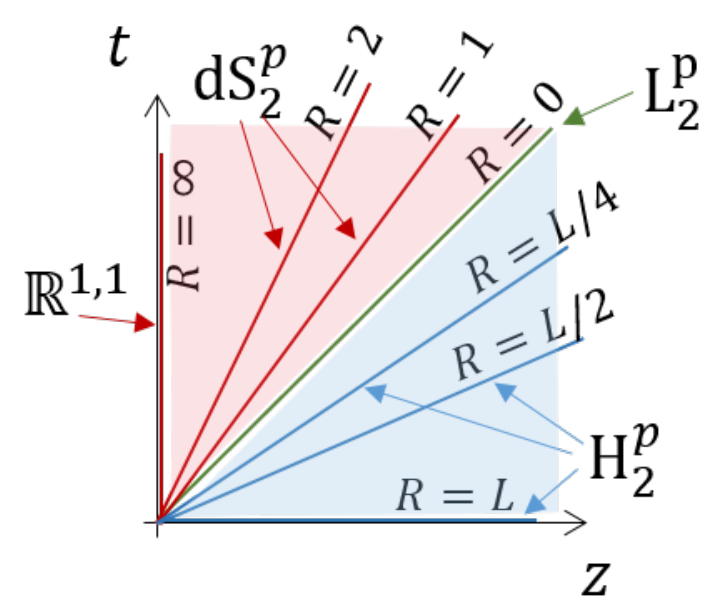}
\caption{
The light sheet geometry L$_2^p$ can be understood as the zero radius limit $\rad \rightarrow 0$ of both the hyperbolic plane H$_2^p$ and Poincare de Sitter space dS$_2^p$, as can be seen when considering these two-dimensional geometries as embedded in Poincare AdS$_3$, where $t$ is the time coordinate, $z$ is the (scale-)space coordinate, and the space coordinate $r$ has been omitted.
\label{fig:limitp} 
}
\end{figure}

Recall that one merit of the Poincare patch AdS$_3^p$ of AdS$_3$ is that the metric \eqref{eqap:AdS3p} explicitly displays its invariance under a subgroup of symmetries of AdS$_3$, namely the Poincare group $SO(1,1)$ of transformations of the Minkowski space $\mathbb{R}^{1,1}$ obtained for each fixed value of the $z$ coordinate, as coordinated by $(t,r)$ and with metric $dl^2 \sim (-dt^2 + dr^2)$. However, the Poincare patch AdS$_3^p$ only covers part (that is, a patch) of AdS$_3$.

In the subsequent discussion it is useful to parametrize these metrics using the scale coordinate $z$ instead of $\eta$, where $z\geq 0$. In the above expressions we redefine $z \equiv \eta/\rad$ as a dimensionless coordinate (where now $\rad \geq 0$ is not bounded by $L$) and obtain
\begin{eqnarray}
dl_{\HH2}^2 &=& \frac{\rad^2dz^2 + dr^2}{z^2},\\
dl_{\ddS2}^2 &=& \frac{-\rad^2dz^2 + dr^2}{z^2},\\
dl_{\LL2}^2 &=& \frac{dr^2}{z^2}.
\end{eqnarray}
These expressions make manifest that the metric of the light sheet L$_2^p$ is the limit $\rad\rightarrow 0$ of either the metric of the hyperbolic plane H$_2^p$ or of the Poincare de Sitter spacetime dS$_2^p$, see Fig. \ref{fig:limitp}. 
 
\subsection{Linear map by path integral II: strip}

\begin{figure}
\includegraphics[width=8.5cm]{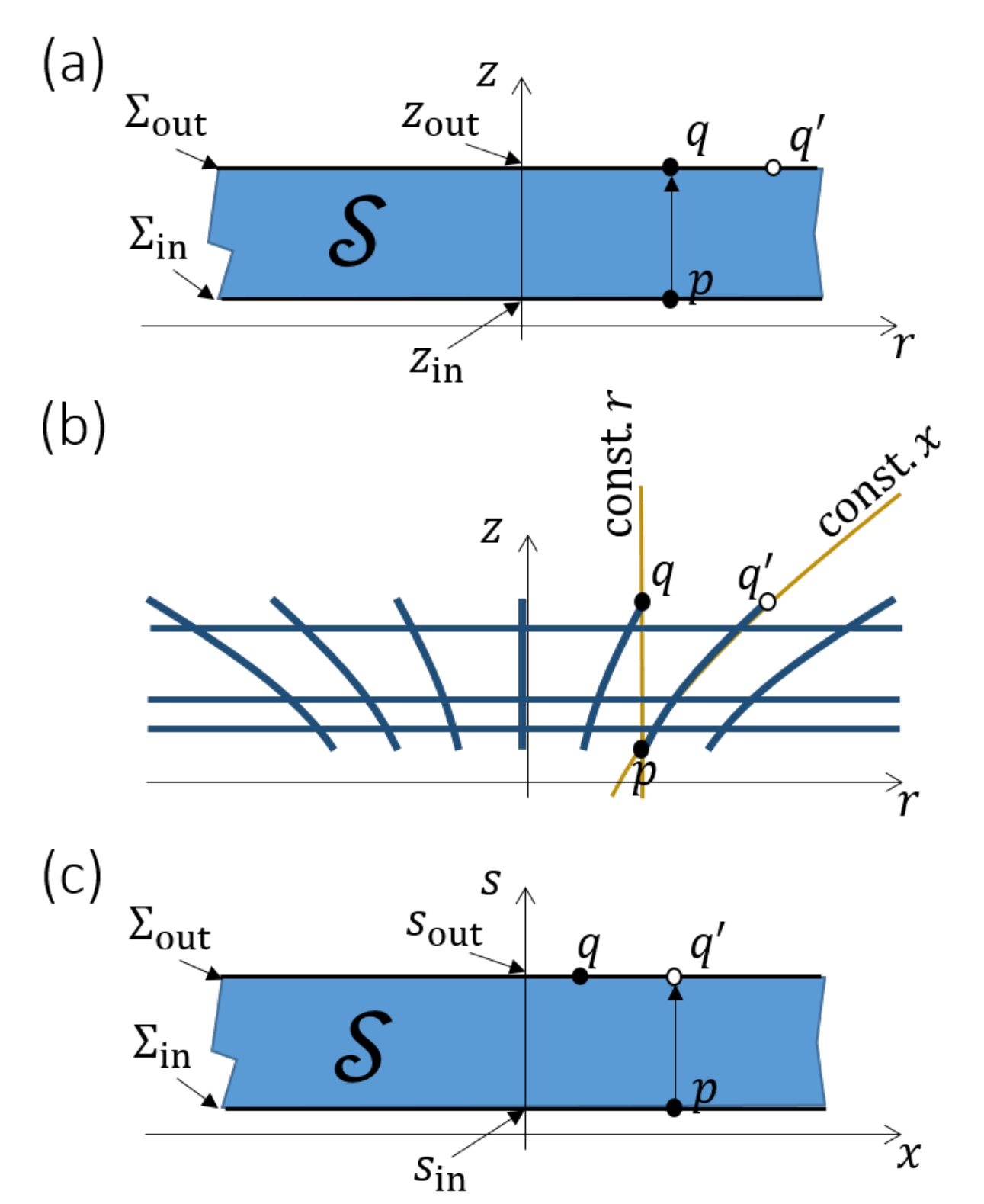}
\caption{
(a) Strip $\mathcal{S}$ with boundaries $\Sigma_{\inn}$ and $\Sigma_{\out}$ corresponding to the lines $z=z_{\inn}$ and $z=z_{\out}$ where $z{\inn} < z_{\out}$. The constant-$r$ identification identifies point $p\in \Sigma_{\inn}$ with point $q \in \Sigma_{\out}$.
(b) Coordinates $(s,x)$ as viewed in coordinates $(z,r)$. Notice that constant $r$ and constant $x$ curves are inequivalent. Therefore the constant-$r$ and constant-$x$ identifications of the boundaries of the strip $\mathcal{S}$ are inequivalent.
(c) The same strip $\mathcal{S}$ with same boundaries $\Sigma_{\inn}$ and $\Sigma_{\out}$ as above, now corresponding to the lines $s=s_{\inn}$ and $s=s_{\out}$ where $s{\inn} < s_{\out}$ (where $s= \log (z)$). The constant-$x$ identification identifies point $p\in \Sigma_{\inn}$ with point $q' \in \Sigma_{\out}$. 
\label{fig:coord_line}
}
\end{figure}

The above geometries H$_2^p$, dS$_2^p$, and L$_2^2$ can all be sliced into slices $\Sigma_z$ defined by a constant value of the scale coordinate $z$. Each slice $\Sigma_z$ corresponds to the real line. Two such real lines with scale coordinate $z = z_{\inn}$ and $z = z_{\out}$, with $z_{\inn} < z_{\out}$, define a horizontal strip $\mathcal{S}$ characterized by $z \in [z_{\inn}, z_{\out}]$, $r \in \mathbb{R}$, see Fig. \ref{fig:coord_line}(a).

Consider a two-dimensional CFT with field $\phi(z,r)$ and action functional $S[\phi(z,r)]$. In any of the three geometries, given $\Sigma_{\inn}$ for $z=z_{\inn}$, we can define the Hilbert space $\mathcal{H}(\Sigma_{\inn})$ of the CFT on that real line in terms of basis states $\ket{\varphi(r)}$. Here $\varphi(r)$, for $r \in \mathbb{R}$, is a field configuration that results from restricting the field $\phi(z,r)$ to the line $\Sigma_{\inn}$. We can similarly define the Hilbert space $\mathcal{H}(\Sigma_{\out})$ of the CFT on a second real line $\Sigma_{\out}$ given by $z = z_{\out}$, again with basis $\ket{\varphi(r)}$. We can then identify the two Hilbert spaces $\mathcal{H}_{\inn}$ and $\mathcal{H}_{\out}$ (from now on simply $\mathcal{H}$) by identifying basis vectors according to
\begin{equation} \label{eqap:identR}
\ket{\varphi(r)}_{\Sigma_{\inn}} \sim \ket{\varphi'(r)}_{\Sigma_{\out}}, ~\mbox{iff} ~\varphi(r) = \varphi'(r) ~~\mbox{for all} ~r.
\end{equation}

For concreteness, from now on we continue the discussion for the hyperbolic plane H$_2^p$, and we postpone the analysis of dS$_2^p$ and L$_2^p$ until the end of this Appendix. Given a strip $\mathcal{S}$ with boundaries $\Sigma_{\inn}$ and $\Sigma_{\out}$ and the above identification between the corresponding Hilbert spaces, we introduce the linear map $\tilde{V}:\mathcal{H} \rightarrow \mathcal{H}$ given by the path integral on that strip $\mathcal{S}$, that is with matrix elements
\begin{equation} \label{eqap:Vline} 
\bra{\varphi'(r)} \tilde{V} \ket{\varphi(r)} = \int [D\phi] e^{-S[\phi(z,r)]},
\end{equation}
where $S[\phi(z,r)]$ is the \textit{euclidean} action of field configurations $\phi(z,r)$ restricted to the strip $\mathcal{S}$ and with boundary conditions
\begin{equation}
\phi(z_{\inn}, r) = \varphi(r),~~~~~\phi(z_{\out}, r) = \varphi'(r).
\end{equation}
 
Above we denoted the linear map $\tilde{V}$ instead of $V$ because later on we will use $V$ to denote a second linear map corresponding to a second set of coordinates $(s,x)$ that are in some sense more natural when describing the euclidean MERA on the real line.

\subsection{Thin strip}
  
In the limit of a thin annulus, when $z_{\out} - z_{\inn} = \epsilon$ for small $\epsilon>0$, we can expand the linear map $\tilde{V} \approx \mathbb{1} - \epsilon \tilde{Q}$ in terms of a generator $\tilde{Q}$. An expression for $\tilde{Q}$ is found by specializing to the current case the general solution of Ref. \cite{TNPathInt}, which is reviewed in Appendix \eqref{sect:linear}. For a diagonal metric of the form
\begin{equation}
dl^2 = \Omega(z)^2 \left(a(z)^2 dz^2 + dr^2 \right)
\end{equation}
this generator is 
\begin{eqnarray}
\tilde{Q} &=& a(z) \int_{-\infty}^{\infty} dr~h(r) = a(z)~H, \\
H &\equiv& \int_{-\infty}^{\infty} dr~h(r).
\end{eqnarray}

For H$_2^p$ in coordinates $(z,r)$ we have a constant $a(z)$, namely
\begin{eqnarray}
a(z) = \rad,\\
\end{eqnarray}
and the metric and generator read:
\begin{eqnarray}  \label{eqap:dlH2p} 
dl^2_{\HH2} &=& \frac{1}{z^2} \left(\rad^2 dz^2 + dr^2\right), \\
\tilde{Q}_{\HH2} &=& \rad~H.  \label{eqap:QH2Sp} 
\end{eqnarray}

\subsection{Thick strip}

The path integral on a thick strip $z\in[z_{\inn},z_{\out}]$ produces the finite linear map 
\begin{eqnarray} 
\tilde{V}_{\HH2} &=& \mathcal{P} \exp \left(-\int_{z_{\inn}}^{z_{\out}} dz ~\tilde{Q}_{\HH2} \right)\\
&=& \exp \left(- \rad H \int_{z_{\inn}}^{z_{\out}} ~dz\right) \\
&=& \exp \left(-\rad\left(z_{\out} - z_{\inn}\right) H \right). \label{eqap:VH2a}
\end{eqnarray}
corresponding to an euclidean time evolution by and amount $\rad(z_{\out} - z_{\inn})$ of euclidean time.

\subsection{Other coordinates}

It is useful to also consider dimensionless coordinates ($s,x$) of the upper half plane given in terms of $z$ and $r$ according to (see Fig. \ref{fig:coord_line}(b))
\begin{equation} \label{eqap:zrTOsx}
s \equiv \log (z),~~~~x \equiv \frac{r}{\aUV ~z},
\end{equation}
or
\begin{equation}
z = e^{s},~~~~r = \aUV~ e^{s} x,
\end{equation}
where $s,x \in \mathbb{R}$ and $\aUV$ is a short-distance length scale that will correspond to the lattice spacing in the tensor network. The metric of H$_2^p$ then reads
\begin{equation}
\label{eqap:dlsx}
dl^2_{\HH2} = \aUV^2 \left(\left[ \frac{\rad^2}{\aUV^2}+x^2\right]ds^2 + 2x~ds ~dx + dx^2\right).
\end{equation}
In these coordinates the metric is no longer diagonal. Given two real lines $\Sigma_{\inn}$ and $\Sigma_{\out}$ corresponding to $s=s_{\inn}$ and $s=s_{\out}$, and corresponding Hilbert spaces $\mathcal{H}_{\inn}$ and $\mathcal{H}_{\out}$, a natural identification between states in $\mathcal{H}_{\inn}$ and $\mathcal{H}_{\out}$ in the coordinates $(s,x)$ is now (see Fig. \ref{fig:coord_line}(c))
\begin{equation} \label{eqap:identX}
\ket{\varphi(x)}_{\Sigma_{\inn}} \sim \ket{\varphi'(x)}_{\Sigma_{\out}}, ~\mbox{iff} ~\varphi(x) = \varphi'(x) ~~\mbox{for all} ~x,
\end{equation}
which is not equivalent to that in Eq. \eqref{eqap:identR}. As a result, the linear map \begin{equation} \label{eqap:Vline2} 
\bra{\varphi'(x)} V \ket{\varphi(x)} = \int [D\phi] e^{-S[\phi(s,x)]},
\end{equation}
where the boundary conditions of the field $\phi(s,x)$ are given by 
\begin{equation}
\phi(s_{\inn}, x) = \varphi(x),~~~~~\phi(s_{\out}, x) = \varphi'(x),
\end{equation}
will have a different generator than $\tilde{V}_{\HH2}$ above. For a thin strip $\mathcal{S}$ with
 $s_{\out} - s_{\inn} = \epsilon$ for small $\epsilon>0$, and metric
\begin{eqnarray}
&&dl^2 = \Omega(s,x)^2 \times \\
 &&~~~\left([a(s,x)^2 + b(s,x)^2] ds^2 + 2b(s,x)^2dsdx + dx^2\right)~~~~~
\end{eqnarray}
the linear map $V \approx \mathbb{1} - \epsilon Q$ is generated by $Q = Q_0 - iQ_1$ (see Appendix \eqref{sect:linear}) with
\begin{eqnarray} \label{eqap:Q0b}
Q_0 &\equiv& \int_{\Sigma_{\inn}} dx ~ a(s,x) ~ h(x), \\
Q_1 &\equiv& \int_{\Sigma_{\inn}} dx ~ b(s,x) ~ p(x), \label{eqap:Q1b}
\end{eqnarray}
which for the current metric \eqref{eqap:dlsx}, with 
\begin{eqnarray}
\Omega(s,x) = \aUV,~~~ a(s,x)  = \frac{\rad}{\aUV}, ~~~b(s,x) = x,~~~
\end{eqnarray}
leads to
\begin{eqnarray} \label{eqap:Q0c}
Q_0 &\equiv& \frac{\rad}{\aUV} \int_{\Sigma_{\inn}} dx  ~h(x) = \frac{\rad}{\aUV} ~H, \\
Q_1 &\equiv& \int_{\Sigma_{\inn}} dx ~ x ~ p(x) = ~ D, \label{eqap:Q1c}
\end{eqnarray}
that is 
\begin{eqnarray}
Q_{\HH2} = \frac{\rad}{\aUV} H - i D,
\end{eqnarray}
where $H$ and $D$ are the CFT Hamiltonian and dilation operators,
\begin{eqnarray}
H \equiv \int_{-\infty}^{\infty} dx~h(x), ~~~~D \equiv \int_{-\infty}^{\infty} dx~x~p(x).
\end{eqnarray}

The path integral on a finite strip $s \in [s_{\inn}, s_{\out}]$, which corresponds to the strip $z \in [z_{\inn} = e^{s_{\inn}},z_{\out}=e^{s_{\out}}]$, produces the linear map
\begin{eqnarray}
V_{\HH2} &=& \mathcal{P} \exp\left(-\int_{s_{\inn}}^{s_{\out}} ds~Q_{\HH2} \right)\\
&=& \exp \left( -\left(\frac{\rad}{\aUV}H - i D\right) \int_{s_{\inn}}^{s_{\out}} ds  \right)\\
&=& \exp \left( -  (s_{\out}-s_{\inn})\left(\frac{\rad}{\aUV}H - i D\right) \right). \label{eqap:VH2b}
\end{eqnarray}
The linear map $V_{\HH2}$ in Eq. \eqref{eqap:VH2b} is not the same as $\tilde{V}_{\HH2}$ in Eq. \eqref{eqap:VH2a}, even though they are both produced by the path integral on the same strip, because they result from a different identification of the Hilbert space at the boundaries of the strip. To relate the two, we need to compensate for the difference in the identifications. Let
\begin{equation}
U(s) \equiv \exp \left(isD \right)
\end{equation}
be the unitary transformation that applies a rescaling by a factor $s$. Then the two finite gates are related by
\begin{equation} \label{eqap:VV}
V_{\HH2} = U(s_{\out}) ~\tilde{V}_{\HH2} ~U(s_{\inn})^{\dagger}
\end{equation}
where $s_{\inn} \equiv \log z_{\inn}$, $s_{\out} \equiv \log z_{\out}$. In words, the gate $V_{\HH2}$ is equivalent to $\tilde{V}_{\HH2}$ after we compensate for two rescalings: one before and one after applying the gate, where these two rescalings are by a different amount, namely $s_{\inn}$ and $s_{\out}$.

Indeed, if we use that $i[D,H] = -H$, that is, 
\begin{eqnarray}
U(s) ~H ~U(s)^{\dagger} &=& e^{is D} H e^{-is D} = He^{-s},\\
U(s) ~e^{-\alpha H} ~U(s)^{\dagger} &=& \exp \left( -\alpha H e^{-s}\right)
\end{eqnarray}
then we first arrive to
\begin{eqnarray}
&&e^{-\beta(\alpha H-iD)} = \left(e^{-\epsilon(\alpha H-iD)}\right)^N\\
&\approx& \left(e^{-\epsilon \alpha H} e^{i\epsilon D}\right)\left(e^{-\epsilon \alpha H} e^{i\epsilon D}\right)\cdots \left(e^{-\epsilon \alpha H} e^{i\epsilon D}\right)~~~~~~\\
&=& e^{-\epsilon \alpha H}\left(e^{i\epsilon D}e^{-\epsilon \alpha H}e^{-i\epsilon D}\right)\left(e^{i2\epsilon D}e^{-\epsilon \alpha H}e^{-i2\epsilon D}\right)~~~   \\
&&\cdots\times \left(e^{i(N-1)\epsilon D}e^{-\epsilon \alpha H}e^{-i(N-1)\epsilon D}\right)e^{iN\epsilon D}\\
&=& \exp\left(-\epsilon \alpha H \right)\exp\left(-\epsilon \alpha e^{-\epsilon}H\right)\exp\left(-\epsilon \alpha e^{-2\epsilon}H\right)   \\
&&\cdots\times \exp\left(-\epsilon \alpha e^{-(N-1)\epsilon}H\right)e^{iN\epsilon D}\\
&=& \exp \left(-\epsilon \alpha \left(1 + e^{-\epsilon} + \cdots +  e^{-(N-1)\epsilon}\right) H \right)e^{i\beta D}~~~~~~~~\\
&=& e^{-\mu H}e^{i\beta D},
\end{eqnarray}
where $\epsilon \equiv \beta/N$ and where, in the limit $\epsilon \rightarrow 0$,
\begin{eqnarray}
\mu &=& \alpha ~\lim_{\epsilon \rightarrow 0} ~\sum_{n=0}^{N-1} ~\epsilon~ e^{-\epsilon n} \\
&=& \alpha \int_{0}^{\beta} d\beta' e^{-\beta'} = \alpha \left(1-e^{-\beta}\right).
\end{eqnarray}
Similarly, we can find
\begin{equation}
e^{-\beta(\alpha H-iD)} = e^{i\beta D}e^{-\mu' H}
\end{equation}
with
\begin{eqnarray}
\mu' &=& \alpha ~\lim_{\epsilon \rightarrow 0} ~  \sum_{n=0}^{N-1} ~\epsilon~ e^{\epsilon n}\\
 &=& \alpha \int_{0}^{\beta} d\beta' e^{\beta'} = \alpha \left(e^{\beta}-1\right).~~~~~~
\end{eqnarray}
Thus, we have seen that
\begin{equation} \label{eqap:conversion}
e^{-\alpha(1-e^{-\beta})H} e^{i\beta D} = e^{-\beta\left(\alpha H-iD\right)} = e^{i\beta D}e^{-\alpha(e^{\beta}-1)H}.
\end{equation}
Then, returning to Eq. \eqref{eqap:VV}, we see that, for $\Delta s \equiv s_{\out} - s_{\inn}$, as announced above we have
\begin{eqnarray}
V_{\HH2} &=& e^{-\Delta s(\alpha H-iD)} = e^{i \Delta s D}e^{-\alpha(e^{\Delta s}-1)H} \\
&=& e^{i \Delta s D} ~U(s_{\inn})^{\dagger} \left(U(s_{\inn}) ~e^{-\alpha(e^{\Delta s}-1)H}~ U(s_{\inn}) \right)U(s_{\inn})^{\dagger}~~\nonumber\\
&=&U(s_{\out})  \exp \left( -\alpha(e^{\Delta s}-1)e^{s_{\inn}} H \right) U(s_{\inn})^{\dagger}\\
&=&U(s_{\out}) \exp \left( -\alpha(e^{s_{\out}}-e^{s_{\inn}}) H \right) U(s_{\inn})^{\dagger}\\
&=& U(s_{\out}) \exp \left(-\alpha (z_{\out}-z_{\inn})H \right) U(s_{\inn})^{\dagger}\\
&=& U(s_{\out}) \tilde{V}_{\HH2} U(s_{\inn})^{\dagger}.
\end{eqnarray}
where we used $\alpha$ for $\rad/\aUV$ to simplify the notation.

\begin{figure}
\includegraphics[width=8.5cm]{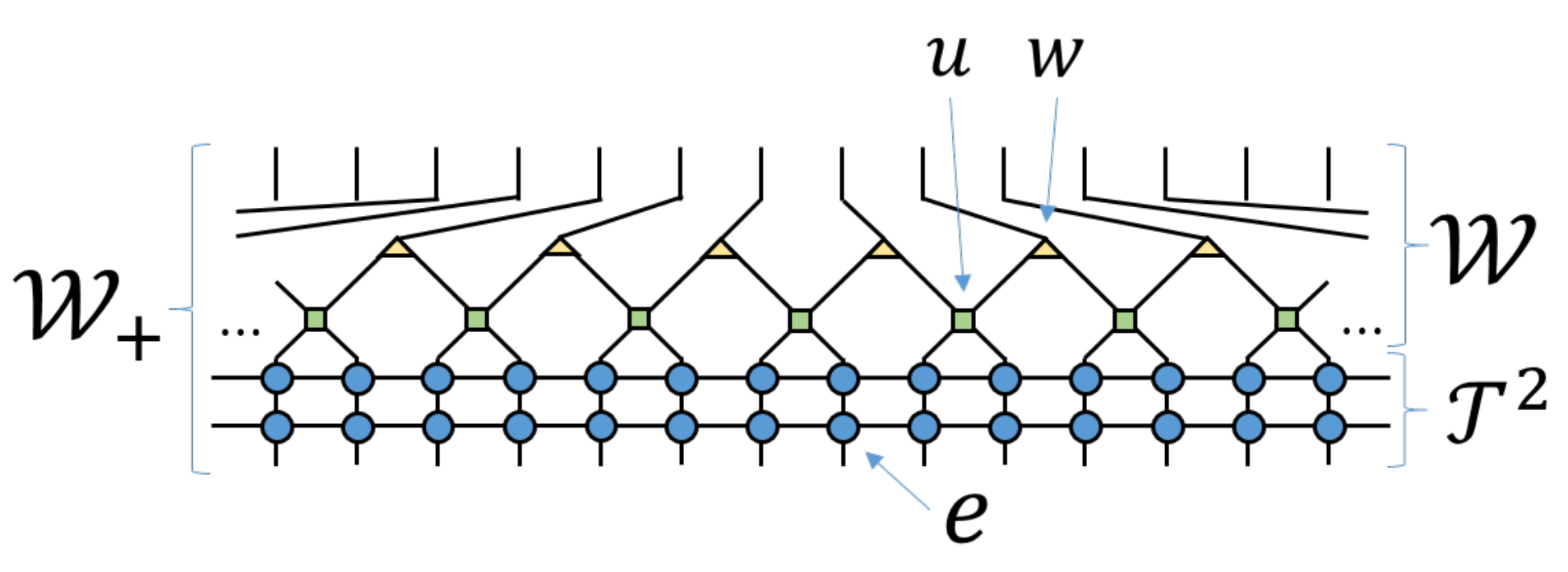}
\caption{
Layer $\mathcal{W}_{+}$ of euclidean MERA on the line. Only part of this infinite layer is shown. $\mathcal{W}_{+}$ is made of the product of $q$ euclidean transfer matrices $\mathcal{T}$ ($q=2$ in the figure) and a layer $\mathcal{W}$ of null or regular MERA, see Eq. \eqref{eqap:Wplusmap}. Each euclidean transfer matrix $\mathcal{T}$ consist of a one-dimensional array of tensors $e$ called euclideons. Each layer $\mathcal{W}$ of null MERA is made of tensors $u$ and $w$ called disentanglers and isometries.
\label{fig:layer_line}
}
\end{figure}

\subsection{Linear map of $\mathcal{W}_{+}$ on the line}

A layer $\mathcal{W}_{+}$ on the line is made of the product of $q$ euclidean transfer matrices $\mathcal{T}$ followed by a layer $\mathcal{W}$ of null MERA,
\begin{equation} \label{eqap:Wplusmap}
\mathcal{W}_{+} \equiv \mathcal{W}~\mathcal{T}^{q},
\end{equation}
see Fig. \ref{fig:layer_line}.

Each euclidean transfer matrix $\mathcal{T}$ is made of an infinite row of euclideons and, by construction, implements the euclidean time evolution
\begin{equation}
\mathcal{T} \approx e^{- H},
\end{equation}
where the spin Hamiltonian $H$ reads
\begin{equation}
 H \equiv \sum_{l=-\infty}^{\infty} h_l.
\end{equation}
On the other hand, which transformation a layer $\mathcal{W}$ of null MERA on the real line implements depends on how we identify the Hilbert spaces of the two spin chains it connects. Following with the above discussion in the continuum, we consider two natural identifications, given by keeping either $r$ or $x$ constant. The constant-$r$ identification in the line is analogous to the constant-$\theta$ identification in the circle discussed in Appendix I. Pairs of contiguous sites in the lower spin chain are mapped into single sites of the upper spin chain. We can then export the result from the circle to the line and conclude that $\mathcal{W}$ (denoted $\tilde{\mathcal{W}}$ in what follows) acts as the identity $\mathbb{1}$ on CFT states on the line,
\begin{equation}
~~~~~~\tilde{\mathcal{W}}  \approx \mathbb{1} ~~~~~~(\mbox{constant-$r$ identification}).
\end{equation} 
In turn, in the constant-$x$ identification, which we will follow here, we identify each spin on the lower spin chain with a spin in the upper spin chain. (This was not possible in the circle because the lower spin chain had twice as many spins as the upper spin chain, but on the line this is not a problem, given that each spin chain has infinitely many spins.) The constant-$x$ identification has been used implicitly in the literature e.g. when extracting conformal data from the scale invariant MERA, in that the identification of local scaling operators is based on diagonalizing a local scaling superoperator that identifies a small number (e.g. three) of sites in one spin chain with the same number of sites in the other spin chain. In this context, a layer of MERA implements a change of scale by a factor $1/2$, that is
\begin{equation}
\mathcal{W} \approx 2^{iD} = \exp \left( i\log (2) D\right) ~~~~~(\mbox{const.-$x$ ident.}).
\end{equation}
Accordingly, the linear map implemented by a layer of euclidean MERA on the line is now:
\begin{equation} \label{eqap:Wplusx}
\mathcal{W}_{+} \approx 2^{iD} e^{-qH} = 2^{-\left(q H-iD\right)},
\end{equation}
where we used Eq. \eqref{eqap:conversion} specialize to $\beta = \log 2$ to obtain
\begin{eqnarray}
e^{-\frac{\alpha}{2}H} 2^{iD} = 2^{-\left(\alpha H-iD\right)} = 2^{iD}e^{- \alpha H}.
\end{eqnarray}

\begin{figure}
\includegraphics[width=\linewidth]{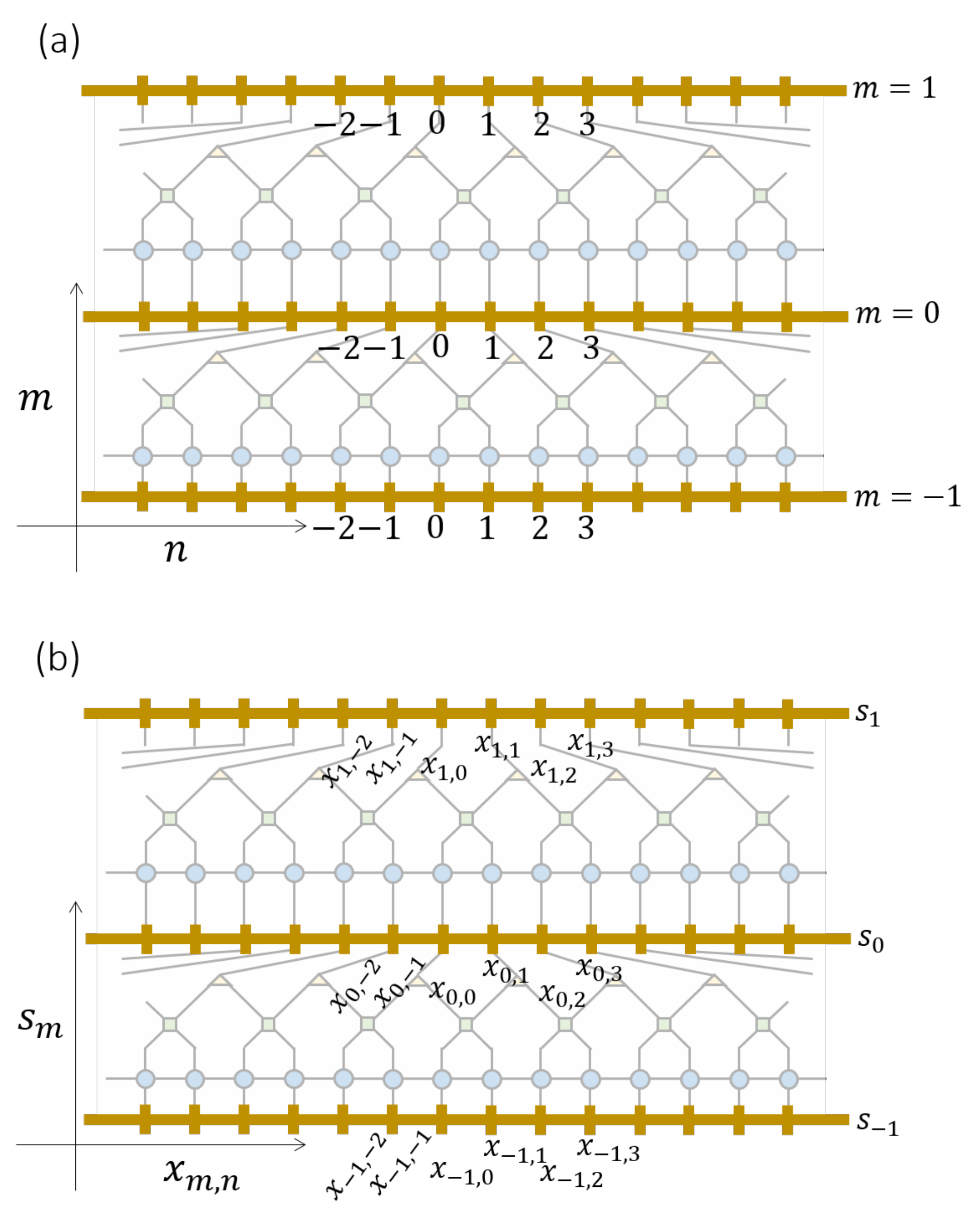}
\caption{
(a) We label each spin chain between two layers of euclidean MERA by an integer $m$, and each sites within that spin chain by the pair $(m,n)$.
(b) We then assign to site $(m,n)$ discrete coordinates $(s_{m},x_{m,n})$ as given in \eqref{eqap:sxdiscretemn}.
\label{fig:sx} 
}
\end{figure}

\subsection{Discrete coordinates on the euclidean MERA on the line}

Consider the euclidean MERA on the line. We use an integer $m$ to label each infinite spin chain between two layers of the euclidean MERA. Each site in spin chain $m$ is labelled by a pair $(m,n)$ of integers, see Fig. \ref{fig:sx}(a). We then associate the discrete coordinates $(s_m,x_{m,n})$ to site $(m,n)$ according to
\begin{equation} \label{eqap:sxdiscretemn}
s_m = m \log 2,~~~~x_{m,n} = n+\frac{1}{2},
\end{equation}
where $\aUV$ is a length scale, see Fig. \ref{fig:sx}(b). Notice that both $s_m$ and $x_{m,n}$ are dimensionless.

A candidate geometry for the euclidean MERA is given by a generic metric in the continuous coordinates $(s,x)$ (see Appendix \eqref{sect:linear}), 
\begin{eqnarray} \label{eqap:sxgeneric}
&&dl^2 = \Omega(s,x)^2 \times \\
 &&~~~\left([\pm a(s,x)^2 + b(s,x)^2] ds^2 + 2b(s,x)^2dsdx + dx^2\right)~~~~~~~~ \nonumber
\end{eqnarray}
where again both $s$ and $x$ are dimensionless, as are functions $a(s,x)$ and $b(s,x)$, whereas the local scale factor $\Omega(s,x)$ has dimensions of length.

\subsection{Rules 1 and 2 of path integral geometry}

Our goal is to determine the signature ($\pm$ sign) and the functions $a(s,x)$, $b(s,x)$, $\Omega(s,x)$ using rules 1 and 2 of Ref. \cite{TNPathInt}.
  
Rule 1 of Ref. \cite{TNPathInt} (\textit{compatibility with path integral}) states that the linear map implemented by a strip $s\in[s_{\inn}, s_{\out}]$ of such geometry, namely
\begin{equation}
V(s_{\inn},s_{\out}) \equiv \mathcal{P} \exp \left(-\int_{s_{\inn}}^{s_{\out}} ds~Q(s) \right)
\end{equation}
where the generator reads
\begin{eqnarray}
Q(s) &=& \int_{-\infty}^{\infty}dx ~\Big(a(s,x) h(x) - i b(s,x) p(x) \Big)
\end{eqnarray}
for euclidean signature ($+$ sign) and
\begin{eqnarray}
Q(s) &=& \int_{-\infty}^{\infty}dx ~\Big(i a(s,x) h(x) - i b(s,x) p(x) \Big)
\end{eqnarray}
for lorentzian signature ($-$ sign), should match, for $s_{\inn} = s_m$ and $s_{\out} = s_{m+1}$, the linear map \eqref{eqap:Wplusx} implemented by the layer $\mathcal{W}_{+}^{(m,m+1)}$ of euclidean MERA between spin chains $m$ and $m+1$. That is, $V(s_m,s_{m+1}) = \mathcal{W}_{+}^{(m,m+1)}$ or
\begin{equation} \label{eqap:rule1sx}
\mathcal{P} \exp \left(-\int_{s_m}^{s_{m+1}} ds~Q(s) \right) = 2^{-\left(q H-iD\right)}.
\end{equation}
This is a constraint on the functions $a(s,x)$ and $b(s,x)$ and the chose of signature ($\pm$ sign), but not on the scale factor $\Omega(s,x)$.

Rule 2 of Ref. \cite{TNPathInt} (\textit{constant lattice spacing}) states that the proper distance between nearest neighbor sites $(m,n))$ and $(m,n+1)$ in spin chain $m$ is the same constant $\aUV$ for all $m$ and $n$. In a constant $s$ cut, the metric \eqref{eqap:sxgeneric} reads $dl^2 = \Omega(s,x)^2 dr^2$ and therefore the distance (within the cut) between points $(s_m,x_{m,n})$ and $(s_m,x_{m,n+1})$ is
\begin{equation}
\int_{x_{m,n}}^{x_{m,n+1}} \sqrt{dl^2(s_m,x)} = \int_{x_{m,n}}^{x_{m,n+1}} dx ~\Omega(s_m,x).
\end{equation}
Rule 2 then implies the constraint
\begin{equation} \label{eqap:rule2sx}
\int_{x_{m,n}}^{x_{m,n+1}} dx ~\Omega(s_m,r) = \aUV.
\end{equation}
This is a constraint on the scale factor $\Omega(s,x)$, and not on the signature or functions $a(s,x)$ and $b(s,x)$ of metric \eqref{eqap:sxgeneric}.

There are now two possible routes to assigning a (path integral) geometry to the euclidean MERA. The first route simply identifies the metric H$_2$ with radius $\rad=\aUV q$ as one that satisfies rules 1 and 2. The second route first restricts the possible geometries on the grounds of discrete symmetries of the network. After that, requiring rules 1 and 2 is seen to completely specify the metric, which is again that of H$_2$ with radius $\rad=\aUV q$.

\subsection{Path integral geometry of euclidean MERA on the line}

Consider the following functions
\begin{eqnarray}\label{eqap:sxABOmega}
a(s,x) = q, ~~~b(s,x) = x,  ~~~\Omega(s,x) = \aUV, ~~~\label{eqap:xsABOmega3} 
\end{eqnarray}
and the choice of euclidean signature ($+$ sign) in metric \eqref{eqap:sxgeneric}. They produce the metric and generator
\begin{eqnarray} \label{eqap:MMplus}
dl^2_{\MMplus} &\equiv& \aUV^2 \left(\left[q^2+x^2\right]ds^2 + 2x~ds ~dx + dx^2\right)\\
Q_{\MMplus} &\equiv&  q   H - i D
\end{eqnarray}
which fulfil rules 1 and 2. 

Indeed, condition \eqref{eqap:rule1sx} is fulfilled since we find
\begin{eqnarray}
-\int_{s_m}^{s_{m+1}} ds ~Q_{\MMplus}~   &=&  -\left(s_{m+1} - s_{m} \right) \left(q   H - i D \right) ~~~~\\
&=& - \log 2\left(q   H - i D \right),
\end{eqnarray}
whereas condition \eqref{eqap:rule2sx} is fulfilled because
\begin{equation}
\int_{x_{m,n}}^{x_{m,n+1}} dx ~\aUV = \aUV (x_{m,n+1} - x_{m,n}) = \aUV.
\end{equation}

We immediately recognize $dl^2_{\MMplus}$ above as the metric \eqref{eqap:dlsx} of the hyperbolic plane H$_2^p$ with radius $\rad=q\aUV$. In the coordinates $(z,r)$ it reads
\begin{equation} \label{eqap:gsymm2}
dl_{\MMplus}^2 =  \frac{1}{z^2} \left((q\aUV)^2dz^2 +  dr^2\right). 
\end{equation}

\subsection{First symmetries, then rules 1 and 2}

Alternatively, we can start again with a general metric specified by a choice of signature and three generic functions $a(s,x)$, $b(s,x)$, and $\Omega(s,x)$ in Eq. \eqref{eqap:sxgeneric} and first impose translation and scale invariance, then rules 1 and 2.

\begin{figure}
\includegraphics[width=\linewidth]{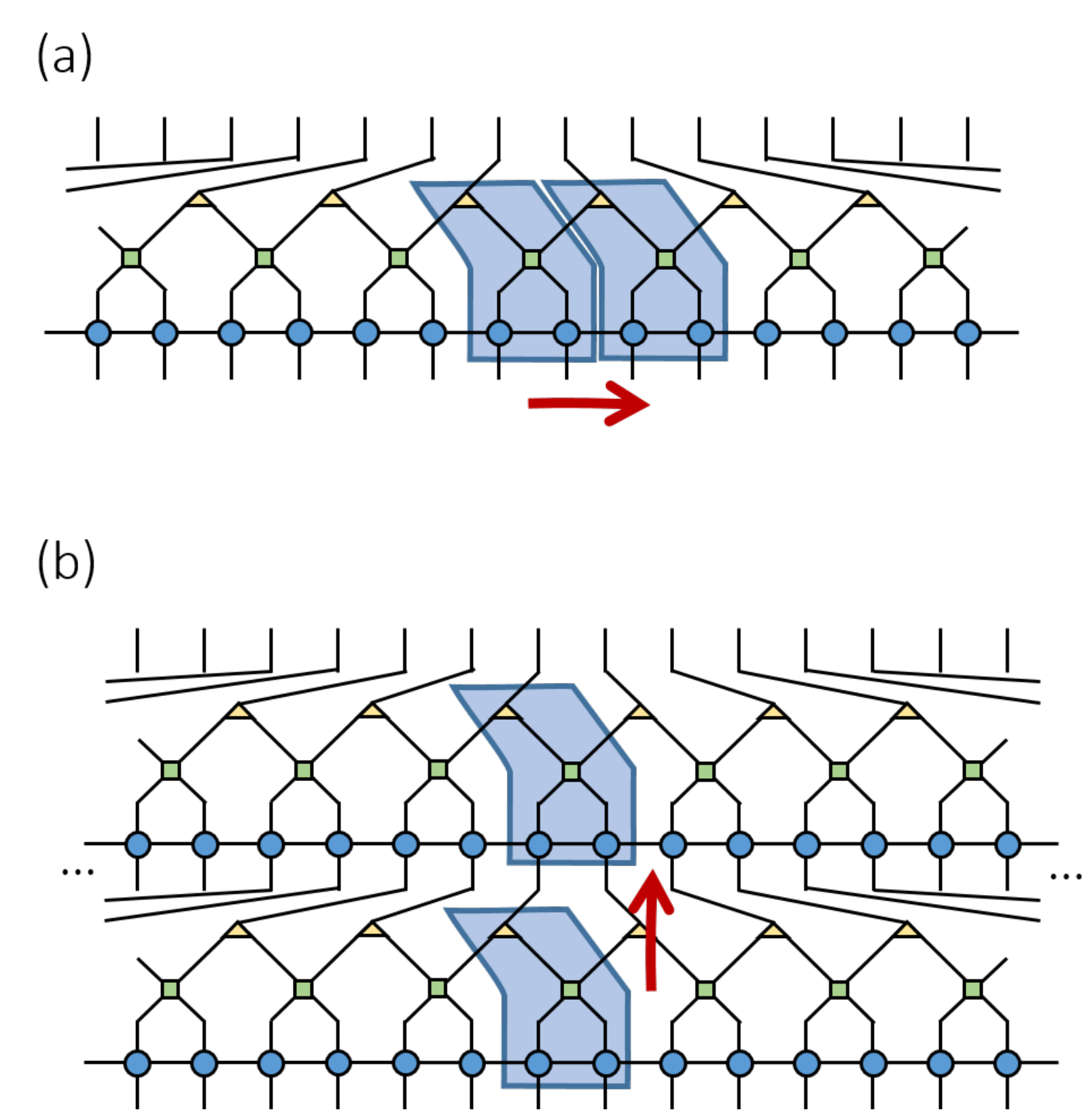}
\caption{
(a) A layer of euclidean MERA on the line, obtained by multiplying a small unit cell of tensors, is invariant under discrete translations.
(b) Since each layer of euclidean MERA is identical, the network is invariant under discrete scale transformations.
\label{fig:sym_line} 
}
\end{figure}

\textit{Translation symmetry.---} A layer $\mathcal{W}_{+}^{(m,m+1)}$ is a tensor network constructed as the infinite product of a basic building block in a way that it is explicitly invariant under discrete translations. Specifically, under translations by one site of the spin chain $m+1$ or, equivalently, two sites of the spin chain $m$, see Fig. \ref{fig:sym_line}(a). This \textit{network} translation invariance of the layer \textit{implies} that the linear map implemented by the layer is also translation invariant. In addition, we may choose (as we do next) to interpret this \textit{network} translation invariance of each layer of euclidean MERA as also implying that the distance between nearest neighbour spins is a constant ${\aUV}_{,m}$ within each spin chain, which may still depend on the integer $m$ that labels different spin chains. 
The translation symmetry of $k$ layers is limited to translations by one site of the spin chain placed above the top layer. Let this top spin chain have label $0$, and let $m \in [0,k]$. The network is invariant under the discrete transformation
\begin{equation}
s_m \rightarrow s_m, ~~~~x_{m,n}  \rightarrow  x_{m,n} + 2^{- m} = x_{m,n} + e^{-s_m}.
\end{equation}

We then \textit{promote} the discrete translation symmetry to continuous translation symmetry of the metric, by demanding that it be invariant under
\begin{equation}
s \rightarrow s, ~~~~x  \rightarrow  x + \lambda e^{-s}
\end{equation}
for an infinitesimal $\lambda$. 

\textit{Scale symmetry.---} The fact that all layers $\mathcal{W}_{+}$ of euclidean MERA are identical to each other, implies a second discrete symmetry of the network, namely discrete scale transformations, see Fig. \ref{fig:sym_line}(b). Indeed, consider the shift 
\begin{eqnarray}
s_m &\rightarrow& s_m + \log 2 = s_{m+1}, \\
x_{m,n} &\rightarrow& x_{m,n} = x_{m+1,n}.
\end{eqnarray}
As we did above with translation invariance, we can now \textit{promote} this discrete scale symmetry of the network to a continuous scale symmetry of the continuous metric by requiring that it be invariant under the transformation
\begin{equation}
s \rightarrow s + \log \lambda,~~~~x \rightarrow x,
\end{equation}
for an infinitesimally small $\lambda$.

We emphasize that these promotions of discrete network symmetries to continuous symmetries can be seen to be compatible with, but are not implied by, rules 1 and 2. It is a way of further restricting the metric. One could consider adding a rule 3 that says that in the presence of a discrete network symmetry, the continuous metric should also have a suitable continuous version of that symmetry. A problem with proceeding this way is that a discrete symmetry does not determine a continuous symmetry uniquely, so such rule 3 would be ambiguous and still require that the discrete network symmetry be promoted to some preferred choice of continuous symmetry.

To enforce the above two symmetries on the metric, it is convenient to use the fact that they correspond to translations $(z,r) \rightarrow (z,r+ \lambda)$ and to rescaling $(z,r) \rightarrow \lambda(z,r)$ in the coordinates $(z,r)$, to produce
\begin{equation} \label{eqap:gsymm}
dl^2 = \frac{C^2}{z^2} \left([\pm A^2 +B^2] dz^2 + 2B dzdr + dr^2\right), 
\end{equation}
which in the $(s,x)$ coordinates corresponds to 
\begin{eqnarray}
a &=& \frac{A}{\aUV},~~~b = x + \frac{B}{\aUV},~~~\Omega = \aUV C,~~~~
\end{eqnarray}
that is
\begin{eqnarray} \label{eqap:dlsxMN}
dl^2 &=& C^2\aUV^2\Bigg(\left[\pm \frac{A^2}{\aUV^2} + \left(x + \frac{B}{\aUV}\right)^2 \right]ds^2  ~~~~~~~~~\\
&& ~~~~~~~~~~~~ + 2\left(x + \frac{B}{\aUV}\right) ds dx + dx^2 \Bigg),~~~~~~~
\end{eqnarray}
with unknown signature ($\pm$ sign) and constants $A$, $B$, and $C$.

Imposing rule 1 through constraint \eqref{eqap:rule1sx} we then obtain euclidean signature (sign $+$), and the values $A=\aUV q$ and $B=0$. Moreover, imposing rule 2 through constraint \eqref{eqap:rule2sx} we obtain $C = 1$. This uniquely leads to $dl_{\MMplus}^2$ in Eq. \eqref{eqap:MMplus}, corresponding to the H$_2^p$ metric in \eqref{eqap:dlsx} with radius $\rad=\aUV q$.

\subsection{Relation to previous derivations}

Our second derivation of a continuous metric for the euclidean MERA on the line, based on symmetries, is similar but not equivalent to previous derivations of a continuous metric for the null MERA \cite{H1,H2,dS1,dS2}. 

Arguing in terms of symmetries alone (including that the lattice spacing $\aUV$ is constant throughout the network) we already arrive at a metric
\begin{equation}
dl^2 = \left(\frac{\aUV}{z}\right)^2 \left([\pm A^2 +B^2] dz^2 + 2B dzdr + dr^2\right), \end{equation}
for unknown signature (sign $\pm$) and constants $A, B$.  In order to determine the signature and set $B=0$ (that is, to recover the metric of H$_2^p$) and to further relate $A$ to $q/\aUV$ (that is, to establish that that the radius $\rad$ of H$_2^p$ is equal to $\aUV q$) we require rule 1. 

As mentioned above, H$_2^p$, dS$_2^p$, and L$_2^p$ have in common the symmetry group they inherit from AdS$_3^p$, which includes both translations and rescalings. Reasoning in terms of symmetries only (that is, without rule 1) we can attach any of the above geometries to the euclidean MERA. Therefore we conclude that rule 1 (or some other rule that goes beyond symmetry considerations) is essential in order to decide which of these three geometries should be assigned to the euclidean MERA.

\subsection{Lorentzian signature: time evolution on the line}
\label{subsect:line_lorentz}

Next we study the linear map $V$ obtained from a path integral on a strip of Poincare de Sitter spacetime dS$_2^p$ and the tensor network geometry of the lorentzian MERA on the line. The analysis is very similar to the one above for hyperbolic space H$_2^p$ and the euclidean MERA, and therefore we will proceed by sketching the argument and highlighting the differences with the previous case, to which we refer for further details.

As in Eq. \eqref{eqap:Vline}, the linear map $\tilde{V}$ is again defined in terms of a path integral on a strip $\mathcal{S}$ with boundaries at $z = z_{\inn}$ and $z=z_{\out}$ according to 
\begin{equation} \label{eqap:Vline} 
\bra{\varphi'(r)} \tilde{V} \ket{\varphi(r)} = \int [D\phi] e^{-i S[\phi(z,r)]},
\end{equation}
where now $S[\phi(z,r)]$ is the \textit{lorentzian} action. For a thin strip, the linear map reads $\tilde{V} \approx \mathbb{1} - \epsilon \tilde{Q}$
For a diagonal metric of the form
\begin{equation}
dl^2 = \Omega(z)^2 \left(-a(z)^2 dz^2 + dr^2 \right)
\end{equation}
this generator is 
\begin{eqnarray}
\tilde{Q} &=& i a(z) \int_{-\infty}^{\infty} dr~h(r) = i a(z)~H, \\
H &\equiv& \int_{-\infty}^{\infty} dr~h(r).
\end{eqnarray}
For dS$_2^p$ in coordinates $(z,r)$ we have a constant $a(z)$, namely
\begin{eqnarray}
a(z) = \rad,\\
\end{eqnarray}
and the metric and generator read:
\begin{eqnarray}  \label{eqap:dldS2p} 
dl^2_{\ddS2} &=& \frac{1}{z^2} \left(-\rad^2 dz^2 + dr^2\right), \\
\tilde{Q}_{\ddS2} &=& i\rad~H.  \label{eqap:QdS2p} 
\end{eqnarray}
The path integral on a thick strip $z\in[z_{\inn},z_{\out}]$ then produces the finite linear map 
\begin{eqnarray} 
\tilde{V}_{\ddS2} &=& \mathcal{P} \exp \left(-\int_{z_{\inn}}^{z_{\out}} dz ~\tilde{Q}_{\ddS2} \right)\\
&=& \exp \left(- i\rad H \int_{z_{\inn}}^{z_{\out}} ~dr\right) \\
&=& \exp \left(-i\rad\left(z_{\out} - z_{\inn}\right) H \right). \label{eqap:VdS2a}
\end{eqnarray}
corresponding to a real time evolution by and amount $\rad(z_{\out} - z_{\inn})$ of time. In terms of the dimensionless coordinates ($s,x$), with  $s = \log (z)$ and $x = r/(\aUV z)$,
the metric of dS$_2^p$ is no longer diagonal and reads
\begin{equation}
\label{eqap:dlsxdS}
dl^2_{\ddS2} = \aUV^2 \left(\left[\frac{-\rad^2}{\aUV^2}+x^2\right]ds^2 + 2xdsdx + dx^2\right).
\end{equation}
With the constant-$x$ identification of Hilbert spaces at the boundaries of the strip, see Eq. \eqref{eqap:identX}, the linear map   
\begin{equation} \label{eqap:Vline2dS} 
\bra{\varphi'(x)} V \ket{\varphi(x)} = \int [D\phi] e^{-iS[\phi(s,x)]},
\end{equation}
for a thin strip ($z_{\out} - z_{\inn} \equiv \epsilon \ll 1$) decomposes as $V \approx \mathbb{1} - \epsilon Q$ with generator $Q = iQ_0 - iQ_1$ for $Q_0$ and $Q_1$ given by Eqs. \eqref{eqap:Q0c}-\eqref{eqap:Q1c}, so that
\begin{eqnarray}
Q_{\ddS2} &=& i\left(\frac{\rad}{\aUV} H - D\right),\\
V_{\ddS2} &=& \mathcal{P} \exp\left(-\int_{s_{\inn}}^{s_{\out}} ds~Q_{\ddS2} \right)\\
&=& \exp \left( -  i(s_{\out}-s_{\inn})\left(\frac{\rad}{\aUV}H - D\right) \right), \label{eqap:VdS2b}
\end{eqnarray}
where (compare with Eq. \eqref{eqap:VV})
\begin{equation} \label{eqap:VVdS}
V_{\ddS2} = U(s_{\out}) ~\tilde{V}_{\ddS2} ~U(s_{\inn})^{\dagger}.
\end{equation}
A layer $\mathcal{W}_{-}$ of lorentzian MERA on the line is made of the product of $q$ lorentzian transfer matrices $\mathcal{T}_{-}$ (each of which has been built to implement a real time evolution $\mathcal{T}_{-} \approx e^{-iH}$)  followed by a layer $\mathcal{W}$ of null MERA,
\begin{equation}
\mathcal{W}_{-} \equiv \mathcal{W}~\mathcal{T}_{-}^{q},
\end{equation}
and implements the linear map
\begin{equation} \label{eqap:Wminx}
\mathcal{W}_{-} \approx 2^{iD} e^{iqH} = 2^{-i\left(q H-D\right)}.
\end{equation}

Given the discrete coordinates $(s_m,x_{m,n})$ from Eq. \eqref{eqap:sxdiscretemn} applied now to to label sites in the spin chains of a lorentzian MERA, and the candidate metric of Eq. \eqref{eqap:sxgeneric}, rule 1 implies
\begin{equation} \label{eqap:rule1sxdS}
\mathcal{P} \exp \left(-\int_{s_m}^{s_{m+1}} ds~Q(s) \right) = 2^{-i\left(q H + D\right)}.
\end{equation}
whereas rule 2 implies Eq. \eqref{eqap:rule2sx} as in the euclidean MERA. Then the same choice of functions $a(s,x)$, $b(s,x)$ and $\Omega(s,x)$ as in Eq. \eqref{eqap:sxABOmega} from the euclidean signature case, but with the choice of a lorentzian signature ($-$ sign) in the metric \eqref{eqap:sxgeneric}, leads to a solution of the constraints \eqref{eqap:rule1sxdS} (rule 1) and \eqref{eqap:rule2sx} with metric and generator
\begin{eqnarray} \label{eqap:MMmin}
dl^2_{\MMmin} &\equiv& \aUV^2 \left(\left[-q^2+x^2\right]ds^2 + 2x~ds ~dx + dx^2\right)\\
Q_{\MMmin} &\equiv&  -  i(q   H - D).
\end{eqnarray}
The metric corresponds to Poincare de Sitter spacetime dS$_2^p$ with radius $\rad = \aUV q$. The same conclusion is reached by first imposing translation and scale invariance (exactly in the same way as in the case of euclidean signature) and then enforce rules 1 and 2.

\subsection{Null signature: no time evolution on the line}
\label{subsect:line_null}

To compute the linear maps corresponding to a path integral on a strip of the light sheet gemoetry L$_2^p$ we will take the zero radius limit, $\rad\rightarrow 0$, of the derivation for either H$_2^p$ or dS$_2^p$. We just list the resulting objects. In the coordinates $(z,r)$ the metric $dl^2_{\LL2}$, generator $\tilde{Q}_{\LL2}$, and linear map $\tilde{V}_{\LL2}$ read
\begin{eqnarray}
dl^2_{\LL2} &=& \frac{dr^2}{z^2},~~~~~~~~~~~\\
\tilde{Q}_{\LL2} &=& 0,\\
\tilde{V}_{\LL2} &=& \mathbb{1}.
\end{eqnarray}
In the coordinates $(s,x)$ the metric $dl^2_{\LL2}$, generator $Q_{\LL2}$, and linear map  $V_{\LL2}$ read
\begin{eqnarray}
dl^2_{\LL2} &=& \aUV^2\left(x^2 ds^2 + 2x~ds~dx + dx^2 \right),~~\\
Q_{\LL2} &=& -iD,\\ 
V_{\LL2}(s_{\inn}, s_{\out}) &=& e^{i(s_{\out}-s_{\inn})D}.
\end{eqnarray}
Finally, the linear map $\mathcal{W}$ implemented by one layer of the null MERA, and the proposed continuous metric $dl^2_{\MMzero}$ for the geometry of the tensor network are
\begin{eqnarray}
\mathcal{W} &=& 2^{iD},\\
dl^2_{\MMzero} &=& \aUV^2\left(x^2 ds^2 + 2x~ds~dx + dx^2 \right).
\end{eqnarray}

We emphasize that, since a layer $\mathcal{W}$ of null MERA does not implement either euclidean nor lorentzian time evolution, from a path integral perspective the geometry of the null MERA is neither the hyperbolic space H$_2$ nor the de Sitter spacetime dS$_2$.

\subsection{Discrete sequence of tensor network geometries (II)}

\begin{figure}
\includegraphics[width=8.5cm]{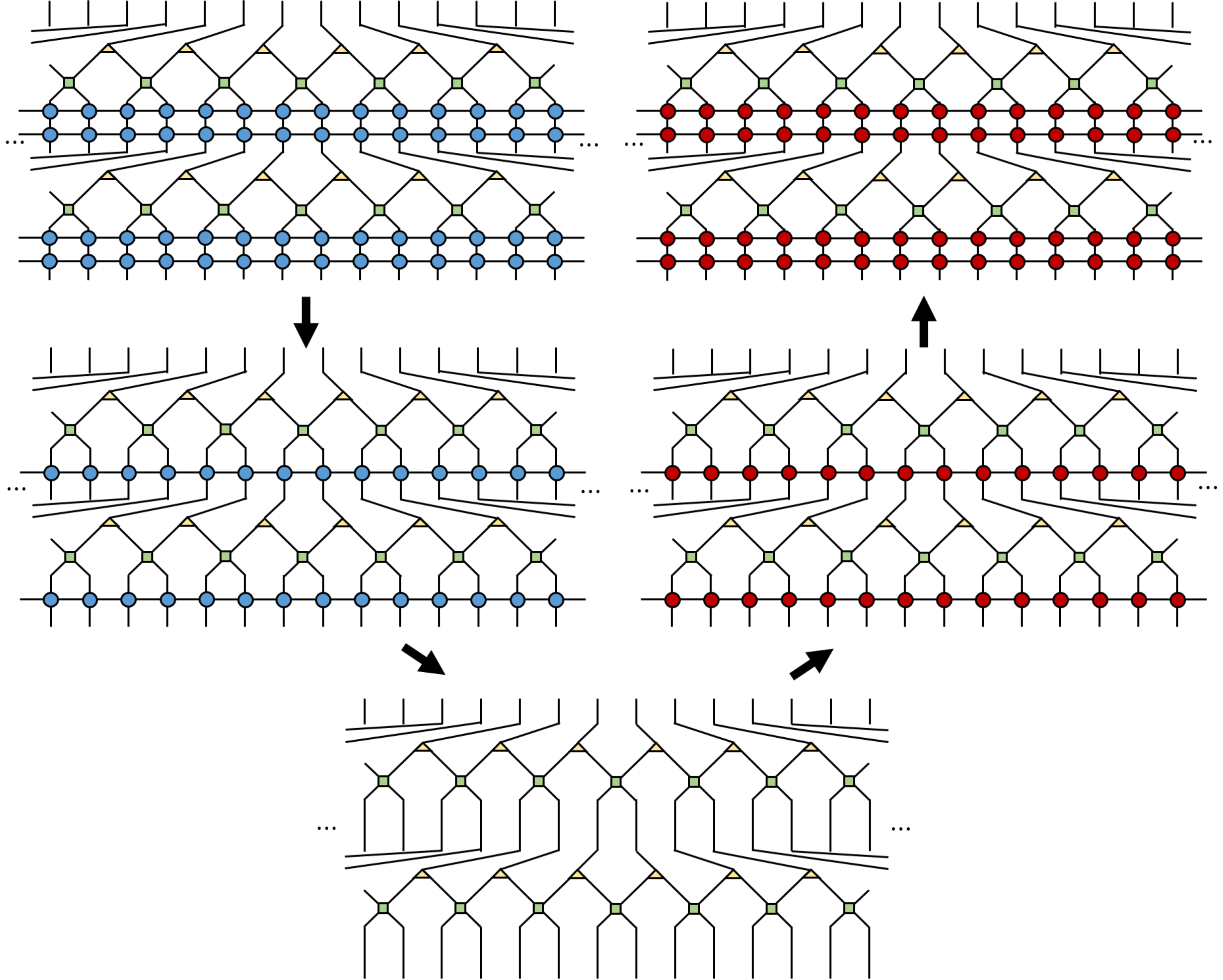}
\caption{
We can consider a sequence of hyperbolic planes H$_2^p$ for decreasing values of the (continuous) radius $\rad$, which has the light sheet L$_2^p$ as its $\rad \rightarrow 0$ limit, then continue with Poincare de Sitter spacetimes dS$_2^p$ for increasing values of the radius $\rad$, see Fig. \ref{fig:limitp}. The above discrete sequence of tensor networks mimics that for discrete values $\rad = \aUV q$ of the radius $\rad$, for $q = 0,1,2,\cdots$.
\label{fig:sequence_line}
}
\end{figure}

To summarize, we have seen that the euclidean, lorentzian, and null MERA tensor networks on the line correspond to a discrete version of a CFT path integral over the hyperbolic plane H$_2^p$, the Poincare de Sitter spacetime dS$_2^p$, and light sheet L$_2^p$. Recall that these three geometries can be embedded in the same ambient space AdS$_3^p$, where both H$_2^p$ and dS$_2^p$ become L$_2^p$ in the limit of a small radius $\rad \rightarrow 0$, see Figs. \eqref{fig:patch} and \eqref{fig:limitp}. We can similarly regard the null MERA $\mathcal{M}$ on the line as a particular case of either the euclidean MERA $\mathcal{M}_{+}$ or lorentzian MERA $\mathcal{M}_{-}$ on the line, namely when there are $q=0$ transfer matrices $\mathcal{T}$ or $\mathcal{T}_{-}$ (implementing euclidean or real time evolution) between layers $\mathcal{W}$, see Fig. \ref{fig:sequence_line}.

\begin{figure}
\includegraphics[width=\linewidth]{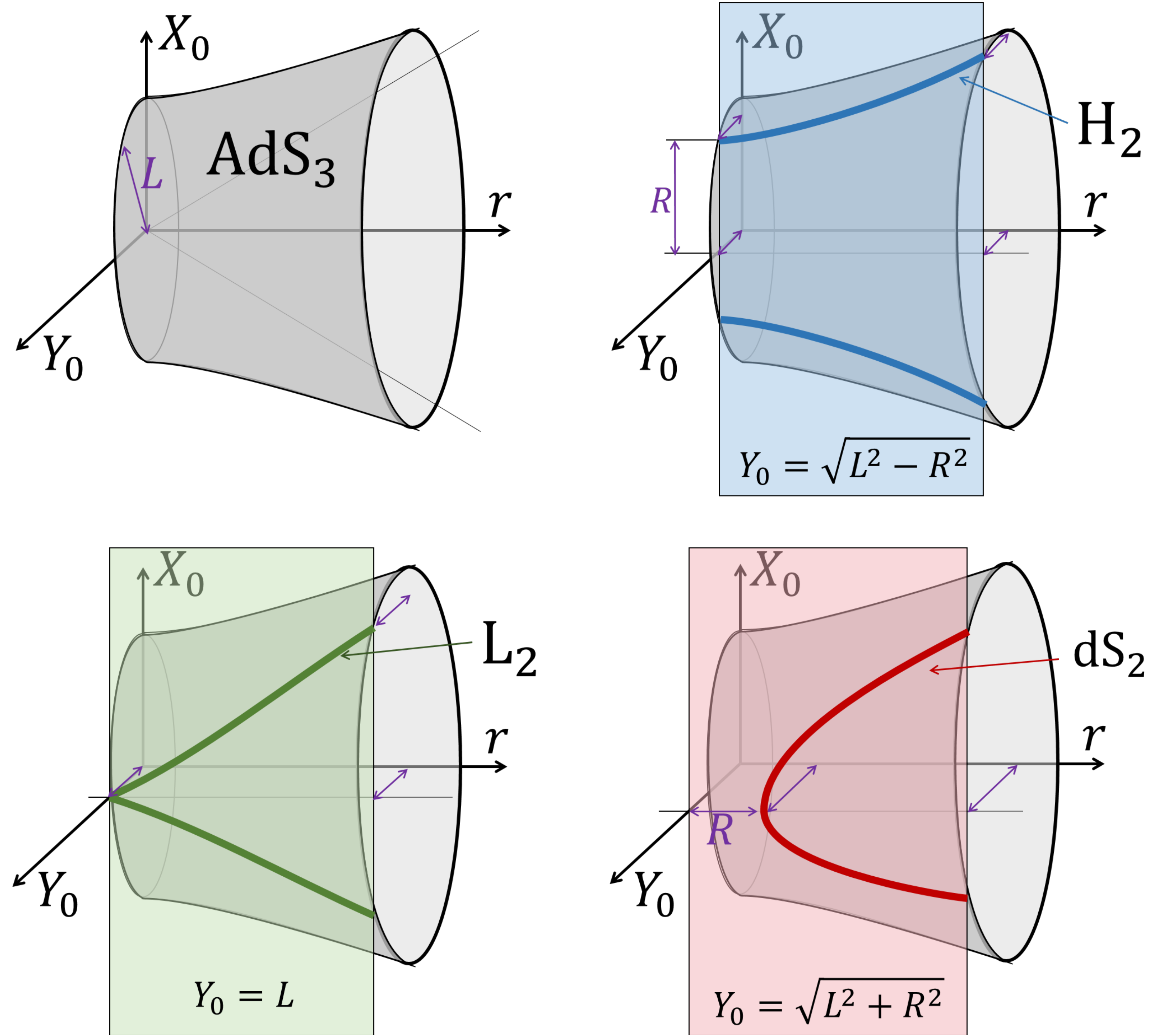}
\caption{
\label{fig:AdS3} 
Anti de Sitter spacetime AdS$_3$ with radius $L$, as embedded in Minkowski $\mathbb{R}^{2,2}$ with coordinates $(X_0,Y_0,X_1,X_2)$ (notice that $r\equiv \sqrt{{X_1}^2+{X_2}^2}$ is shown instead of $X_1$ and $X_2$, so as to obtain a three dimensional projection of $\mathbb{R}^{2,2}$; each point should be augmented to a circle of radius $r$ parameterized by $\theta \in [0,2\pi)$). AdS$_3$ corresponds to the constraint $-{X_0}^2 - {Y_0}^2 + r^2 = -L^2$. The plot also shows H$_2$, L$_2$ and dS$_2$ as the intersections of AdS$_3$ with hyperplanes defined by a constant value of $Y_0$, namely $Y_0 = \sqrt{L^2-\rad^2}$, $Y_0=L$, and $Y_0= \sqrt{L^2 + \rad^2}$, respectively. Each of these hyperplanes corresponds to a copy of Minkowski $\mathbb{R}^{1,2}$ as parameterized by $(X_0,X_1,X_2)$.
}
\end{figure}

\section{Appendix: Embedding in global anti de Sitter}
\label{sect:AdS}

In the context of the AdS/CFT correspondence, one may ask whether tensor networks such as MERA can be understood as a lattice realization of the holographic principle, in the sense of representing a two-dimensional slice of anti de Sitter spacetime AdS$_3$ \cite{H1,H2, dS1,dS2, dS3, dS4}. 

The three two-dimensional geometries under consideration ---hyperbolic plane H$_2$ (or hyperbolic disk), the light sheet L$_2$ (or light cone), and the de Sitter spacetime dS$_2$--- can also be embedded in AdS$_3$, which in turn can be embedded in flat $\mathbb{R}^{2,2}$, as briefly reviewed below. Therefore, MERA and the proposed euclidean and lorentzian versions of MERA can be understood as discrete versions of 2d CFT path integrals on different types of 2d slices of AdS$_3$. 

Notice, however, that there is no additional, emergent, holographic dimension in a two-dimensional representation of a path integral of a two-dimensional CFT! Therefore it is unclear that the ability to embed the two-dimensional geometries H$_2$, L$_2$, and dS$_2$ attached to the above MERA tensor networks in AdS$_3$ should be interpreted as evidence that the tensor networks are a realization of holography.

Let $X_0$ and $Y_0$ denote time coordinates and $X_1$ and $X_2$ denote space coordinates, such that the metric of $\mathbb{R}^{2,2}$ reads
\begin{equation}
dl^2 = - {dX_0}^2 - {dY_0}^2 + {dX_1}^2 + {dX_2}^2.
\end{equation}
Then AdS$_3$ is defined by the constraint
\begin{equation} \label{eq:AdS}
-{X_0}^2 - {Y_0}^2 + {X_1}^2 + {X_2}^2 = -L^2,
\end{equation}
where $L$ is the AdS radius, see Fig. \ref{fig:AdS3}.

Let us consider a parameterization of $\mathcal{R}^{2,2}$ in terms of new coordinates $(L,t,r,\theta)$ given by
\begin{eqnarray}  
X_0 &=& \sqrt{r^2+L^2} \sin (t/L), \label{eqap:AdS3a}\\
Y_0 &=& \sqrt{r^2+L^2} \cos (t/L), \label{eqap:AdS3b}\\
X_1 &=& r \cos (\theta), \label{eqap:AdS3c}\\
X_2 &=& r \sin (\theta), \label{eqap:AdS3d}.
\end{eqnarray}
Notice that for any fixed value of $L$,
\begin{eqnarray}
&&-{X_0}^2 - {Y_0}^2 + {X_1}^2 + {X_2}^2 \\
&=& -(r^2 + L^2)(\cos^2\theta + \sin^2\theta) + r^2 (\cos^2\theta + \sin^2\theta)~~~~~~\\
&=& -(r^2 + L^2) + r^2 = -L^2.
\end{eqnarray}
and thus this parameterization is consistent with the constraint \eqref{eq:AdS} defining AdS$_3$ of radius $L$. The induced metric reads
\begin{equation}
dl^2 = -\left((r/L)^2+1\right) dt^2 + \frac{dr^2}{(r/L)^2+1} + r^2 d\theta^2.
\end{equation}
As embedded in $\mathbb{R}^{2,2}$, the time direction $t$ of AdS$_3$ is compact, namely $t/L \sim t/L + 2\pi$. By removing such identification, we then obtain a time coordinate $t$ that can take any real value, $t \in (-\infty, \infty)$.

For visualization purposes, we also consider replacing the radial coordinate $r \in [0, \infty)$ in parameterization \eqref{eqap:AdS3a}-\eqref{eqap:AdS3d} with a new radial coordinate $\rho \in [0,\pi/2]$ given by
\begin{equation}
r/L = \tan(\rho),
\end{equation}
in terms of which the metric becomes
\begin{equation}
dl^2 = \frac{1}{\cos^2 \rho}\left(-dt^2 + d\rho^2 + L^2 \sin^2 \rho ~d\theta^2 \right).
\end{equation}
This is the \textit{soup can} representation of AdS$_3$, see Fig. \ref{fig:soup}. 

\subsection{Family of H$_2$ slices}

Let us introduce an alternative parameterization of $\mathbb{R}^{2,2}$ in terms of coordinates $(L,\rad, r, \theta)$ given by:
\begin{eqnarray}  
X_0 &=& \sqrt{\rad^2+r^2}, \label{eqap:H2a}\\
Y_0 &=& \sqrt{L^2-\rad^2},\label{eqap:H2b}\\
X_1 &=& r \cos (\theta), \label{eqap:H2c}\\
X_2 &=& r \sin (\theta), \label{eqap:H2d}, 
\end{eqnarray}
with $L \geq \rad > 0$, $r \geq 0$, $\theta \in [0,2\pi)$.
Notice that, again, a fixed value of $L$ is still compatible with constraint \eqref{eq:AdS} defining AdS$_3$ with radius $L$:
\begin{eqnarray}
&&-{X_0}^2 - {Y_0}^2 + {X_1}^2 + {X_2}^2 \\
&=& -(\rad^2 + r^2) - (L^2 - \rad^2) + r^2(\cos^2\theta + \sin^2\theta)~~~~\\
&=& -L^2.
\end{eqnarray}
Then, for a fixed value of $L$ and $\rad$ (with $\rad < L$), we also notice the time coordinate $Y_0$ in Eq. \eqref{eqap:H2b} is just a constant and that the remaining coordinates $(X_0,X_1,X_2)$ describe three-dimensional Minkowski spacetime $\mathbb{R}^{1,2}$ as in the previous section. Moreover, Eq. \eqref{eqap:H2a} can be interpreted as a constraint ${X_0}^2 = \rad^2 + r^2$ in $\mathbb{R}^{1,2}$, namely the constraint \eqref{eqap:embedding1b} defining H$_2$. Therefore, for each pair ($L,\rad$) with $\rad<L$, we have obtained a hyperbolic plane H$_2$  with radius $\rad$, with metric \eqref{eqap:dl1}, embedded in AdS$_3$ with radius $L$, see Fig. \ref{fig:AdS3}.

\begin{figure}
\includegraphics[width=\linewidth]{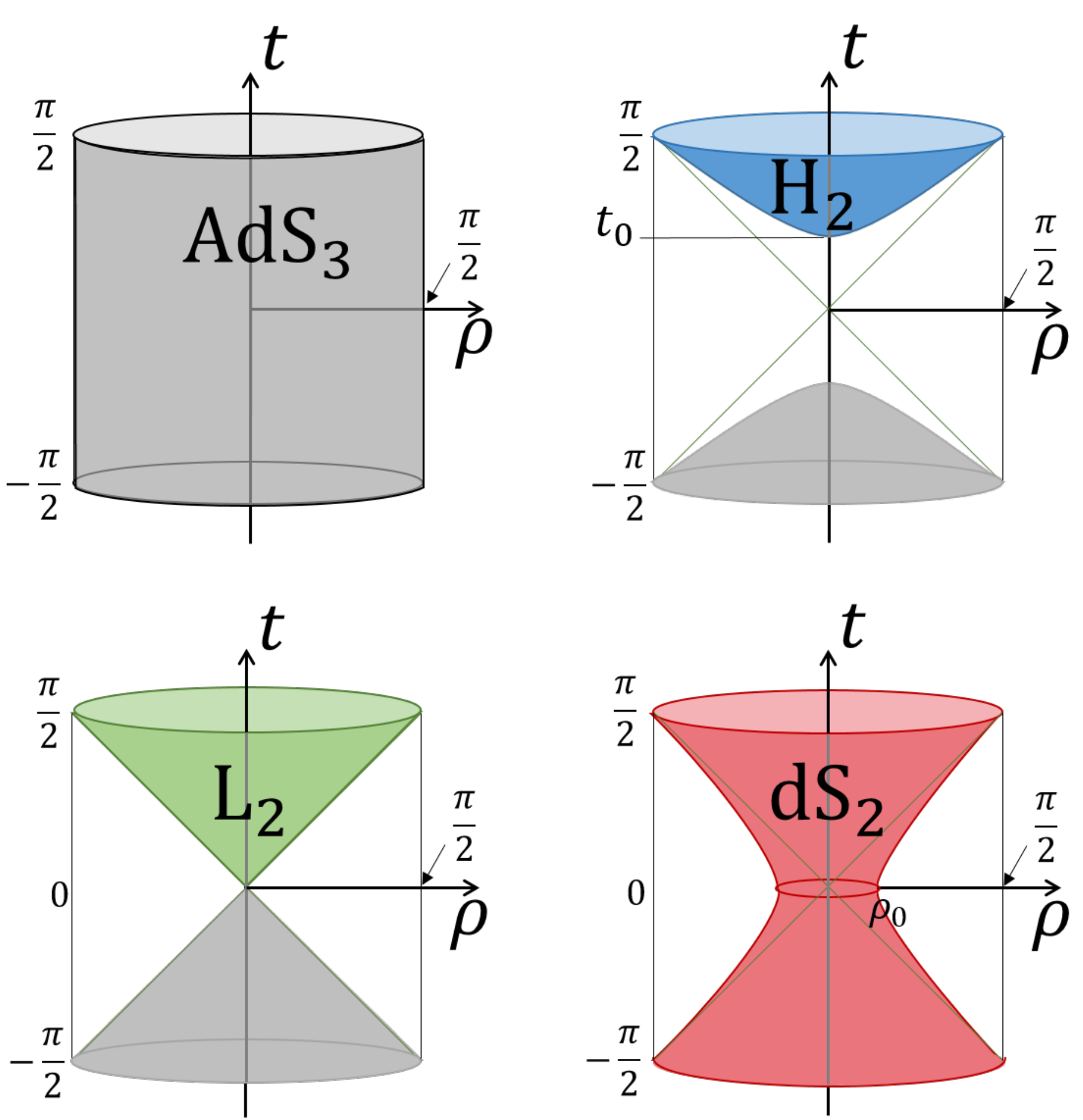}
\caption{AdS$_3$ in coordinates $(t,\rho,\theta$, where $t \in (-\infty,\infty)$ is the uncompactified time coordinate, $\rho \in [0,\pi/2)$ is a radial space coordinate, and $\theta \in [0,2\pi)$ is an angular space coordinate.
H$_2$, L$_2$, and dS$_2$ slices that hit the AdS$_3$ boundary $\rho=\pi/2$ at time $t=\pi/2$.
\label{fig:soup} 
}
\end{figure}

We can compare the two AdS$_3$ parameterizations \eqref{eqap:AdS3a}-\eqref{eqap:AdS3d} and \eqref{eqap:H2a}-\eqref{eqap:H2d} to conclude that
\begin{equation}
(r^2+L^2)\cos^2 (t/L) = L^2 - \rad^2
\end{equation}
or, in terms of the radial coordinate $\rho$,
\begin{eqnarray} \label{eqap:H2trho}
\cos (t/L) = \sqrt{1 - (\rad/L)^2} \cos  \rho.
\end{eqnarray}
The corresponding surface is indicated in Fig. \ref{fig:soup}. Notice that for $\rad=L$ we obtain $\cos(t/L) = 0$, that is, the time slice $t=\pi/2$.

\subsection{L$_2$ slice}

Let us introduce the parametrization
\begin{eqnarray}  
X_0 &=& \pm |r|, \label{eqap:L2a}\\
Y_0 &=& L, \label{eqap:L2b}\\
X_1 &=& r \cos (\theta), \label{eqap:L2c}\\
X_2 &=& r \sin (\theta), \label{eqap:L2d}, 
\end{eqnarray}
where we immediately recognize a three-dimensional manifold with coordinates $(L,r,\theta)$ embedded in $\mathbb{R}^{2,2}$ or, ignoring $Y_0=L$ in Eq. \eqref{eqap:L2b}, a two-dimensional manifold with coordinates $(r,\theta)$ embedded in $\mathbb{R}^{1,2}$, with \eqref{eqap:L2a} being equivalent to the constraint \eqref{eqap:embedding3} defining the light cone L$_2$, see Fig. \ref{fig:AdS3}. 

We can compare the two parameterizations \eqref{eqap:AdS3a}-\eqref{eqap:AdS3d} and \eqref{eqap:L2a} - \eqref{eqap:L2d} to conclude that
\begin{equation}
(r^2+L^2)\cos^2 (t/L) = L^2
\end{equation}
or, in terms of the radial coordinate $\rho$,
\begin{eqnarray} \label{eqap:L2trho}
\cos (t/L) = \cos  \rho.
\end{eqnarray}
The corresponding surface is indicated in Fig. \ref{fig:soup}. 

\subsection{Family of dS$_2$ slices}

We now introduce yet another parameterization of Minkowski $\mathbb{R}^{2,2}$ in terms of new coordinates $(L, \rad, r, \theta)$ this time given by
\begin{eqnarray}  
X_0 &=& \sqrt{r^2-\rad^2}, \label{eqap:dS2a}\\
Y_0 &=& \sqrt{L^2+\rad^2}, \label{eqap:dS2b}\\
X_1 &=& r \cos (\theta), \label{eqap:dS2c}\\
X_2 &=& r \sin (\theta), \label{eqap:dS2d}
\end{eqnarray}
where  $r \geq \rad > 0$, $\theta \in [0,2\pi)$. We can again check that a fixed value of $L$ defines AdS$_3$ with radius $|L|$ (constraint \eqref{eq:AdS}):
\begin{eqnarray}
&&-{X_0}^2 - {Y_0}^2 + {X_1}^2 + {X_2}^2 \\
&=& -(r^2 - \rad^2) - (L^2 + \rad^2) + r^2\cos^2\theta + r^2 \sin^2\theta~~~~\\
&=& -L^2.
\end{eqnarray}
Then, for a fixed value of $L$ and $\rad$, we also notice the time coordinate $Y_0$ in Eq. \eqref{eqap:dS2b} is just a constant and that the remaining coordinates $(X_0,X_1,X_2)$ describe three-dimensional Minkowski spacetime $\mathbb{R}^{1,2}$ as in the previous section. Moreover, Eq. \eqref{eqap:dS2a} can be interpreted as a constraint ${X_0}^2 = \rad^2 - r^2$ in $\mathbb{R}^{1,2}$, namely the constraint \eqref{eqap:embedding3b} defining dS$_2$. Therefore, for each pair ($L,\rad$) we have obtained a hyperbolic plane H$_2$ with radius $|\rad|$ and metric \eqref{eqap:dl3}, embedded in AdS$_3$ with radius $|L|$, see Fig. \ref{fig:AdS3}.  

We can compare the two AdS$_3$ parameterizations \eqref{eqap:AdS3a}-\eqref{eqap:AdS3d} and \eqref{eqap:dS2a}-\eqref{eqap:dS2d} to conclude that
\begin{equation}
(r^2+L^2)\cos^2 (t/L) = L^2 + \rad^2
\end{equation}
or, in terms of the radial coordinate $\rho$,
\begin{eqnarray} \label{eqap:dS2trho}
\cos (t/L) = \sqrt{1 + (\rad/L)^2} \cos  \rho.
\end{eqnarray}
The corresponding surface is indicated in Fig. \ref{fig:soup}. Notice that for $\rad=\infty$ we obtain $\cos(\rho) = 0$, that is, the AdS$_3$ boundary $\rho=\pi/2$.

\subsection{L$_2$ as the $\rad\rightarrow 0$ limit of H$_2$, dS$_2$}

Notice that the light cone geometry L$_2$ is recovered as the $\rad\rightarrow 0$ limit of both H$_2$ and dS$_2$. Indeed, for $\rad=0$, both the parametrization \eqref{eqap:H2a}-\eqref{eqap:H2d} of H$_2$ and the parametrization \eqref{eqap:dS2a}-\eqref{eqap:dS2d} of dS$_2$ coincide with the parametrization \eqref{eqap:L2a}-\eqref{eqap:L2d} of L$_2$, and the relation \eqref{eqap:H2trho} of H$_2$ and \eqref{eqap:dS2trho} of dS$_2$ agree with the relation \eqref{eqap:dS2trho} of dS$_2$. See Fig. \ref{fig:limit}(c).


\section{Generator of the linear map obtained through a path integral }
\label{sect:linear}

In this appendix we briefly review the parameterization of metric and the generator of linear map used in the appendices I and II above. These expressions have been justified in the Appendix of \cite{TNPathInt}.

\subsection{Euclidean}

Consider the generic euclidean metric
\begin{eqnarray}\label{eqap:ge1}
g_{\mu\nu}(\tau,x) &=& \left(
\begin{array}{cc}
g_{00}(\tau,x) & g_{01}(\tau,x) \\
g_{10}(\tau,x) & g_{11}(\tau,x)
\end{array} \right)\\
&=&\Omega^{2}(\tau,x)\left(\begin{array}{cc} a(\tau,x)^2+b(\tau,x)^2 & b(\tau,x)\\ b(\tau,x) & 1 \end{array} \right)~~~~~~\label{eqap:ge2} 
\end{eqnarray}
Then the path integral on a thin strip with boundaries $\Sigma$ and $\Sigma'$ at $\tau=\tau_0$ and $\tau= \tau_0+ \epsilon$ produces a linear map $V:\mathcal{H}_{\Sigma} \rightarrow \mathcal{H}_{\Sigma'}$ between the QFT Hilbert spaces $\mathcal{H}_{\Sigma}$ and $\mathcal{H}_{\Sigma'}$ at $\Sigma$ and $\Sigma'$. When point $(\tau_0,x) \in \Sigma$ is identified with point $(\tau_0+\epsilon, x) \in \Sigma'$, then the two Hilbert spaces $\mathcal{H}_{\Sigma}$ and $\mathcal{H}_{\Sigma'}$ can be identified accordingly and we refer to them simply as $\mathcal{H}$. The linear map $V:\mathcal{H} \rightarrow \mathcal{H}$ for small $\epsilon$ reads 
\begin{equation}
V \approx \mathbb{1} - \epsilon (Q_0 -  i Q_1) 
\end{equation}
where
\begin{eqnarray} \label{eqap:Q0}
Q_0 \equiv \int_{\Sigma} dx~a(\tau,x)~ h_L(\tau,x),\\
Q_1 \equiv \int_{\Sigma} dx~b(\tau,x)~ p_L(\tau,x),\label{eqap:Q1}
\end{eqnarray}
and where $h_L$ and $p_L$ are the Lorentzian Hamiltonian and momentum densities \textit{in flat spacetime}. For instance, for the free boson they read
\begin{eqnarray}
h_L &\equiv& \frac{1}{2}\left((\partial_t \phi)^2 + (\partial_x \phi)^2 \right),\\
p_L &\equiv& -\partial_t \phi \partial_x \phi.
\end{eqnarray}

Specializing to $a(\tau,x) = a$ and $b(\tau,x) = b$ we have
\begin{eqnarray}
Q_0 &=& a \int_{\Sigma} dx~h_{L} \equiv a ~H,~~~ (\mbox{Hamiltonian operator})~~~\\
Q_1 &=& b \int_{\Sigma} dx~p_L \equiv b ~P, ~~~(\mbox{momentum operator})~~~
\end{eqnarray}
and a finite map reads
\begin{equation}
V = \exp \left(-a H + i b P \right),
\end{equation}
as we expected.
Moreover, for $a=0$ and $b(\tau,x) = \alpha x$ we have
\begin{eqnarray}
Q_0 &=& 0,\\
Q_1 &=& b \int_{\Sigma} dx~x~p_L \equiv \alpha D ~~~ (\mbox{dilation operator})~~~
\end{eqnarray}
and a finite map reads
\begin{equation}
V = \exp \left(i \alpha D \right).
\end{equation}

\subsection{Lorentzian}

Consider now the generic lorentzian metric
\begin{eqnarray}\label{eqap:gl1}
g_{\mu\nu}(\tau,x) &=& \left(
\begin{array}{cc}
g_{00}(t,x) & g_{01}(t,x) \\
g_{10}(t,x) & g_{11}(t,x)
\end{array} \right)\\
&=&\Omega^{2}(t,x)\left(\begin{array}{cc} -a(t,x)^2+b(t,x)^2 & b(t,x)\\ b(\tau,x) & 1 \end{array} \right)~~~~~\label{eqap:gl2} 
\end{eqnarray}
Then the path integral on a thin strip with boundaries $\Sigma$ and $\Sigma'$ at $t=t_0$ and $t= t_0+ \epsilon$ produces a linear map $V:\mathcal{H}_{\Sigma} \rightarrow \mathcal{H}_{\Sigma'}$ between the QFT Hilbert spaces $\mathcal{H}_{\Sigma}$ and $\mathcal{H}_{\Sigma'}$ at $\Sigma$ and $\Sigma'$. When point $(t_0,x) \in \Sigma$ is identified with point $(t_0+\epsilon, x) \in \Sigma'$, then the two Hilbert spaces $\mathcal{H}_{\Sigma}$ and $\mathcal{H}_{\Sigma'}$ can again be identified accordingly and we refer to them simply as $\mathcal{H}$. The linear map $V:\mathcal{H} \rightarrow \mathcal{H}$ for small $\epsilon$ now reads 
\begin{equation}
V \approx \mathbb{1} - i\epsilon Q_0 + i \epsilon Q_1  
\end{equation}
where $Q_0$ and $Q_1$ are as defined above in the euclidean case.

Specializing to $a(\tau,x) = a$ and $b(\tau,x) = b$ the finite map reads
\begin{equation}
V = \exp \left(-i a H + i b P \right),
\end{equation}
as we expected. For $a=0$ and $b(\tau,x) = \alpha x$ we again have
\begin{equation}
V = \exp \left(i \alpha D \right).
\end{equation}

Notice that the generator $Q_1$ appears identically in both the euclidean and lorentzian cases, that is $V \approx \mathbb{1} + iQ_1$, whereas the generator $Q_0$ appears as
\begin{eqnarray}
V_E \approx \mathbb{1} - \epsilon Q_0,~~~V_L \approx \mathbb{1} - i \epsilon Q_0,
\end{eqnarray}
which is consistent with 
\begin{eqnarray}
V_E = \exp(-\tau H), ~~~~V_L = \exp(-i t H),
\end{eqnarray}
for the case of Hamiltonian evolution under the relation $\tau = i t$, as expected.

\section{Appendix: Low energy identification between spin chains of different size}
\label{sect:experiment}

A layer $\mathcal{W}$ of optimized MERA defines a linear map (also denoted by $\mathcal{W}$) between the Hilbert spaces of two periodic spin chains, one of size $N$ (bottom chain) and the other of size $N/2$ (top chain). We would like to characterize this linear map $\mathcal{W}$ as a discrete version of a linear map $V$ acting on the Hilbert space of a CFT on the circle, as a means to relate the layer $\mathcal{W}$ to a CFT path integral on a strip geometry compatible with the linear map $V$. 

Fortunately, we can use the techniques of Refs. \cite{SpinChain1, SpinChain2, SpinChain3}, which are based on work by Cardy \cite{Cardy} and by Koo and Saleur \cite{KooSaleur}, to relate a basis of \textit{low energy} states $\{\ket{\phi^{N}_{\alpha}}\}$ and $\{\ket{\phi^{N/2}_{\alpha}}\}$ on each of the two spin chains with CFT states $\ket{\phi_{\alpha}^{\CFT}}$, so that the matrix elements $\bra{\phi^{N/2}_{\beta}}\mathcal{W}\ket{\phi^{N}_{\alpha}}$  can be directly compared to the corresponding matrix elements $\bra{\phi^{\CFT}_{\beta}}V\ket{\phi^{\CFT}_{\alpha}}$ on the CFT. Below we explain how this is accomplished, step by step. Importantly, the three conjectured possible forms of $V$, namely
\begin{equation}
V = \left\{ \begin{array}{ll}
e^{-\frac{R}{r}H} = \sum_{\alpha} e^{-\frac{R}{r}E_{\alpha}} \proj{\phi^{\CFT}_{\alpha}} & (\mbox{H$_2$})\\
&\\
e^{0} =\sum_{\alpha} \proj{\phi^{\CFT}_{\alpha}} &(\mbox{L$_2$})\\
&\\
e^{-i\frac{R}{r}H} = \sum_{\alpha} e^{-i\frac{R}{r}E_{\alpha}} \proj{\phi^{\CFT}_{\alpha}} & (\mbox{dS$_2$})
\end{array}
\right.
\end{equation}
are all diagonal in the basis $\ket{\phi^{\CFT}_{\alpha}}$ of simultaneous eigenvectors of the Hamiltonian and momentum operators. In particular, in order to distinguish between the last two options, which only differ by complex phases, special care is needed to properly fix the arbitrary complex phase of each state $\ket{\phi_{\alpha}^{N}}$ and $\ket{\phi_{\alpha}^{N/2}}$ (which unavoidably appears as a result of the diagonalization of the Hamiltonian and momentum operators on $N$ and $N/2$ sites).

\subsection{Operator-state correspondence}

The operator-state correspondence of conformal field theory \cite{CFT1,CFT2,CFT3} implies that the scaling operators $\phi^{\CFT}_{\alpha}(x)$ of a given CFT are in one-to-one correspondence with the states $\ket{\phi^{\CFT}_{\alpha}}$ of the Hilbert space of that CFT on the circle. 

Let $\Delta_{\alpha}$ and $S_{\alpha}$ be the scaling dimension and conformal spin of the scaling operator $\phi^{\CFT}_{\alpha}$ and let $E_{\alpha}^{\CFT}$ and $P_{\alpha}^{\CFT}$ be the energy and momentum of a simultaneous eigenstate $\ket{\phi^{\CFT}_{\alpha}}$ of the Hamiltonian operator $H^{\CFT}$ and momentum operator $P^{\CFT}$
\begin{eqnarray} \label{eq:HP_CFT}
H^{\CFT} \equiv \int_{0}^{L} dx  ~h(x), ~~~~~P^{\CFT} \equiv \int_{0}^{L} dx  ~p(x), ~~~
\end{eqnarray}
on a circle of length $L$, where $h(x)$ and $p(x)$ are the hamiltonian and momentum densitites of the CFT. Then a first consequence of the operator-state correspondence is that
\begin{eqnarray} \label{eq:EP_CFT}
E^{\CFT}_{\alpha} = \frac{2\pi}{L}\left(\Delta_{\alpha} - \frac{c}{12} \right),~~~~~~
P^{\CFT}_{\alpha} = \frac{2\pi}{L} S_{\alpha}.~~~
\end{eqnarray} 
The above spectral relations are sometimes already sufficient in order to identify the state $\ket{\phi^{\CFT}_{\alpha}}$ with the corresponding operator $\phi^{\CFT}_{\alpha}$. However, the spectra of energies and momenta (equivalently, of scaling dimensions and conformal spins) generically contains degeneracies, that is, there are pairs $(E_{\alpha}^{\CFT},P_{\alpha}^{\CFT}) = (E_{\alpha'}^{\CFT},P_{\alpha'}^{\CFT})$ for $\alpha\not = \alpha'$. In order to correctly identify each state with its corresponding operator we can then use that the states $\ket{\phi^{\CFT}_{\alpha}}$ are organized in irreducible representations of the conformal group, or conformal towers, with Virasoro generators given by
\begin{eqnarray} \label{eq:Ln}
L^{\CFT}_n \equiv \frac{L}{(2\pi)^2} \int_0^{L} dx ~e^{+inx\frac{2\pi}{L}} T(x) + \frac{c}{24}\delta_{n,0},\\
\bar{L}^{\CFT}_n \equiv \frac{L}{(2\pi)^2} \int_0^{L} dx ~e^{-inx\frac{2\pi}{L}} \bar{T}(x) + \frac{c}{24}\delta_{n,0}, \label{eq:Lnbar}
\end{eqnarray}
where $T(x)$ and $\bar{T}(x)$ are the holomorphic and antiholomorphic components of the stress tensor,
\begin{equation}
T(x) \equiv 2\pi\frac{h(x) + p(x)}{2},~~~~~~~~~\bar{T}(x) \equiv 2\pi \frac{h(x) - p(x)}{2},
\end{equation}
or $h(x) = (T(x)+\bar{T}(x))/2\pi$ and $p(x)=(T(x)-\bar{T}(x))/2\pi$, and $c$ is the central charge of the CFT. For example, the ground state/identity state $\ket{\mathbb{1}^{\CFT}}$ relates to the stress tensor states $\ket{T^{\CFT}}$ and $\ket{\bar{T}^{\CFT}}$ through 
\begin{equation}\label{eq:L2I_CFT}
L_{-2}^{\CFT} \ket{\mathbb{1}^{\CFT}} = \sqrt{\frac{c}{2}}\ket{T^{\CFT}},~~~\bar{L}_{-2}^{\CFT} \ket{\mathbb{1}^{\CFT}} = \sqrt{\frac{c}{2}}\ket{\bar{T}^{\CFT}}.
\end{equation}
Thus we can use this expression in order to unambiguously identify states $\ket{T^{\CFT}}$ and $\ket{\bar{T}^{\CFT}}$ in the list $\{\ket{\phi_{\alpha}^{\CFT}}\}$ of simultaneous eigenstates of $H^{\CFT}$ and $P^{\CFT}$ and, importantly for our purposes, remove any spurious relative complex phase between states $\ket{\mathbb{1}^{CFT}}$, $\ket{T^{\CFT}}$, and $\ket{\bar{T}^{\CFT}}$. Indeed, notice that if $e^{i\varphi}$ is a random complex phase, both $\ket{T^{\CFT}}$ and $e^{i\varphi}\ket{T^{\CFT}}$ are equally valid simultaneous eigenvectors of $H^{\CFT}$ and $P^{\CFT}$, but only the choice $e^{i\varphi}=1$ is compatible with Eq. \eqref{eq:L2I_CFT}. Thus we can use Eq. \eqref{eq:L2I_CFT} to set the relative complex phase $e^{i\varphi}$ between $\ket{\mathbb{1}^{\CFT}}$ and $\ket{T^{\CFT}}$ to $1$ (and similarly for the relative complex phase between $\ket{\mathbb{1}^{\CFT}}$ and $\ket{\bar{T}^{\CFT}}$).
More generally, any descendant state $\ket{\phi^{\CFT}_{\alpha}}$ in the conformal tower of the ground state/identity primary state $\ket{\mathbb{1}^{\CFT}}$ is equal to some known linear combination of strings of operators $L_n$'s and $\bar{L}_{n}$'s acting on $\ket{\mathbb{1}^{\CFT}}$, which we can use to unambiguously identify the descendant state and remove any spurious relative complex phase within this conformal tower. 

Moreover, given any other primary operator $\chi^{\CFT}$, the Virasoro generators $L_n$ and $\bar{L}_{n}$ similarly connect the corresponding primary state $\ket{\chi^{\CFT}}$ with all its descendant states $\ket{\phi^{\CFT}_{\alpha}}$, which leads to their unambiguous identification with operators $\phi^{\CFT}_{\alpha}$ and the removal of spurious relative complex phases within the conformal tower. 
For instance, for the Ising CFT, this allow us to identify each state $\ket{\phi^{\CFT}_{\alpha}}$ as a concrete descendant of one of its three primary states: the identity primary state $\ket{\mathbb{1}^{\CFT}}$, the spin primary state $\ket{\sigma^{\CFT}}$ and the energy density primary state $\ket{\epsilon^{\CFT}}$. Moreover, any spurious relative complex phase within each of the conformal towers has been removed.

The generators of the conformal group, the Virasoro generators $L_n$ and $\bar{L}_n$ connect states within each conformal tower and that allowed us to unambiguously identify each state $\ket{\phi_{\alpha}^{\CFT}}$ with its corresponding scaling operators $\phi^{\CFT}_{\alpha}$ as well as to eliminate a spurious complex phase (introduced during the diagonalization of $H^{\CFT}$ and $P^{\CFT}$) \textit{within} each conformal tower. However, we are left with a spurious complex phase relative of any conformal tower with respect to the identity tower. Given a primary operator $\chi^{\CFT}$, we can remove the remaining spurious complex phase $e^{i\phi_{\chi}}$ between the ground state $\ket{\mathbb{1}^{\CFT}}$ (and its descendant states) and the primary state $\ket{\chi^{\CFT}}$ (and its descendant states) using known CFT matrix elements such as \cite{SpinChain3}
\begin{eqnarray} \label{eq:2p_CFT}
\bra{\chi^{\CFT}} \chi^{\CFT,0} \ket{\mathbb{1}^{\CFT}} = 
\left(\frac{2\pi}{L}\right)^{\Delta_{\chi}},~~~~~~
\end{eqnarray}
where we assumed for simplicity that $\chi^{\CFT}$ has conformal spin $S_{\alpha}=0$, and where
\begin{equation}
\chi^{\CFT,0}_{\alpha} \equiv \frac{1}{L} \int_0^L dx~\chi^{\CFT}_{\alpha}(x)
\end{equation}
is the zero Fourier mode of the primary operator $\chi^{\CFT}_{\alpha}$. In this way, all complex phases between all the states in the CFT have been fixed, and we are only left with a global complex phase.

\subsection{ $H_n^{\CFT}$ instead of $L_n^{\CFT}$} 

For later reference, we point out that the relations in the example of Eq. \eqref{eq:L2I_CFT} are equivalent to the less standard expressions
\begin{equation} \label{eq:H2I_CFT}
H_{-2}^{\CFT} \ket{\mathbb{1}^{\CFT}} = \sqrt{\frac{c}{2}}\ket{T^{\CFT}},~~~H_{2}^{\CFT} \ket{\mathbb{1}^{\CFT}} = \sqrt{\frac{c}{2}}\ket{\bar{T^{\CFT}}},
\end{equation}
in terms of the Fourier mode $H_n^{\CFT}$ of the Hamiltonian density $h(x)$, 
\begin{eqnarray} \label{eq:Hn_CFT}
H_n^{\CFT} &\equiv& \frac{L}{2\pi} \int_0^{L} dx ~e^{+inx\frac{2\pi}{L}} h(x)\label{eq:Hn}\\
&=& L_n + \bar{L}_{-n}  -\frac{c}{12}\delta_{n,0},
\end{eqnarray}
where we note that
\begin{equation}
H_0^{\CFT} = L_0^{\CFT}+\bar{L}_{0}^{\CFT} -\frac{c}{12} = \frac{L}{2\pi} H^{\CFT}.
\end{equation}
More generally, the identification of states $\ket{\phi_{\alpha}^{\CFT}}$ with their corresponding operators $\phi^{\CFT}_{\alpha}$ and the removal of spurious relative  complex phases within a conformal tower can be conducted using the Fourier modes $H_n^{\CFT}$ of the Hamiltonian density $h(x)$ instead of the Fourier modes $L_n$ and $\bar{L}_n$ of the stress tensor components $T(x)$ and $\bar{T}(x)$.

Here we notice that although $H_n^{\CFT}$ in Eq. \eqref{eq:Hn} acts as a concrete linear combination of the two Virasoro generators $L_n^{\CFT}$ and $\bar{L}^{\CFT}_{-n}$, we can always split the result of acting with $H_n^{\CFT}$ on a given energy/momentum eigenstate $\ket{\phi_{\alpha}^{\CFT}}$ into \textit{higher energy} and \textit{lower energy} contributions, which for $n < 0$ will correspond to having acted on $\ket{\phi_{\alpha}^{\CFT}}$ with $L_n$ and $\bar{L}_{-n}$, respectively (and for $n>0$ will correspond to having acted on $\ket{\phi_{\alpha}^{\CFT}}$ with $\bar{L}_{-n}$ and $L_{n}$, respectively). Indeed, both $L_n$ and $\bar{L}_{n}$ lower the energy (by $n 2\pi/L$) for $n>0$ and increase the energy (by $n 2\pi/L$) for $n<0$.

\subsection{One spin chain}

All the above manipulations referred to the CFT in the continuum. On the lattice, we can also identify each simultaneous eigenstate $\ket{\phi_{\alpha}}$ of a critical quantum spin chain Hamiltonian $H = \sum_{j=1}^N h_j$ and one-site translation $T=e^{-iP}$ operators,
\begin{eqnarray}
H \ket{\phi_{\alpha}} = E_{\alpha} \ket{\phi_{\alpha}},~~~T \ket{\phi_{\alpha}} = e^{-iP_{\alpha}}\ket{\phi_{\alpha}},
\end{eqnarray}
with its corresponding CFT scaling operators $\phi_{\alpha}^{\CFT}$, and eliminate spurious complex phases, as we did in the continuum. For that purpose we will use \textit{approximate} lattice versions of several continuum expressions mentioned above. First, as pointed out by Cardy \cite{Cardy}, on the lattice the low energy spectra of energies and momenta resemble the CFT result \eqref{eq:EP_CFT} by replacing the length $L$ of the circle with the number of sites $N$ in the periodic quantum spin chain, 
\begin{eqnarray} \label{eq:EP}
E_{\alpha} \approx \frac{2\pi}{N}\left(\Delta_{\alpha} - \frac{c}{12} \right),~~~~~~
P_{\alpha} = \frac{2\pi}{N} S_{\alpha}.~~~
\end{eqnarray} 
Here $\approx$ indicates that we are neglecting non-universal sub-leading contributions which typically scale as $O(N^{-1-\gamma})$ for some $\gamma>0$. In some cases, this already allows us to relate a lattice state $\ket{\phi_{\alpha}}$ with its corresponding scaling operator $\phi^{\CFT}_{\alpha}$. More generally, however, we need to resort to the Koo-Saleur formula \cite{KooSaleur}, which provides a lattice version of the Virasoro generators $L_n^{\CFT}$ and $\bar{L}_n^{\CFT}$ in Eqs. \eqref{eq:Ln}-\eqref{eq:Lnbar} or, following Refs. \cite{SpinChain1, SpinChain2}, the equivalent but more convenient lattice version $H_n$ of the Fourier modes $H_n^{\CFT}$ of the Hamiltonian density in Eq. \eqref{eq:Hn_CFT}, namely
\begin{eqnarray} \label{eq:Hn}
H_n  &\equiv& \frac{N}{2\pi} \sum_{j=1}^{N} e^{+inj\frac{2\pi}{N}} h_j, ~~~ H_0 = \frac{N}{2\pi}H.
\end{eqnarray}
We can use this lattice Fourier modes $H_n$ similarly as in the continuum. For instance, the approximate lattice version of Eq. \eqref{eq:H2I_CFT} is
\begin{equation} \label{eq:H2I}
H_{-2}  \ket{\mathbb{1}} \approx \sqrt{\frac{c}{2}}\ket{T },~~~H_{2}  \ket{\mathbb{1}} \approx \sqrt{\frac{c}{2}}\ket{\bar{T}},
\end{equation}
which we can use to unambiguously identify the lattice states $\ket{T}$ and $\ket{\bar{T}}$ by acting on the ground state $\ket{\mathbb{1}}$, while at the same time removing any spurious complex phase between these states. In this way we are only left with a spurious complex phase $e^{i\varphi_\chi}$ between the identity tower and the tower of each other primary state $\ket{\chi}$. The next step is to follow Ref. \cite{SpinChain3} in order to identify a lattice version $\chi_j$ (acting around site $j$) of each CFT primary operator $\chi^{\CFT}(x)$, then build its zero Fourier mode
\begin{equation}
\chi^{0} \equiv \frac{1}{N} \sum_{j=1}^N \chi_{j},
\end{equation}
where again for simplicity we assumed $S_{\chi}=0$, and then use the approximate, lattice version of Eq. \eqref{eq:2p_CFT},
\begin{eqnarray} \label{eq:2p}
\bra{\chi } \chi^{0} \ket{\mathbb{1} } \approx 
\left(\frac{2\pi}{L}\right)^{\Delta_{\chi}},~~~~~~
\end{eqnarray}
to remove the complex phase $e^{i\varphi_{\chi}}$. For instance, for the Ising model, the lattice version of the spin primary operator $\sigma^{\CFT}(x)$ and of the energy density primary operator $\epsilon^{\CFT}(x)$ are, approximately, given by \cite{SpinChain3}
\begin{eqnarray}
\sigma_j \approx \sigma^{x}_j,~~~~\epsilon_{j+1/2} \approx \sigma^{x}_j \sigma^x_{j+1} - (\sigma^z_j + \sigma^z_{j+1})/2,
\end{eqnarray}
and we eliminate spurious complex phases $e^{i\varphi_{\sigma}}$ and $e^{i\varphi_{ \epsilon}}$ by requiring that the matrix elements $\bra{\sigma} \sigma^{0} \ket{\mathbb{1}}$ and $\bra{\epsilon} \epsilon^{0} \ket{\mathbb{1}}$ are approximately as expected in the CFT, namely
\begin{equation}
\bra{\sigma } \sigma^{0} \ket{\mathbb{1} } = \left(\frac{2\pi}{L}\right)^{1/8},~~~~~~\\
\bra{\epsilon } \epsilon^{0} \ket{\mathbb{1} } = \left(\frac{2\pi}{L}\right)^{1}.
\end{equation}

\subsection{Two spin chains}

Finally, let us consider two spin chains of size $N$ and $N/2$. Using the lattice Fourier modes $H_n$ on each chain, we can identify simultaneous eigenstates $\ket{\phi_{\alpha}^N}$ of $H$ and $P$ on $N$ sites with simultaneous eigenstates $\ket{\phi_{\alpha}^{N/2}}$ of $H$ and $P$ on $N/2$ sites 
\begin{equation}
 \ket{\phi_{\alpha}^{N}} \sim \ket{\phi_{\alpha}^{N/2}}.
\end{equation}
through the identifications $\ket{\phi_{\alpha}^{N}} \sim \ket{\phi^{\CFT}_{\alpha}}$ and $\ket{\phi_{\alpha}^{N/2}} \sim \ket{\phi^{\CFT}_{\alpha}}$. We can also use $H_n$ and the matrix elements of the type \eqref{eq:2p} to eliminate all relative complex phases within the low energy states of $N$ sites and of $N/2$ sites separately. Then there is still one relative complex phase $e^{i\varphi^{N\rightarrow N/2}}$ (independent of $\alpha$ and $\beta$) that affects our identification of states $\ket{\phi_{\alpha}^{N}}$ on $N$ spins with states $\ket{\phi_{\alpha}^{N/2}}$ of $N/2$ spins. As a result, the matrix element
\begin{equation}
\mathcal{W}_{\alpha\beta} \equiv \bra{\phi^{N}_\alpha} \mathcal{W} \ket{\phi^{2N}_{\beta }}
\end{equation}
is proportional to this arbitrary complex phase $e^{i\varphi^{N\rightarrow N/2}}$, which we eliminate by demanding that $\mathcal{W}_{\mathbb{1}\mathbb{1}} = \bra{\mathbb{1}^{N/2}}\mathcal{W}\ket{\mathbb{1}^{N}}$ be positive. 

A numerical exploration then shows that, after a consistent choice of reference frame across different spin chains (see below), the linear map $\mathcal{W}$ acts on low energy states as the identity map,
\begin{equation} \label{eq:Walphabeta}
\mathcal{W}_{\alpha\beta} \approx  \delta_{\alpha\beta}.
\end{equation}

Specifically, Table I shows the order of magnitude of $|\mathcal{W}_{\alpha\beta} - \delta_{\alpha \beta}|$ for the 17 lowest energy eigestates of a coarse-grained version of the Ising model, when $\mathcal{W}$ connects the low energy states of a spin chain with $N=8$ sites to low energy states of a spin chain with $N/2=4$ sites. Here, each site is described by a vector space of dimension $\chi=8$ and effectively represents dozens of spins of the original Ising model.

\begin{widetext}
\begin{table}[b]
\begin{tabular}{|r|lllllllllllllllll|}
\hline
  & $I$ & $\sigma$ & $\varepsilon$ & $\partial\sigma$ & $\overline{\partial}\sigma$ & $T$ & $\partial\varepsilon$ & $\overline{\partial}\varepsilon$ & $\overline{T}$ & $\partial\overline{\partial}\sigma$ & $\partial^2\sigma$ & $\overline{\partial}^2 \sigma$ & $\partial\overline{\partial}\varepsilon$ & $\partial T$ & $\partial^2 \varepsilon$ & $\overline{\partial}^2\varepsilon$ & $\overline{\partial}\overline{T}$ \\
\hline
$I$ & $10^{-5}$ & $0$ & $10^{-2}$ & $0$ & $0$ & $0$ & $0$ & $0$ & $0$ & $0$ & $0$ & $0$ & $10^{-3}$ & $0$ & $0$ & $0$ & $0$ \\ 
$\sigma$ & $0$ & $10^{-5}$ & $0$ & $0$ & $0$ & $0$ & $0$ & $0$ & $0$ & $10^{-3}$ & $0$ & $0$ & $0$ & $0$ & $0$ & $0$ & $0$ \\ 
$\varepsilon$ & $10^{-2}$ & $0$ & $10^{-5}$ & $0$ & $0$ & $0$ & $0$ & $0$ & $0$ & $0$ & $0$ & $0$ & $10^{-5}$ & $0$ & $0$ & $0$ & $0$ \\ 
$\partial\sigma$ & $0$ & $0$ & $0$ & $10^{-5}$ & $0$ & $0$ & $0$ & $0$ & $0$ & $0$ & $0$ & $0$ & $0$ & $0$ & $0$ & $0$ & $0$ \\ 
$\overline{\partial}\sigma$ & $0$ & $0$ & $0$ & $0$ & $10^{-5}$ & $0$ & $0$ & $0$ & $0$ & $0$ & $0$ & $0$ & $0$ & $0$ & $0$ & $0$ & $0$ \\ 
$T$ & $0$ & $0$ & $0$ & $0$ & $0$ & $10^{-4}$ & $0$ & $0$ & $0$ & $0$ & $0$ & $0$ & $0$ & $0$ & $0$ & $10^{-3}$ & $0$ \\ 
$\partial\varepsilon$ & $0$ & $0$ & $0$ & $0$ & $0$ & $0$ & $10^{-4}$ & $0$ & $0$ & $0$ & $0$ & $0$ & $0$ & $0$ & $0$ & $0$ & $10^{-3}$ \\ 
$\overline{\partial}\varepsilon$ & $0$ & $0$ & $0$ & $0$ & $0$ & $0$ & $0$ & $10^{-4}$ & $0$ & $0$ & $0$ & $0$ & $0$ & $10^{-3}$ & $0$ & $0$ & $0$ \\ 
$\overline{T}$ & $0$ & $0$ & $0$ & $0$ & $0$ & $0$ & $0$ & $0$ & $10^{-4}$ & $0$ & $0$ & $0$ & $0$ & $0$ & $10^{-3}$ & $0$ & $0$ \\ 
$\partial\overline{\partial}\sigma$ & $0$ & $10^{-3}$ & $0$ & $0$ & $0$ & $0$ & $0$ & $0$ & $0$ & $10^{-5}$ & $0$ & $0$ & $0$ & $0$ & $0$ & $0$ & $0$ \\ 
$\partial^2\sigma$ & $0$ & $0$ & $0$ & $0$ & $0$ & $0$ & $0$ & $0$ & $0$ & $0$ & $10^{-3}$ & $10^{-3}$ & $0$ & $0$ & $0$ & $0$ & $0$ \\ 
$\overline{\partial}^2 \sigma$ & $0$ & $0$ & $0$ & $0$ & $0$ & $0$ & $0$ & $0$ & $0$ & $0$ & $0$ & $10^{-3}$ & $0$ & $0$ & $0$ & $0$ & $0$ \\ 
$\partial\overline{\partial}\varepsilon$ & $10^{-3}$ & $0$ & $10^{-5}$ & $0$ & $0$ & $0$ & $0$ & $0$ & $0$ & $0$ & $0$ & $0$ & $10^{-4}$ & $0$ & $0$ & $0$ & $0$ \\ 
$\partial T$ & $0$ & $0$ & $0$ & $0$ & $0$ & $0$ & $0$ & $10^{-3}$ & $0$ & $0$ & $0$ & $0$ & $0$ & $10^{-3}$ & $0$ & $0$ & $0$ \\ 
$\partial^2 \varepsilon$ & $0$ & $0$ & $0$ & $0$ & $0$ & $0$ & $0$ & $0$ & $10^{-3}$ & $0$ & $0$ & $0$ & $0$ & $0$ & $10^{-3}$ & $0$ & $0$ \\ 
$\overline{\partial}^2\varepsilon$ & $0$ & $0$ & $0$ & $0$ & $0$ & $10^{-3}$ & $0$ & $0$ & $0$ & $0$ & $0$ & $0$ & $0$ & $0$ & $0$ & $10^{-3}$ & $0$ \\ 
$\overline{\partial}\overline{T}$ & $0$ & $0$ & $0$ & $0$ & $0$ & $0$ & $10^{-3}$ & $0$ & $0$ & $0$ & $0$ & $0$ & $0$ & $0$ & $0$ & $0$ & $10^{-3}$ \\ 
\hline
\end{tabular} 
\caption{Order of magnitude of matrix elements of $V - I$, labelled by CFT operators. Matrix elements $\le 10^{-10}$ are shown as ``$0$''. The asymmetry in the $\partial^2 \sigma, \overline{\partial}^2\sigma$ sector is due to using one of these matrix elements to break a momentum degeneracy in the coarse system.}
\end{table}

\end{widetext}

\subsection{Consistent reference frame across spin chains}

\begin{figure}
\includegraphics[width=\linewidth]{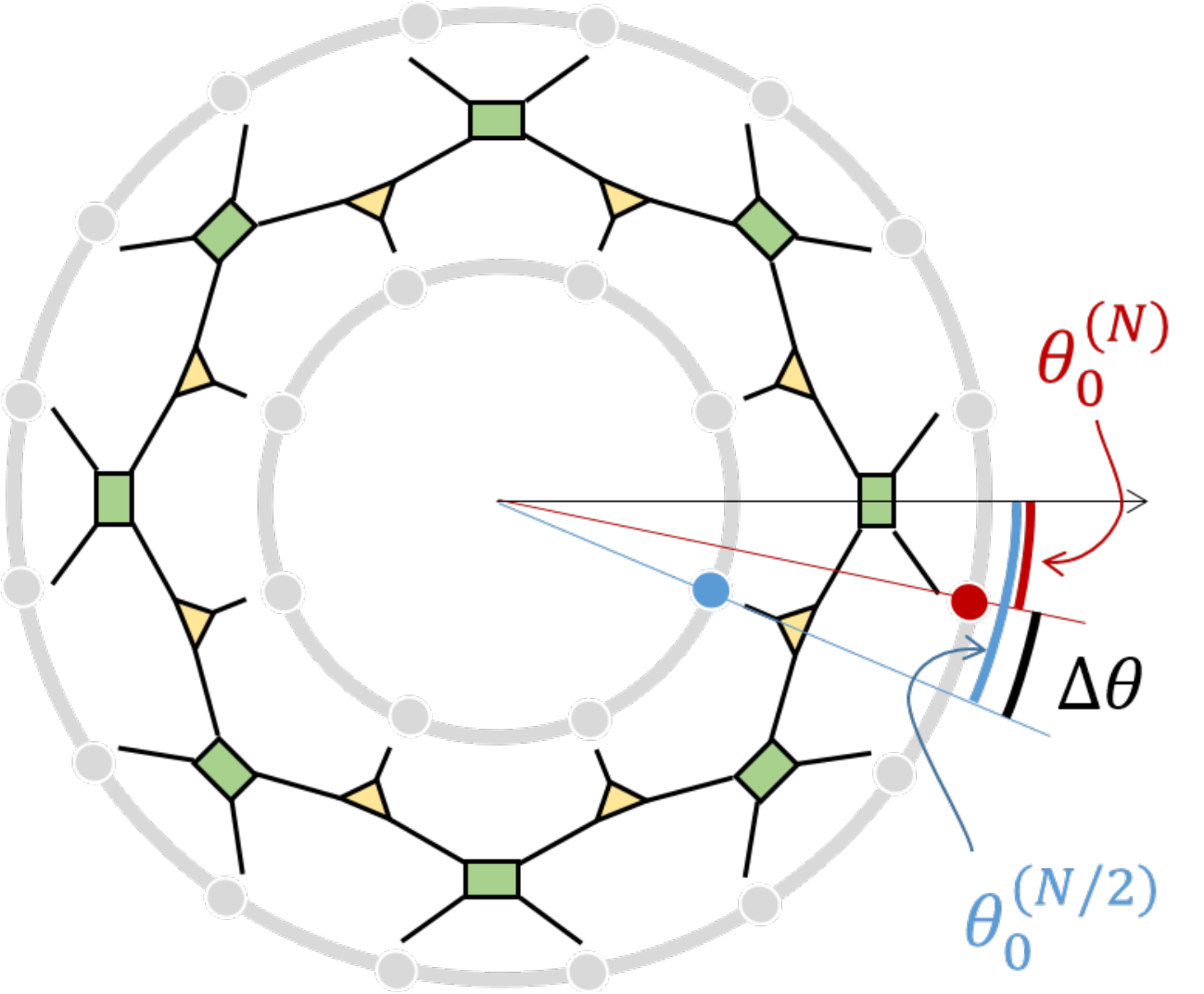}
\caption{
The difference $\Delta \theta$ in the reference angles $\theta_N$ and $\theta_{N/2}$ for the $N$ spin and the $N/2$ spin chains is of $1/2 \times 2\pi/N$ (half of a one-site translation in the $N$ spin chain). 
\label{fig:angle} 
}
\end{figure}

In order to characterize the action of one layer $\mathcal{W}$ of optimized MERA as a map between the low energy subspaces of two critical quantum spin chain of sizes $N$ and $N/2$, eventually resulting in \eqref{eq:Walphabeta} above, we made a convenient choice of the origin of the angle $\theta$ measuring the position of the spins in the chains. For a spin chain of size $N = 2^{T}$ for some integer $T>2$, we chose the position of the spins to be at angles
\begin{equation}
\theta^{(N)}_j = \frac{2\pi}{N} \left( j + \frac{1}{2} \right),
\end{equation}
see Figs. \eqref{fig:layer} and \eqref{fig:angle}, so that the Hamiltonian Fourier modes $H_n$ are not actually taken as in Eq. \eqref{eq:Hn} but as
\begin{eqnarray} \label{eq:Hnhalf}
H_n^{(N)}  &\equiv& \frac{N}{2\pi} \sum_{j=1}^{N} e^{+in\theta^{(N)}_j} h_j\\
&=&\frac{N}{2\pi} \sum_{j=1}^{N} e^{+in\left(j+ \frac{1}{2}\right)\frac{2\pi}{N}} h_j.
\end{eqnarray}
The reason for this choice, which places the origin $\theta = 0$ half way between two lattice sites, it that it can be chosen consistently for chains of size $N$ and $N/2$. 
Indeed, had we chosen $H_n^{(N)} = \frac{N}{2\pi} \sum_{j=1}^{N} e^{+in j\frac{2\pi}{N}} h_j$, and thus $H_n^{(N/2)} = \frac{N}{4\pi} \sum_{j=1}^{N/2} e^{+in j\frac{4\pi}{N}} h_j$, which amounts to considering that the $j=0$ sites is at angle $0$ for both spin chains, then we would observe that $\mathcal{W}$ effectively implements a rotation by angle $\Delta \theta \equiv 1/2 \times 2\pi/N$ in mapping the spin chain of size $N$ to the spin chain of size $N/2$, in such a way that its matrix elements now would be
\begin{equation} \label{eq:Walphabeta2}
\mathcal{W}_{\alpha\beta} \approx \delta_{\alpha\beta} e^{i\Delta\theta S_{\alpha}},
\end{equation}
where $S_{\alpha}$ is the conformal spin of state $\ket{\phi^{N}_{\alpha}}$.

To summarize, if we choose a consistent reference frame for $\theta$ across spin chains of different sizes, then the layer $\mathcal{W}$ of optimized MERA is seen to act as the identity map between low energy subspaces. If, instead, in going from one spin chain to another one we implicitly apply a translation by an angle $\Delta\theta$ (by changing the reference frame), then the same layer $\mathcal{W}$ of optimized MERA implements this translation by an angle $\Delta\theta$. Importantly, in none of these cases, the linear map $\mathcal{W}$ adds a Boltzmann weight $e^{-\mu E_{\alpha}}$ or a complex phase $e^{-i\mu E_{\alpha}}$ to state $\ket{\phi_{\alpha}}$ (for any measurable $\mu >0)$, as it would happen in an euclidean or lorentzian path integral.


\begin{thebibliography}{99}
\bibitem{normalization} The local Hamiltonian term $h_n$ is normalized (by a constant shift and a rescaling, $h_n \rightarrow a + b ~h_n$, where the constants $a$ and $b$ are independent of the size $N$ of the spin chain) such that the low energy spectrum matches that of a CFT on a circle of length $N\aUV$, or radius $r=N\aUV/2\pi$, where $\aUV$ is the lattice spacing.
\bibitem{complex} If $\ket{\psi^{N}_{\alpha}}$ is a normalized eigenstate of $H^{(N)}$, with $H^{(N)}\ket{\psi^{N}_{\alpha}} = E^{N}_{\alpha}\ket{\psi^{N}_{\alpha}}$, then so is $e^{i\varphi_{\alpha}^N}\ket{\psi^{N}_{\alpha}}$ for any angle $\varphi_{\alpha}^N \in [0,2\pi)$. Therefore by diagonalizing $H^{(N)}$ we obtain (normalized) energy eigenvectors $\ket{\psi^{N}_{\alpha}}$ that are only defined up to an arbitrary complex phase $e^{i\varphi_{\alpha}^N}$.
\end{thebibliography}
\end{document}